\documentclass{ametsocV6.1}

\nolinenumbers
\let\internallinenumbers\relax

\usepackage{enumitem} 
\usepackage{dsfont}

\title{Explainable AI for Geoscientific Regression: Evaluating Sensitivity and Attribution Methods with the Lorenz-63 System and Applications\thanks{This Work has been submitted to {\it Artificial Intelligence for the Earth Systems}. Copyright in this Work may be transferred without further notice.}}

\authors{Ieuan Higgs\aff{a,b}\correspondingauthor{Ieuan Higgs, i.c.higgs@reading.ac.uk}
Todd Jones,\aff{a} 
Kieran Hunt,\aff{b,c} 
Anna-Louise Ellis\aff{d} 
}

\affiliation{\aff{a}{Department of Computer Science, University of Reading, UK}\\
\aff{b}{Department of Meteorology, University of Reading, UK}\\
\aff{c}{National Centre for Atmospheric Science (NCAS), UK}\\
\aff{d}{Met Office, UK}}

\abstract{
As artificial intelligence (AI) moves from research prototypes to operational tools in Earth system science, trust in its predictions becomes increasingly important. Explainable AI (XAI) offers insight into how AI models make predictions. However, many influential XAI methods were developed or popularised for classification, and their interpretation for regression, which dominates geoscience, is less well established. A central obstacle is the lack of ground truth for judging explanations.\
We show that, for a dynamical system with known governing equations, matched references can be built directly from the physics. We extend established sensitivity methods to provide physical references for both sensitivity and attribution, separately for methods with implicit and explicit baselines, all computed from the governing equations independently of the network. Applied to an AI emulator of the Lorenz-63 system, these references show that method families answer distinct questions rather than approximating one ``true'' explanation. The emulator's sensitivities and attributions can depart most from their physical references where it forecasts best, so predictive skill is not a substitute for the physical realism of what a model has learned.\
We then review XAI applications to Earth system regression, organised by method family, extending to issues an idealised system cannot address: computational scalability, strongly correlated high-dimensional predictors, and reliance on expert judgement where no physical reference exists. We also discuss how an XAI method can be mathematically valid yet only partially relevant to a given stakeholder group. We close with future directions and practical recommendations for responsible XAI use in geoscientific modelling and forecasting.
}

\begin{document}

\maketitle

%
%
%
%
%
%

\statement Artificial intelligence (AI) is increasingly used in weather and climate forecasting, yet the reasoning behind its predictions is difficult to interpret. Explainable AI methods aim to reveal which inputs drive a prediction, but there is rarely an objective way to verify their explanations. Using the Lorenz-63 model, a classic chaotic system with known equations, we derive physics-based reference explanations against which these methods can be tested. Different methods prove to answer different questions, and explanations can be least reliable where the AI forecasts best, so forecast accuracy alone cannot justify trust in an explanation. A review of Earth science applications extends these lessons to larger systems. Future work should benchmark explanation methods against physical references in operational forecasting models.

%








\section{Introduction}
The application of artificial intelligence (AI)\footnote{Artificial intelligence (AI) and machine learning (ML) are often used interchangeably in meteorology and public discourse. Strictly, AI is an umbrella term encompassing designed intelligences such as rule-based and expert systems as well as machine-learning algorithms. ML is a subset of AI in which algorithms automatically identify and generalise patterns from data, producing models that make predictions, classifications, or other inferences on new, unseen inputs without being explicitly programmed with task-specific rules.} 
in geosciences and meteorology is now widespread \citep{zhao2024artificial}. AI techniques have been integrated into, or even fully emulate, tasks ranging from numerical weather prediction \citep[NWP;][]{lang2024aifs}, data assimilation \citep[DA;][]{bonavita2020machine, bocquet2024accurate}, nowcasting \citep{pirone2023short}, climate modelling \citep{nguyen2023climax}, and hazard detection \citep{jones2023ai, cheng2024deep}. 
As AI transitions from research settings to operational forecasting \citep{moldovan2026aifs}, establishing confidence and trust in model outputs becomes increasingly desirable. Reflecting this, the EU AI Act imposes transparency obligations on high-risk AI systems, encouraging the adoption of explainable AI (XAI) techniques to ensure accountability and regulatory compliance \citep{eu_ai_act_2023}.
Towards building this confidence and trust, XAI is increasingly recognised as a desirable emerging trend for Earth system sciences \citep{boukabara2021outlook, sun2022review, zhao2024artificial, schiller2025artificial}, allowing stakeholders to interrogate model behaviours in support of high-stakes decision-making.


The development of AI models differs fundamentally from that of traditional physics-based systems.
A traditional system is built from components primarily grounded in physical laws and first principles, represented in a form amenable to discretisation and efficient numerical solution \citep{coiffier2011fundamentals}.
When this is impractical, due to limited resolution, computational constraints, or insufficient process understanding, parameterisations approximate phenomena occurring below the model's grid spacing, such as cloud formation, convection, and boundary-layer turbulence \citep[e.g.,][]{christensen2022parametrization}. 
In contrast, AI models ``learn'' patterns directly from data, often without explicit representation of the underlying physical processes \citep{zhang2025machine}. Although their inputs and outputs are observable, their internal decision-making is often not readily interpretable owing to the complexity and scale of their underlying architectures. This limitation is not unique to AI and is arguably characteristic of many sufficiently complex models exhibiting emergent behaviour, including NWP systems.
However, due to their complexity and difference in development approach, AI models are frequently regarded as ``black boxes.'' 
This is particularly concerning for high-stakes decision-making contexts common in Earth system applications: issuing storm warnings \citep{economou2016use}, coordinating responses to extreme weather \citep{coughlan2016action}, and projecting agricultural yields \citep{corcoran2023current} all rely heavily on trustworthy model output. 
When replacing or augmenting these high-stakes systems with AI, it is highly desirable to ensure that the AI component is at least explainable, if not directly interpretable.

The field of explainable AI (XAI) emerged from the need to better understand AI decision-making, shifting focus from evaluating prediction errors/outputs to understanding the functional relationships and internal mechanisms which link inputs to outputs.
To date, much XAI development has concentrated on classification tasks \citep[e.g.,][]{mathews2019explainable, letzgus2022towards, mamalakis2022investigating, gevaert2022explainable}, such as image recognition \citep[e.g., object detection in medical imaging,][]{pintelas2020explainable}, natural language processing \citep[e.g., sentiment analysis,][]{bodria2020explainability}, and structured tabular data \citep[e.g., credit scoring and fraud detection,][]{talaat2024toward, awosika2024transparency}.
While XAI methods for classification can be useful in meteorological applications with a natural categorical framing \citep[e.g., storm severity category or event detection,][]{mcgovern2019making, kondylatos2022wildfire, liu2023evaluation}, many core meteorological tasks are regression problems (i.e., they predict a continuous variable such as a temperature field).

Interpreting XAI outputs in Earth system contexts is challenging for both classification and regression, since models must represent nonlinear interactions between coupled physical processes from high-dimensional data with complex spatio-temporal dependencies \citep{silva2024limitations}. However, classification offers two clear advantages for interpretation: First, the implicit knowledge associated with each class provides a discrete context, allowing model behaviour to be understood in terms of decision boundaries \citep{ebert2020evaluation, letzgus2022towards}; 
Second, class probabilities provide an indication of model confidence that can give context to an explanation, although they should not necessarily be interpreted as calibrated uncertainty estimates.

In contrast to classification, regression models have no class boundaries or discrete decisions to anchor an explanation. Their continuous outputs respond directly to changes in the inputs, so the magnitude and direction of the model response become interesting objects to interpret. This is particularly relevant when outputs are spatially and temporally structured physical quantities with well-defined units. Explanations must therefore characterise not only which inputs influence a prediction, but also how changes in those inputs relate to changes in the predicted quantity. Many general-purpose and classification-oriented XAI methods carry over to regression mathematically, but several difficulties become consequential for effective interpretation in a regression setting: baseline choice, correlated or off-manifold inputs, aggregation across space or time, and loss of sign information can all distort the reported magnitude or direction, issues we return to later. Forecasting applications add a further complication in autoregressive settings, where the predicted state is fed back as the input for the next step, so errors and biases can propagate through space and time, affecting both predictions and explanations. Regression is central to Earth system modelling, yet there is little guidance on how to interpret explanations of the AI models used for it.

This article explores the landscape of XAI methods applied to regression tasks in Earth system sciences, using a chaotic system as a testbed to ground methodological concepts before examining how these techniques have been applied in practice. We structure the article as follows:
Section~\ref{sec:XAI-method_overview} overviews terminology and definitions for a selection of prominent XAI techniques.
Section~\ref{sec:XAI-L63} introduces the AI emulation of the Lorenz-63 (L63) model, an archetypal chaotic system widely used to investigate non-linear dynamics.
Section~\ref{sec:illustrate_XAI_methods} details several input feature-based XAI methods and demonstrates their application to the L63 system. Section~\ref{sec:eval_metrics} shows a novel evaluation of XAI method agreement to a set of physically derived analogues. 
Section~\ref{sec:XAI-review} reviews a wider set of XAI applications used in  regression-focused Earth system contexts, and offers a perspective on the varying needs of differing stakeholders.
Finally, Section~\ref{sec:trends_and_future_directions} summarises the current state of XAI usage for regression in Earth science, discusses key challenges and future directions, and concludes with practical recommendations.

\section{Overview of XAI nomenclature}
\label{sec:XAI-method_overview}
Explainable Artificial Intelligence (XAI) encompasses methods developed to improve the understanding of AI and ML model behaviour, which is particularly useful for modern deep learning models whose complex internal structure make decision-making difficult to interpret directly \citep{mamalakis2020explainable}.

The terminology around ``interpretability'' and ``explainability'' remains neither standardised nor fixed, despite widespread use. The two terms are often used interchangeably, and distinctions drawn between them vary across fields, with many differences reflecting terminology choices as much as substantive conceptual disagreements \citep{bacsaugaouglu2022review, jiang2024interpretable}. 
In the interest of adopting consistent working definitions, we follow \citet{flora2024machine}: interpretability refers to the degree to which a model and its components can be understood directly, whereas explainability refers to the degree to which a black-box model can be approximately understood through post-hoc methods. 
This work focuses on post-hoc methods for understanding emergent behaviour in complex regression models, making explainability the more suitable framing here. Post-hoc benchmarks and evaluation also offer greater stability given rapidly developing AI models, as they can often be readily applied to newly published models in a timely, systematic manner \citep{brocker2026verification}.

Beyond these definitions, numerous authors stress that these terms do not refer to a single, unified notion \citep{lipton2018mythos}, and that meaning often depends on the application domain and stakeholder \citep{rudin2019stop}. 
The suitability of XAI methods also depends on the modelling task. Many widely used techniques were developed or popularised in classification settings, where their outputs have commonly been interpreted in relation to properties such as discrete decision boundaries, class-specific evidence, or well-defined ``other'' or ``null'' class \citep{letzgus2022towards}. 
Regression settings (e.g., dynamical forecast systems) lack this discrete structure and instead require explanations that characterise how continuous-valued variables contribute to model behaviour across multiple spatial and temporal scales within high-dimensional input and output spaces.
In Earth science, explainability is further shaped by physical constraints and domain knowledge, which provide context for judging whether a model's behaviour, or an explanation of it, appears ``physically reasonable.'' However, such assessments can still be subjective: an explanation aligning with a scientist's prior understanding may still fail to reflect the model's true decision-making process, and reasonable interpretations may differ between experts, especially in complex settings \citep{mamalakis2022investigating}.

XAI techniques are commonly categorised according to several characteristics, including the degree of interpretability (model-intrinsic or post-hoc approaches), the scope of explanations (global vs. local), and the underlying methodology used to generate explanations:

\begin{itemize}
    \item {\bf Transparent (model-intrinsic) or post-hoc (applied after training)} approaches distinguish approaches by whether a model's behaviour can be read from its trained structure, or whether a further method must be applied to expose it. Intrinsic methods, such as linear regression or decision trees, are inherently transparent since their weights or rules can be directly inspected and understood. Post-hoc methods instead operate externally on already-trained black-box models, approximating input-output relationships without exposing internal structure, and thus enhance explainability rather than interpretability \citep[following][]{flora2024machine}.
    \item {\bf Global vs. local} explanations capture different aspects of behaviour. Global explanations describe the model's overall behaviour, identifying general patterns and key features driving predictions across the dataset. Local explanations focus on individual predictions, clarifying why the model made a specific decision for a particular instance. The perspectives are complementary: global explanations aid system-level understanding, while local explanations give actionable insight at the level of individual predictions.
    \item {\bf Underlying methodological distinctions} are important. 
    Sensitivity-based methods assess how small input changes affect the output, revealing where the model is most responsive to perturbations.
    Attribution methods assign scores to input features reflecting their relative contribution to the predicted output.
    Other approaches include surrogate models (which approximate complex models with simpler, interpretable ones), and counterfactual explanations (which identify minimal input changes that would alter a prediction). 
\end{itemize}

Many XAI methods combine these characteristics. For example, an attribution method might provide both local and global insights depending on how results are aggregated, or a surrogate model (with intrinsic interpretability) may be used post-hoc to explore feature sensitivity. Which characteristics matter most depends on the type of insight sought and what is feasible given the ML model's (lack of) direct interpretability.

\section{Setup for an illustrative chaotic dynamical model for regression emulation and XAI}
\label{sec:XAI-L63}
To ground the subsequent review, we first illustrate the contrasting behaviour of several standard XAI methods in a controlled, shared setting (Sect.~\ref{sec:illustrate_XAI_methods}), allowing their properties to be compared directly (Sect.~\ref{sec:eval_metrics}) before surveying the wider literature (Sect.~\ref{sec:XAI-review}). 

We adopt a simple regression setup: a neural network trained to forecast continuous state variables in a dynamical system, reflecting a common framing in geoscientific applications. We use the L63 \citep{lorenz1963deterministic} system as our physical-system testbed\footnote{The L63 system was initially derived as a truncated model of Rayleigh-Bénard convection in a two-dimensional fluid flow.}, as it continues to be widely used in weather and climate science as the archetypal chaotic system providing a clear, tractable, deterministic demonstration of sensitivity to initial conditions and non-linear, aperiodic behaviour \citep{carrassi2018data}.
The instantaneous state of the L63 system is described by three state variables:
\begin{equation}
\label{eq:l63_state}
    \mathbf{x} = \begin{pmatrix} x \\ y \\ z \end{pmatrix},
\end{equation}
and the L63 model is given by three nonlinear ODEs: 
\begin{equation}
    \begin{aligned}
    \frac{dx}{dt} &= \sigma (y - x), \\
    \frac{dy}{dt} &= x(\rho - z) - y, \\
    \frac{dz}{dt} &= xy - \beta z.
    \end{aligned}
    \label{eq:L63}
\end{equation}
These have been implemented with parameters: $\sigma = 10$, $\rho = 28$ and $\beta = \frac{8}{3}$ \citep[as in ][]{lorenz1963deterministic}, integrated using a fourth-order Runge-Kutta scheme with time step $\Delta t = 0.01$.

\begin{figure}[h!]
    \centering
    \includegraphics[width=0.99\textwidth]{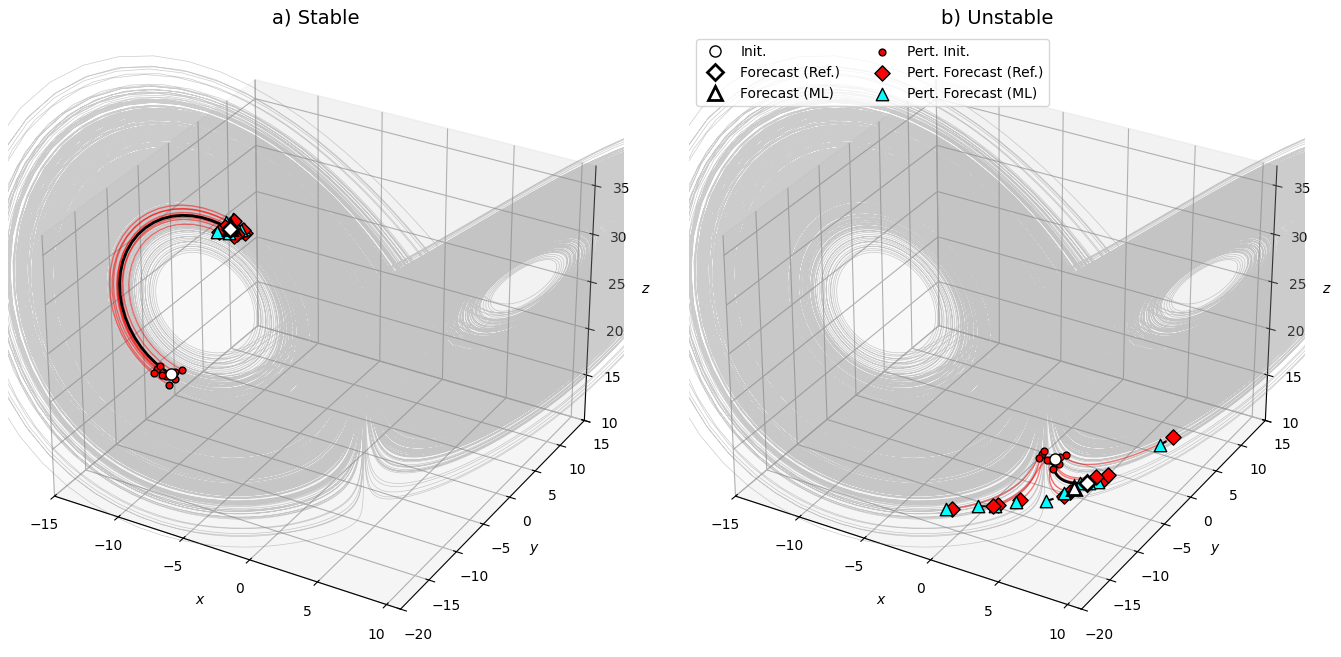}   
    \caption{L63 forecasts in phase space, from (a) a stable initial condition on a ``lobe'' of the attractor, and (b) at an unstable initial condition near the centre of the attractor.
    White/red circles show the initial condition and an ensemble of perturbations from it.
    Black/red solid lines are the corresponding trajectories, each over 25 integration steps of the reference model (Eq.~\ref{eq:L63}).
    White/red diamonds are the reference forecast states; white/cyan triangles are the ML forecasts.
    Grey lines/shadow show a long model trajectory, representing the ``plausible'' states in phase space. 
    }
    \label{fig:l63_combine}
\end{figure}

In the L63 system, the initial condition (some given values for Eq.~\ref{eq:l63_state}) completely determines the trajectory (and forecast state) after any fixed lead time\footnote{A trajectory is the path a point traces through state space over time. In discretised form, the computed trajectory also depends on the numerical integration scheme; throughout this work we use the fourth-order Runge-Kutta method exclusively.}. Here, the forecast lead time is set to \(t=0.25\), providing a finite period over which initially small perturbations can evolve and grow \citep{lorenz1963deterministic}. Such finite-time perturbation growth is a key feature of predictability in the L63 system, where error growth is transient and varies across the attractor \citep{trevisan1995transient, boffetta1998extension}. Previous ensemble-prediction studies have similarly considered perturbation growth over finite lead times when examining predictability in L63 \citep{feng2014application}. At \(t=0.25\), these differences in trajectory evolution are evident in Fig.~\ref{fig:l63_combine} (and later, Fig.~\ref{fig:diagnostic_forecast_error_vs_anisotropy}): for trajectories within a single attractor lobe\footnote{The attractor of a dynamical system is the set of states towards which trajectories converge over time. In L63, this takes the form of a butterfly-shaped strange attractor.}, small perturbations generally produce only minor deviations (Fig.~\ref{fig:l63_combine}a), whereas near the central region, perturbations in \(x\) and \(y\) can lead to substantial trajectory deviations while perturbations in \(z\) have comparatively little impact (Fig.~\ref{fig:l63_combine}b). 

To emulate the L63 system, we implement a fully-connected feed-forward artificial neural network (ANN), using the state at time $t$ as input to forecast the system state at time $t+0.25$ ($25$ integration steps).
The ANN is implemented with PyTorch \citep{paszke2019pytorch} and consists of three fully-connected hidden layers (widths are 128, 64 and 64), with ReLU activations. Training data is generated over a long trajectory ($2,000$ time units, i.e., $200,000$ integration steps), split sequentially into training, validation and test sets at a $64:16:20$ ratio, and standardised by subtracting the mean and scaling to unit variance. The Adam optimiser \citep{kingma2014adam} is used with a learning rate of $0.001$, mean square error loss, and early stopping with patience of 10 epochs. 
The network achieves $R^2 = 0.9999$ (NRMSE $= 0.67\%$) on the held-out chronological test set; all subsequent XAI evaluations, however, are computed on points drawn from a separate, independently generated 500-time-unit trajectory (40{,}000 points after a 100-time-unit burn-in), sampled at $n=4000$ unless stated otherwise in a figure's caption.
The aim of this work is not to construct an optimal L63 emulator, but to compare XAI methods under a consistent and competent predictive baseline. 

\section{Illustrative XAI methods on a continuous dynamical system}
\label{sec:illustrate_XAI_methods}
We now apply the methods outlined in Section~\ref{sec:XAI-method_overview} to the L63 forecasting model of Section~\ref{sec:XAI-L63}. All are post-hoc, explaining a pre-trained model that takes the L63 state $\mathbf{x}_t$ as input and predicts $\mathbf{x}_{t+0.25}$ (i.e., the state after 25 integration steps at $\Delta t=0.01$). Each yields local sensitivity or attribution scores quantifying each input's influence on each output. Because L63 is low-dimensional, these can be visualised directly in phase space, showing both individual explanations and the global patterns across the attractor; in higher-dimensional applications such as operational AIWP, aggregation, averaging, or compositing becomes essential.

The methods fall into two top-level families, sensitivity and attribution, with attribution further divided according to whether its reference is implicit in the method or explicitly selected by the user (cf.\ Sect.~\ref{sec:XAI-review}\ref{sec:apps_attribution}): \textit{sensitivity}, a local derivative (Saliency; Sect.~\ref{sec:illustrate_XAI_methods}\ref{sec:saliency}); \textit{implicit-baseline attribution}, a derivative weighted by the input state, with any reference fixed by the method (Input$\times$Gradient and LRP; Sects.~\ref{sec:illustrate_XAI_methods}\ref{sec:inputxgradient}--\ref{sec:illustrate_XAI_methods}\ref{sec:lrp_illustration}); and \textit{explicit-baseline attribution}, defined relative to a user-chosen baseline (Integrated Gradients and SHAP; Sects.~\ref{sec:illustrate_XAI_methods}\ref{sec:integrated_gradients}--\ref{sec:illustrate_XAI_methods}\ref{sec:shap_illustration}). Here we inspect aggregated importance qualitatively; Section~\ref{sec:eval_metrics} then evaluates the full, signed explanations against physically derived analogues of each family, and Section~\ref{sec:XAI-review} reviews their application in the wider literature.

\subsection{Output targets from state-to-state models}
\label{sec:scalar_objective}
The forecasting model maps an input state to a predicted output state, $\mathcal{M} : \mathbb{R}^d \rightarrow \mathbb{R}^d$, where $d$ is the number of elements in the state vector. Most XAI methods, however, are defined with respect to a differentiable scalar output target $f(\mathbf{x}) \in \mathbb{R}$, such as a single output element, $f(\mathbf{x}) = \mathcal{M}(\mathbf{x})_j$, or any differentiable reduction over several output elements (e.g., a sum or mean)\footnote{The state-to-state formulation is used for convenience; the model may include additional inputs (e.g., system forcings) or predict only part of the state. The key requirement is a differentiable scalar quantity of interest $f(\mathbf{x})$.}.

To summarise input ``importance'' across outputs, we compute attributions separately for each output $f_j(\mathbf{x}) = \mathcal{M}(\mathbf{x})_j$ and aggregate their absolute values,
\begin{equation}
\label{eq:aggregation}
    \bar{A}_i(\mathbf{x}) = \sum_{j=1}^{d} \left| A_i^{(j)}(\mathbf{x}) \right|, \quad i = 1, \dots, d,
\end{equation}
where $A_i^{(j)}(\mathbf{x})$ denotes the sensitivity or attribution score assigned to input $i$ with respect to output $j$ by a given method. This aggregation is applied consistently below for visualisation. It discards the sign of influence, which Section~\ref{sec:eval_metrics} shows is what most sharply distinguishes the method families; non-aggregated pairwise scores are given in Appendix~A.

\subsection{Saliency (Vanilla Gradients)}
\label{sec:saliency}
Saliency, the sensitivity method considered here, measures how sensitive a model's output is to perturbations in its inputs \citep{baehrens2010explain, simonyan2013deep}, defined as the gradient of the output target (Sect.~\ref{sec:illustrate_XAI_methods}\ref{sec:scalar_objective}):
\begin{equation}
\label{eq:saliency}
    \text{Saliency:}~A_{i}(\mathbf{x}) = \frac{\partial f(\mathbf{x})}{\partial \mathbf{x}_i}, \quad i = 1, \dots, d.
\end{equation}
Eq.~\eqref{eq:saliency} is the ML analogue of adjoint sensitivity in NWP, where the adjoint model (the transpose of the tangent-linear model) propagates the sensitivity of a scalar output back to the initial conditions; in ML models, the same quantity is obtained efficiently via automatic differentiation. This equivalence underpins the physically grounded benchmark of Section~\ref{sec:eval_metrics}.

\begin{figure}[h!]
    \centering
    \includegraphics[width=0.99\textwidth]{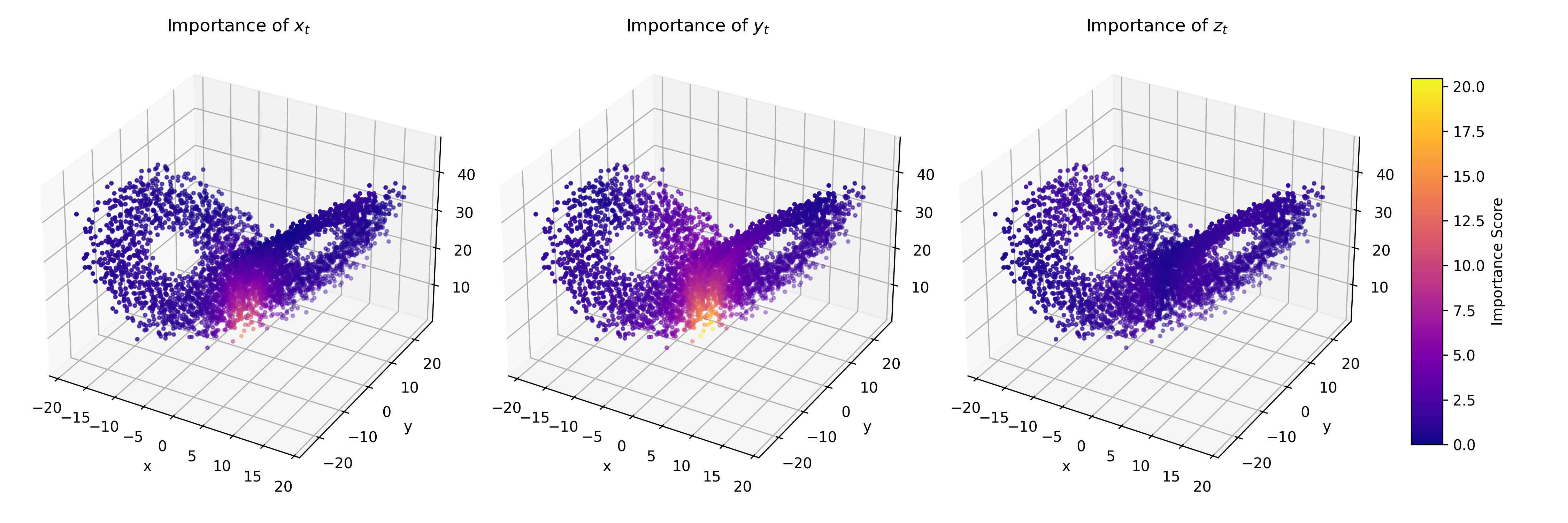}
    \caption{Saliency-derived importance scores $\bar{A}_i(\mathbf{x})$ for each input feature of the L63 system, computed via Eqs.~\eqref{eq:aggregation} and~\eqref{eq:saliency}. Each panel shows the L63 attractor, with input states coloured by the importance attributed to the corresponding input variable.
    }
    \label{fig:attr_Saliency}
\end{figure}

As expected from the L63 dynamics, aggregated saliency (Fig.~\ref{fig:attr_Saliency}) reflects known sensitivities of the attractor \citep{sparrow2012lorenz}: importance is weak on each lobe and peaks near the diverging region between lobes, where trajectories near $x\simeq y\simeq0$ easily switch lobes under small perturbations (cf.\ Fig.~\ref{fig:l63_combine}). Input $y$ is most important there, followed by $x$, while $z$ shows comparatively little importance (see also Appendix~A).

For high-dimensional systems, iterating over every output and aggregating afterwards, as here, is computationally expensive\footnote{Similar storage constraints are common in Earth system modelling: background error covariance matrices, full model state output at every time step, and full ensemble member sets are rarely stored in their entirety.}. A common alternative, used in many of the sensitivity studies reviewed in Sect.~\ref{sec:XAI-review}\ref{sec:apps_sensitivity}, is to reduce the outputs to a single scalar (e.g., a sum, bounded-area mean, or weighted combination) \emph{before} backpropagation. This assumes the reduction is a suitable proxy for the phenomenon of interest; comparing Fig.~\ref{fig:attr_Saliency} with Fig.~\ref{fig:app_saliency_detail} reveals finer structures that such a combined scalar could miss. Retaining the sign of the gradients also enables sensitivity-informed perturbation experiments: amplifying a feature highlighted by saliency (e.g., intensifying a low-pressure system) tests whether the model responds as physically expected (e.g., increased surface windspeeds in a target region).

\subsection{Input $\times$ Gradient}
\label{sec:inputxgradient}
\begin{figure}[h!]
    \centering
    \includegraphics[width=0.99\textwidth]{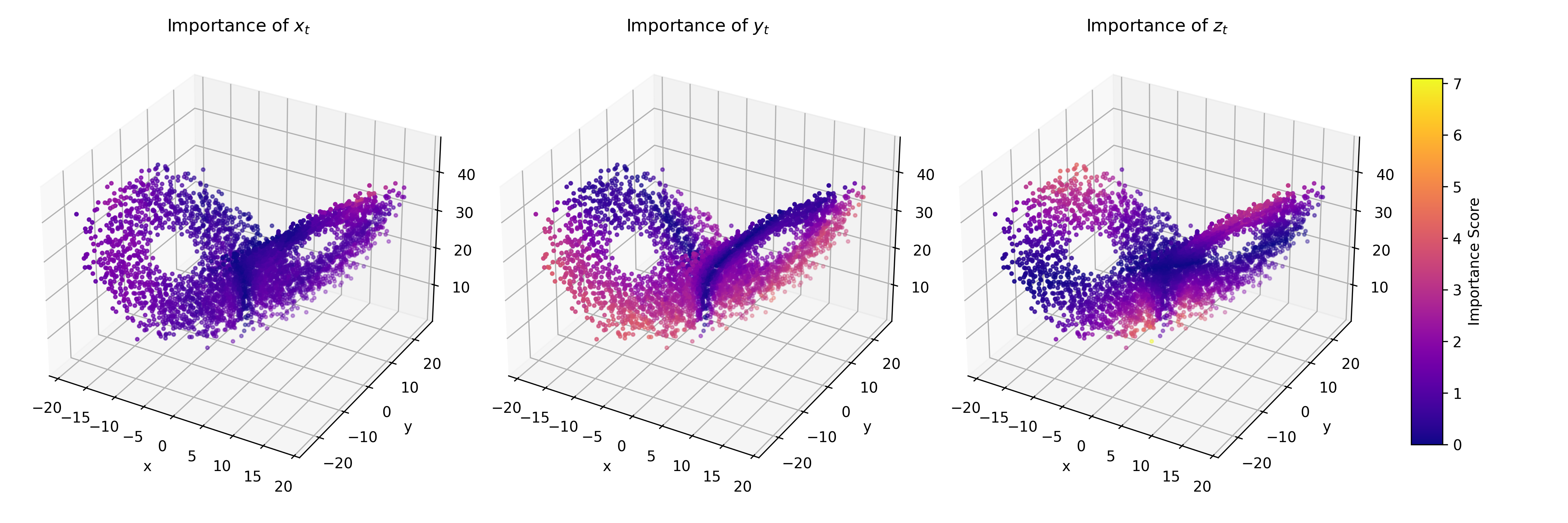}
    \caption{Input$\times$Gradient-derived importance scores $\bar{A}_i(\mathbf{x})$ for each input feature of the L63 system, computed via Eqs.~\eqref{eq:aggregation}~and~\eqref{eq:inputxgradient}. Each panel shows the L63 attractor, coloured by the scores attributed to the corresponding input variable.
    }
    \label{fig:attr_IG}
\end{figure}

Input$\times$Gradient \citep{shrikumar2017learning} multiplies each input feature element-wise by its saliency gradient. Mathematically a simple extension of saliency, it is conceptually an attribution method, and the simplest implicit-baseline one: rather than asking how sensitive the output is to each input, it asks how much each input contributes to the output. The intuition comes from linear models, where this product is exactly a feature's contribution:
\begin{equation}
\label{eq:inputxgradient}
\text{Input}\times\text{Gradient:}~A_i(\mathbf{x}) = \mathbf{x}_i \cdot \frac{\partial f(\mathbf{x})}{\partial \mathbf{x}_i} , \quad i = 1, \dots, d.
\end{equation}
More generally, a first-order Taylor expansion gives $f(\mathbf{x}) - f(\mathbf{0}) \approx \mathbf{x}\cdot\nabla f(\mathbf{x})$, so the method attributes the difference between the prediction and that at an all-zero reference. Each attribution thus combines the model's local response with the input's distance from that reference, and inputs close to the reference receive little attribution ($A_i \to 0$ as $\mathbf{x}_i \to 0$). In L63 (Fig.~\ref{fig:attr_IG}), $x$ and $y$ receive little attribution near the attractor centre ($x \simeq y \simeq 0$), where they are close to the reference, while $z$ receives large attributions near the upper attractor, where it is furthest from zero.

The interpretability of these attributions therefore rests on whether the reference is meaningful. In image classification, a zero input is a natural reference, as near-zero pixels are often uninformative background \citep{smilkov2017smoothgrad}. In raw L63 coordinates, however, the all-zero state is the system's unstable fixed point at the origin, which lies off the attractor, so attributions explain departures from a state the system never visits. For normalised inputs, the implicit reference instead becomes the mean state, and attributions explain departures from the mean prediction: a variable near its mean then rightly receives little attribution, reflecting the limited predictive leverage of near-mean conditions, although, as with explicit baselines (Sects.~\ref{sec:illustrate_XAI_methods}\ref{sec:integrated_gradients},\ref{sec:shap_illustration}), the mean itself may not be a realisable state.

\subsection{Layer-wise Relevance Propagation}
\label{sec:lrp_illustration}
Like Input$\times$Gradient, LRP is an implicit-baseline attribution method but, rather than backpropagating local sensitivities, redistributes the prediction itself through the network in proportion to each neuron's forward activation contribution. This gives a conservation property: relevance is neither created nor destroyed across layers, yielding a direct decomposition of the prediction onto the input features, with positive (negative) relevance indicating that a feature pushes the prediction higher (lower). LRP is therefore suited less to diagnosing the system's intrinsic instability than to auditing the model's learned representation: which state variables the network relies on in different regions of phase space, and whether this aligns with physical expectations.

LRP is a family of propagation rules \citep{montavon2019layer}, and rule selection for each layer is a design choice that should be justified; we illustrate three commonly used rules. Let \( R_k^{(l+1)} \) be the relevance of neuron \( k \) in layer \( l+1 \), and \( z_{jk} = a_j^{(l)} w_{jk} \) the contribution from neuron \( j \) in layer \( l \) to neuron \( k \), where \( a_j^{(l)} \) is the activation of neuron \( j \) and \( w_{jk} \) the connection weight. Relevance is conditioned on $\mathbf{x}$ through the forward-pass activations, and the attribution for input feature $i$ is the relevance at the input layer:
\begin{equation}
\label{eq:lrp_attribution}
    \text{LRP:} \quad A_i(\mathbf{x}) = R_i^{(0)}, \quad i = 1, \dots, d.
\end{equation}

\subsubsection{Epsilon $\epsilon$ rule}
\label{sec:lrp_epsilon}
\begin{figure}[h!]
    \centering
    \includegraphics[width=0.99\textwidth]{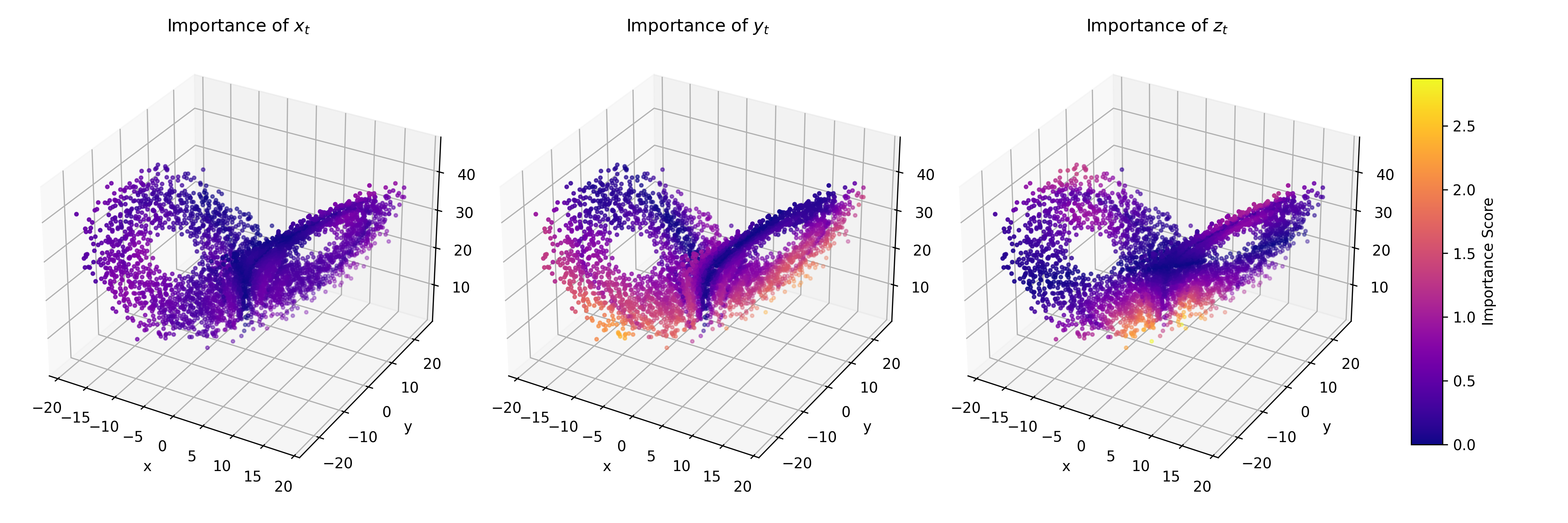}
    \caption{LRP-derived importance scores $\bar{A}_i(\mathbf{x})$ for each input feature of the L63 system, computed via Eq.~\eqref{eq:aggregation}. Each panel shows the L63 attractor, coloured by the score attributed to the corresponding input variable. The $\epsilon$-rule ($\epsilon=0.1$) is applied at every layer.}
    \label{fig:attr_LRP_epsilon}
\end{figure}

The $\epsilon$-rule stabilises the z-rule \citep[see][]{bach2015pixel} by adding a small positive term to the denominator, preventing near-zero pre-activations (the weighted sum input to a neuron before the activation function) from inflating relevance:
\begin{equation}
\label{eq:lrp_epsilon}
R_j^{(l)} = \sum_k \frac{z_{jk}}{\epsilon + \sum_{0,j} z_{jk} } \, R_k^{(l+1)}.
\end{equation}
In ReLU feedforward networks (as here), LRP-z reduces to an Input$\times$Gradient-like decomposition \citep{kindermans2016investigating}, so the LRP-$\epsilon$ importances (Fig.~\ref{fig:attr_LRP_epsilon}) resemble Input$\times$Gradient and share its implicit zero reference; the relatively large $\epsilon$ (0.1) used here damps propagation through neurons with small pre-activations, giving a smoother redistribution.

\subsubsection{Alpha 1 Beta 0 ($\alpha1\beta0$) rule}
\label{sec:lrp_alphabeta}
\begin{figure}[h!]
    \centering
    \includegraphics[width=0.99\textwidth]{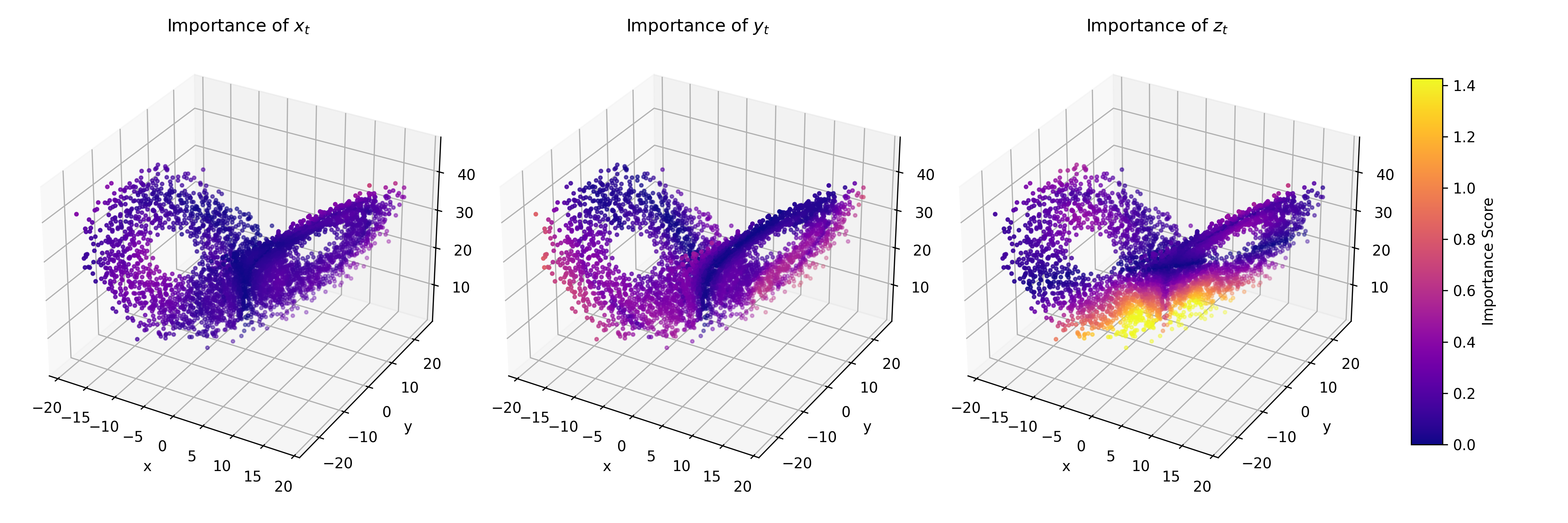}
    \caption{LRP-derived importance scores $\bar{A}_i(\mathbf{x})$ for each input feature of the L63 system, computed via Eq.~\eqref{eq:aggregation}. Each panel shows the L63 attractor, coloured by the score attributed to the corresponding input variable. The $\epsilon$-rule ($\epsilon=0.1$) is applied at every layer except the penultimate layer, where the $\alpha1\beta0$-rule is used.}
    \label{fig:attr_LRP_alpha1beta0}
\end{figure}

The alpha-beta rule treats positive and negative contributions asymmetrically, under the constraint $\alpha-\beta=1$:
\begin{equation}
\label{eq:lrp_alphabeta}
R_j^{(l)} = \sum_k \left( \alpha \cdot \frac{z_{jk}^+}{\sum_{0,j} z_{jk}^+} - \beta \cdot \frac{z_{jk}^-}{\sum_{0,j} z_{jk}^-} \right) R_k^{(l+1)},
\end{equation}
where \( z_{jk}^+ = \max(z_{jk}, 0) \) and \( z_{jk}^- = \min(z_{jk}, 0) \). For $\alpha=1$, $\beta=0$, only positive contributions propagate. This produces informative explanations in classification, where the goal is identifying evidence supporting a predicted class \citep{mamalakis2022investigating}, but discards the distinction between supporting and opposing evidence \citep{mamalakis2022neural}, which is often physically meaningful in regression. Figure~\ref{fig:attr_LRP_alpha1beta0} shows the resulting imbalance: relative to the $\epsilon$-rule, the lower parts of the lobes ($z<20$) are emphasised and the upper parts ($z>20$) de-emphasised, as regions with positive increments in the predicted state are systematically favoured.

\subsubsection{Gamma $\gamma$ rule}
\label{sec:lrp_gamma}
\begin{figure}[h!]
    \centering
    \includegraphics[width=0.99\textwidth]{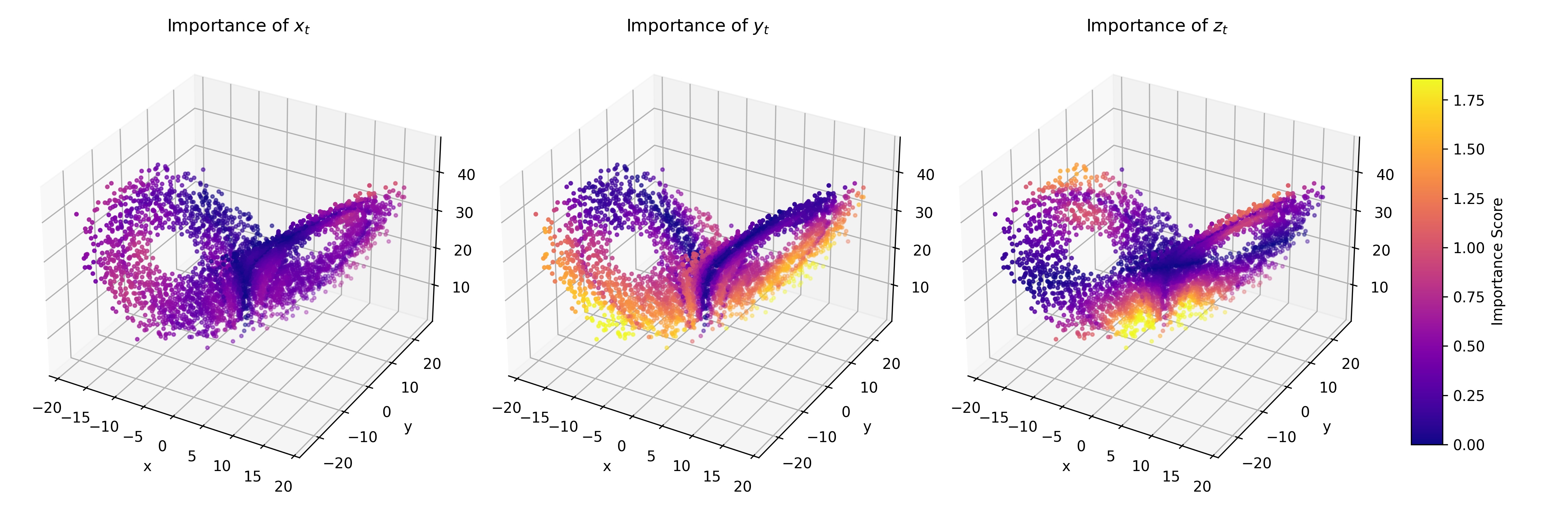}
    \caption{LRP-derived importance scores $\bar{A}_i(\mathbf{x})$ for each input feature of the L63 system, computed via Eq.~\eqref{eq:aggregation}. Each panel shows the L63 attractor, coloured by the score attributed to the corresponding input variable. The $\epsilon$-rule ($\epsilon=0.1$) is applied at every layer except the penultimate layer, where the $\gamma$-rule is used.
    }
    \label{fig:attr_LRP_gamma}
\end{figure}

The $\gamma$-rule \citep{montavon2019layer} amplifies positive contributions by a factor $(1 + \gamma)$ on positive weights, reducing to LRP-z as $\gamma \rightarrow 0$:
\begin{equation}
\label{eq:lrp_gamma}
    \begin{aligned}
    z_{jk}^{(\gamma)} &= a_j^{(l)} \left( w_{jk} + \gamma \cdot \max(w_{jk}, 0) \right),\\
    R_j^{(l)} &= \sum_k \frac{z_{jk}^{(\gamma)}}{\sum_{0,j} z_{jk}^{(\gamma)}} \, R_k^{(l+1)}.
    \end{aligned}
\end{equation}
It thus sits conceptually between the symmetric $\epsilon$/z rules and the asymmetric $\alpha\beta$-rule, emphasising positive contributions without discarding negative evidence. Like $\alpha1\beta0$, it emphasises the lower parts of the lobes, particularly for inputs $y$ and $z$ where $z<20$ (Fig.~\ref{fig:attr_LRP_gamma}).

Because weather and climate dynamics reflect a balance between competing tendencies (e.g., forcing and dissipation, or stabilising and destabilising processes), selectively amplifying positive contributions can distort the signed balance of effects in regression. Such rules may nonetheless suit identifying inputs that actively drive extreme events, such as heatwaves or heavy rainfall, so rule choice should follow the scientific goal; Section~\ref{sec:eval_metrics} quantifies how far each rule departs from its physical analogue.

\subsection{Integrated Gradients}
\label{sec:integrated_gradients}
\begin{figure}[h!]
    \centering
    \includegraphics[width=0.99\textwidth]{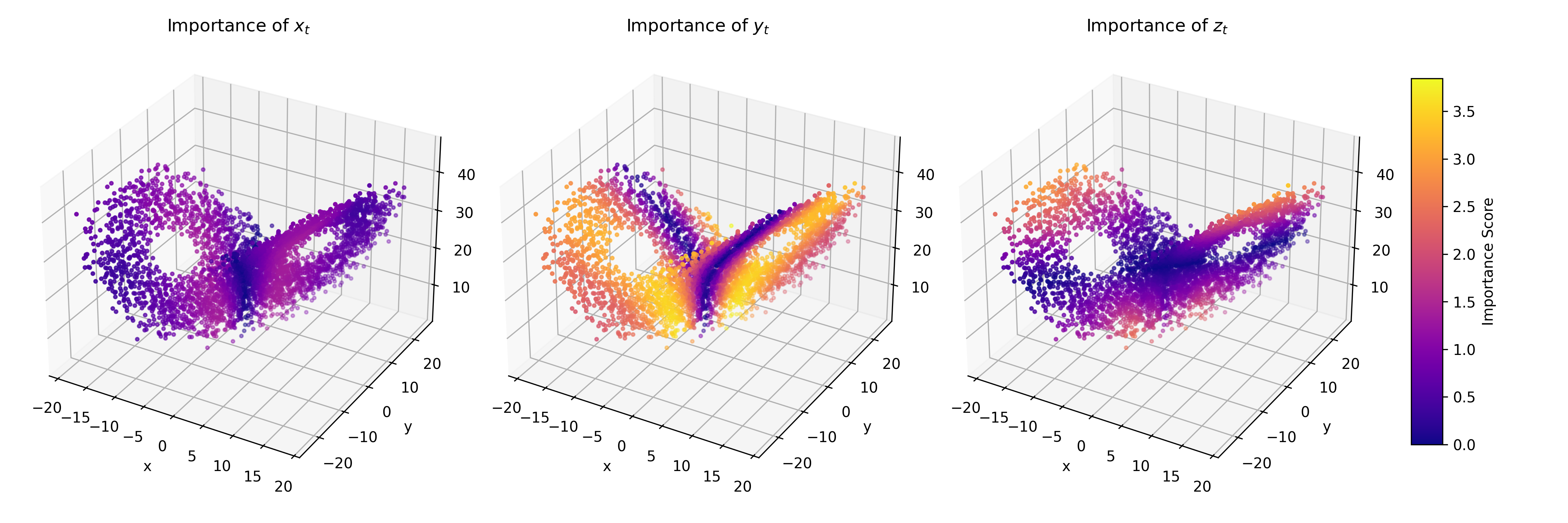}
    \caption{IntG-derived importance scores $\bar{A}_i(\mathbf{x})$ for each input feature of the L63 system, computed via Eqs.~\eqref{eq:aggregation}~and~\eqref{eq:integratedgradients}. Each panel shows the L63 attractor, coloured by the scores attributed to the corresponding input variable. The baseline state for an individual explanation is the mean system state.
    }
    \label{fig:attr_IntG}
\end{figure}

Integrated Gradients \citep[IntG,][]{sundararajan2017axiomatic} is an explicit-baseline method addressing two issues of gradient-based attribution: reliance on a first-order approximation at a single point, and saturation (near-zero gradients despite strong influence). Given a baseline input state $\mathbf{x}'$ and the input of interest $\mathbf{x}$,
\begin{equation}
\label{eq:integratedgradients}
\text{Integrated Gradient: }~A_i(\mathbf{x}) = (\mathbf{x}_i - \mathbf{x}'_i) \cdot \int_{\alpha=0}^{1} \frac{\partial f\!\left( \mathbf{x}' + \alpha (\mathbf{x} - \mathbf{x}') \right)}{\partial \mathbf{x}_i} \, d\alpha , \quad i = 1, \dots, d.
\end{equation}
The integral accumulates the gradient along the straight-line path from $\mathbf{x}'$ to $\mathbf{x}$ (approximated by a Riemann sum), so IntG summarises the model's sensitivity to feature $i$ along this path, scaled by that feature's change from its baseline. With the mean state as baseline (as in standard implementations; Fig.~\ref{fig:attr_IntG}), $y$ is again the most globally important input, though all variables show little importance around the attractor centre, and both $x$ and $z$ show moderate importance on each lobe.
 
An explicit baseline can be an advantage, as it states the question being asked \citep{mamalakis2023carefully}, but choosing one is non-trivial: for dynamical systems a ``null'' state is ambiguous and, depending on the aim of the explanation, a daily-varying climatological mean, a reference forecast, or another ensemble member could be appropriate. A non-physical baseline, such as a mean state that is never realised (common in high dimensions), may be problematic for models trained on realistic states, and even between physical endpoints the integration path may traverse regions lacking physical meaning, introducing unrealistic sensitivities \citep{zhang2026baselines}.

\subsection{SHAP (SHapley Additive exPlanations)}
\label{sec:shap_illustration}
SHAP \citep{lundberg2017unified}, also an explicit-baseline method, attributes a prediction to input features using Shapley values from cooperative game theory: the average marginal contribution of feature $i$ across all feature subsets \( S \subseteq N \setminus \{i\} \),
\begin{equation}
\label{eq:shap}
\text{SHAP}:~A_i(\mathbf{x}) = \sum_{S \subseteq N \setminus \{i\}}
    \frac{|S|! \, (|N| - |S| - 1)!}{|N|!}
    \left[ f_{S \cup \{i\}}(\mathbf{x}_{S \cup \{i\}}) - f_S(\mathbf{x}_S) \right],
\end{equation}
where \( N = \{1, \dots, d\} \), and \( f_S(\mathbf{x}_S) \) denotes the model output when only the features in \( S \) are known, with the remainder replaced by values from a reference baseline: a single state (e.g., a zero or mean state) or a set of states (e.g., a sample of training data). SHAP values satisfy three properties \citep{lundberg2017unified}: local accuracy, missingness, and consistency.\footnote{\textit{Local accuracy}: the attributions sum to the difference between the model's prediction and its expected output under the reference. \textit{Missingness}: a feature that is already marked as missing in the input being explained (i.e., absent from its simplified binary representation) receives zero attribution. \textit{Consistency}: if a model changes such that a feature's marginal contribution increases or stays the same for every subset of the other features, that feature's attribution does not decrease.} Exact computation scales exponentially with feature count, so approximations such as KernelSHAP, TreeSHAP, and DeepSHAP \citep{lundberg2017unified, lundberg2020local} are used in practice, each with model-specific assumptions which significantly affect scalability and fidelity.
 
\begin{figure}[h!]
    \centering
    \includegraphics[width=0.99\textwidth]{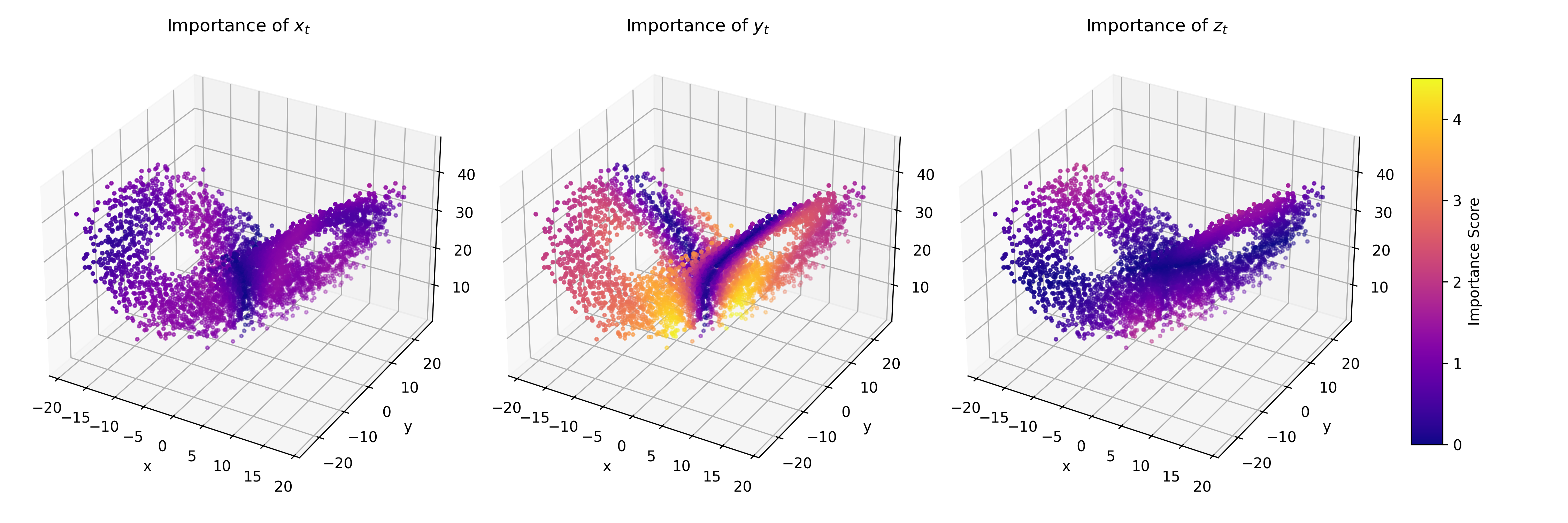}
    \caption{SHAP-derived importance scores $\bar{A}_i(\mathbf{x})$ for each input feature of the L63 system, computed via Eqs.~\eqref{eq:aggregation}~and~\eqref{eq:shap}. Each panel shows the L63 attractor, coloured by the scores attributed to the corresponding input variable. The reference state used for missing features is the mean system state.
    }
    \label{fig:attr_KSHAP}
\end{figure}

Marginalising ``missing'' features independently assumes feature independence; for L63 states, as for most dynamical systems, the resulting combinations will typically be off-manifold, potentially yielding misleading explanations \citep{frye2020shapley}. With the same mean-state reference, SHAP (Fig.~\ref{fig:attr_KSHAP}) and IntG pose the same question and give similar answers: $y$ is most important, while $x$ and $z$ are substantially less so. The detailed output (Fig.~\ref{fig:app_shap_detail}) shows that $y$'s importance arises largely from its influence on output $z$, a pattern also seen in IntG, and one obscured by aggregation.

\subsection{Further explanation methods}
Beyond these three families, a growing body of bespoke approaches generates explanations for Earth system regression tasks, including latent-space analysis, counterfactual generation, links between high-complexity and simpler interpretable models, occlusion, and methods encoding domain knowledge directly into model representations. These do not conform to standard input feature-based XAI methods and are reviewed in Section~\ref{sec:XAI-review}\ref{sec:assorted_other}.

The qualitative differences above, between sensitivity, implicit-baseline and explicit-baseline methods, and between LRP rules within a single family, are quantified against physically derived analogues in Section~\ref{sec:eval_metrics}.

\section{Evaluating XAI methods against physical references}
\label{sec:eval_metrics}

We provide a quantitative comparison of the XAI methods against physically derived reference explanations for the L63 system, each constructed to match the mathematical definition of a method family, extending the qualitative comparison above. This should not be mistaken for a general solution to the XAI evaluation problem: it exploits properties specific to models with a physically derived sensitivity model.
A central challenge is the absence of a ground truth for explanations \citep{seth2025bridging}. \citet{mamalakis2022neural} address this with a synthetic regression benchmark whose additively separable response function makes each input's true contribution known by construction, demonstrating that sensitivities and attributions are distinct quantities. We take a complementary approach: L63 carries nonlinear feature interactions and dynamics, and its known governing equations allow a physical reference to be matched to each method family.

In Earth sciences, saliency has a direct physical interpretation, equivalent to adjoint sensitivity propagation, providing a partial bridge to established frameworks. Here we extend this bridge beyond saliency alone: because L63’s governing equations, and hence its exact tangent-linear model (TLM), are fully known, we can construct independent, physically exact references for sensitivity and for each attribution family. Each is a ground truth only for the method family it is constructed to match; for the other methods it serves as a common reference point, since no single true explanation exists. This construction relies on L63’s dynamics being known exactly. Realistic Earth system models often have tangent-linear or adjoint models, but these approximate unknown true dynamics and carry compounding sources of error and bias, so comparable references would be physically grounded but approximate.

Beyond the plain TLM, whose use as a sensitivity diagnostic is well established in data assimilation and predictability studies, we construct two further physically derived quantities, StateWeightedTLM and PathIntegratedTLM\footnote{Each reference applies the formula of an established XAI method, with the physically derived TLM in place of the learned network gradient: the TLM corresponds to Saliency (Eq.~\eqref{eq:saliency}), StateWeightedTLM to Input$\times$Gradient (Eq.~\eqref{eq:inputxgradient}), and PathIntegratedTLM to Integrated Gradients (Eq.~\eqref{eq:integratedgradients}).}, which correspondingly fall into the implicit-baseline (Input$\times$Gradient, LRP) and explicit-baseline (Integrated Gradients, Shapley) attribution families.
All three quantities are computed directly from the governing equations and are entirely independent of the trained network, unlike, say, using a smoothed or ensemble-based variant of the network's own gradient as a pseudo ground truth, which avoids any circularity between the object under evaluation and the standard it is evaluated against. Only the plain TLM, however, is a standard, named diagnostic with substantial precedent in the literature; StateWeightedTLM and PathIntegratedTLM are, to our knowledge, novel constructions introduced here to give each method family a ground truth matching its own definition, rather than established physical quantities in their own right, and we refer to them throughout as ``TLM-based physical attribution'' to avoid overstating their provenance.

\begin{figure}[h!]
    \centering
    \includegraphics[width=0.99\textwidth]{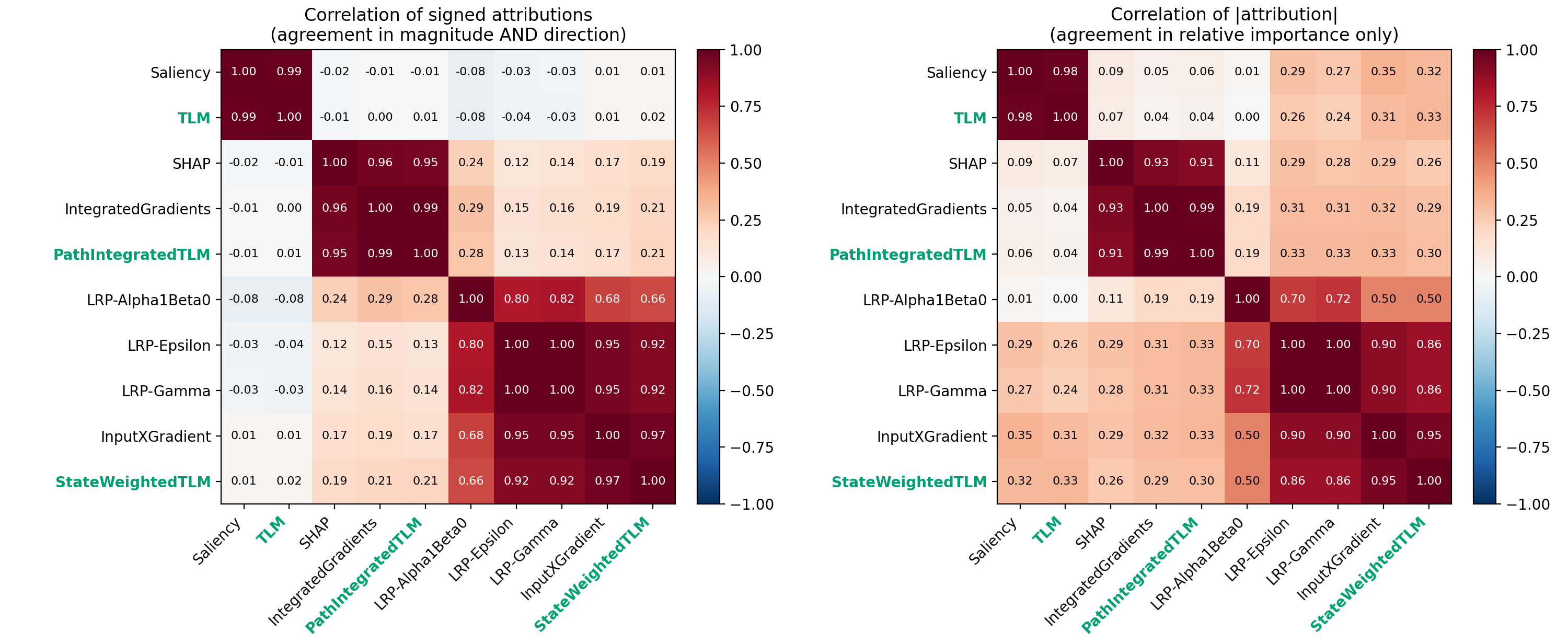}   
    \caption{Pairwise correlation between all 10 methods (the 7 XAI attribution methods plus the three TLM-based physical attribution objects,
    TLM, StateWeightedTLM, and PathIntegratedTLM, shown in green bold), pooled by flattening the full $500 \times 3 \times 3$ (sample-point $\times$ output
    $\times$ input) attribution tensor to a single vector per method, so the
    correlation reflects agreement across the entire input-output sensitivity
    structure at every point. Left: correlation of signed attributions (agreement in both magnitude and direction). Right: correlation of
    $|\text{attribution}|$ (agreement in relative importance only).}
    \label{fig:method-correlation}
\end{figure}

The pairwise correlation matrix resolves into three groups rather than ten independent methods: a sensitivity group (Saliency, alongside its physical analogue TLM), an implicit-baseline attribution group (Input$\times$Gradient and the LRP variants, alongside StateWeightedTLM), and an explicit-baseline attribution group (Integrated Gradients and exact Shapley, alongside PathIntegratedTLM). Here, “implicit-baseline” refers to the absence of an explicitly selected reference input: in standard InputXGradient and LRP formulations, any reference is effectively determined by the formulation or propagation rule, rather than chosen separately for each explanation, in contrast to explicit-baseline methods such as Integrated Gradients and SHAP. 
Each group corresponds to a distinct mathematical construction (a local derivative, a derivative weighted by the input state, and a derivative integrated along a path from a baseline, respectively), and each has a purely physics-derived analogue, built directly from the governing equations and independent of the trained network, that lands inside the group it was constructed to match. The correspondence is not perfect, however: within the implicit-baseline group, Input$\times$Gradient correlates far more strongly with the two LRP rules known to reduce to a gradient-times-input-like propagation on ReLU networks (LRP-Epsilon, LRP-Gamma; $r=0.95$ signed, $0.90$ magnitude) than with LRP-Alpha1Beta0 ($r=0.68$ signed, $0.50$ magnitude), whose relevance-propagation rule visibly weakens the correspondence even within a single, ostensibly homogeneous method family; this divergence should only grow as rule choice and model complexity increase. 
We also note that the three groups separate considerably more cleanly when magnitude and sign are considered jointly (signed correlation) than when only relative importance is compared ($|$attribution$|$ correlation): the sensitivity and implicit-baseline groups, for instance, are close to uncorrelated under signed correlation ($r=-0.08$ to $0.02$) but show a much weaker separation under magnitude-only correlation ($r$ up to $0.35$), indicating that the direction of attributed influence, not just its relative ranking, most sharply distinguishes the three families.

\subsection{Multi-criteria comparison with physical references}
\label{sec:faithfulness-metrics}

For a physically derived reference sensitivity tensor $T \in \mathbb{R}^{N \times 3 \times 3}$ and an attribution tensor $A \in \mathbb{R}^{N \times 3 \times 3}$ from a given method
(indexed by sampled attractor point $p = 1,\dots,N$, output component $o \in \{x,y,z\}$, and input component $i \in \{x,y,z\}$), we report four complementary metrics, each targeting a distinct notion of agreement between $T$ and $A$. 

\paragraph{Sign agreement.}
Measures whether the attribution correctly recovers the \emph{direction} of
each input's influence:
\begin{equation}
S_{\text{sign}} = \frac{1}{|\mathcal{M}|} \sum_{(p,o,i) \in \mathcal{M}}
\mathds{1}\!\left[\operatorname{sign}(T_{p,o,i}) = \operatorname{sign}(A_{p,o,i})\right],
\end{equation}
where $\operatorname{sign}(\cdot) \in \{-1,0,+1\}$ and the index set
$\mathcal{M} = \{(p,o,i) : |T_{p,o,i}| \geq \tau_q\}$ excludes entries where
the true sensitivity is exactly zero (sign undefined) and, with
$q = 0.25$, the smallest-magnitude quartile of $|T_{p,o,i}|$, whose sign is
dominated by numerical noise rather than genuine physical influence.
$\tau_q$ is the $q$-th quantile of $|T_{p,o,i}|$ over its nonzero entries.

\paragraph{Dominant-variable accuracy.}
Measures whether the attribution identifies the correct \emph{most
influential} input for each output at each point:
\begin{equation}
S_{\text{dom}} = \frac{1}{3N} \sum_{p=1}^{N} \sum_{o \in \{x,y,z\}}
\mathds{1}\!\left[\operatorname*{arg\,max}_{i} |T_{p,o,i}| =
\operatorname*{arg\,max}_{i} |A_{p,o,i}|\right].
\end{equation}

\paragraph{Mean pointwise rank correlation.}
Measures whether the attribution reproduces the correct \emph{relative
ranking} of input importance, independent of absolute scale:
\begin{equation}
\bar{\rho}_s = \frac{1}{|\mathcal{P}|} \sum_{(p,o) \in \mathcal{P}}
\rho_s\!\left(\, |T_{p,o,\cdot}|,\ |A_{p,o,\cdot}| \,\right),
\end{equation}
where $\rho_s(\cdot,\cdot)$ denotes the Spearman rank correlation
coefficient computed over the three input components $|T_{p,o,\cdot}| =
(|T_{p,o,x}|, |T_{p,o,y}|, |T_{p,o,z}|)$ at fixed $(p,o)$, and $\mathcal{P}$
excludes point/output pairs for which either vector is constant (rank
correlation undefined for $n_{\text{in}} = 3$ with ties).

\paragraph{Magnitude correlation.}
Measures whether attribution \emph{magnitude} tracks the true sensitivity's magnitude, pooling across the entire sampled attractor:
\begin{equation}
r_{\text{mag}} = \operatorname{corr}\!\left( \operatorname{vec}(|T|),\ \operatorname{vec}(|A|) \right),
\end{equation}
the Pearson product-moment correlation coefficient computed over all $3N \times 3$ (point, output, input) entries pooled together, where $\operatorname{vec}(\cdot)$ denotes flattening the tensor to a single vector. A Spearman-rank variant of this metric is also computed for robustness checks but is not used as a primary result.

Together, the four metrics compare each XAI method with the physically derived references across distinct properties of a sensitivity or an attribution: sign agreement tests the direction of each input's influence, dominant-variable accuracy whether the most influential input is identified, mean rank correlation whether the full ordering of inputs is recovered, and magnitude correlation whether attribution magnitudes track the reference response across the attractor.

\begin{figure}[h!]
    \centering
    \includegraphics[width=0.99\textwidth]{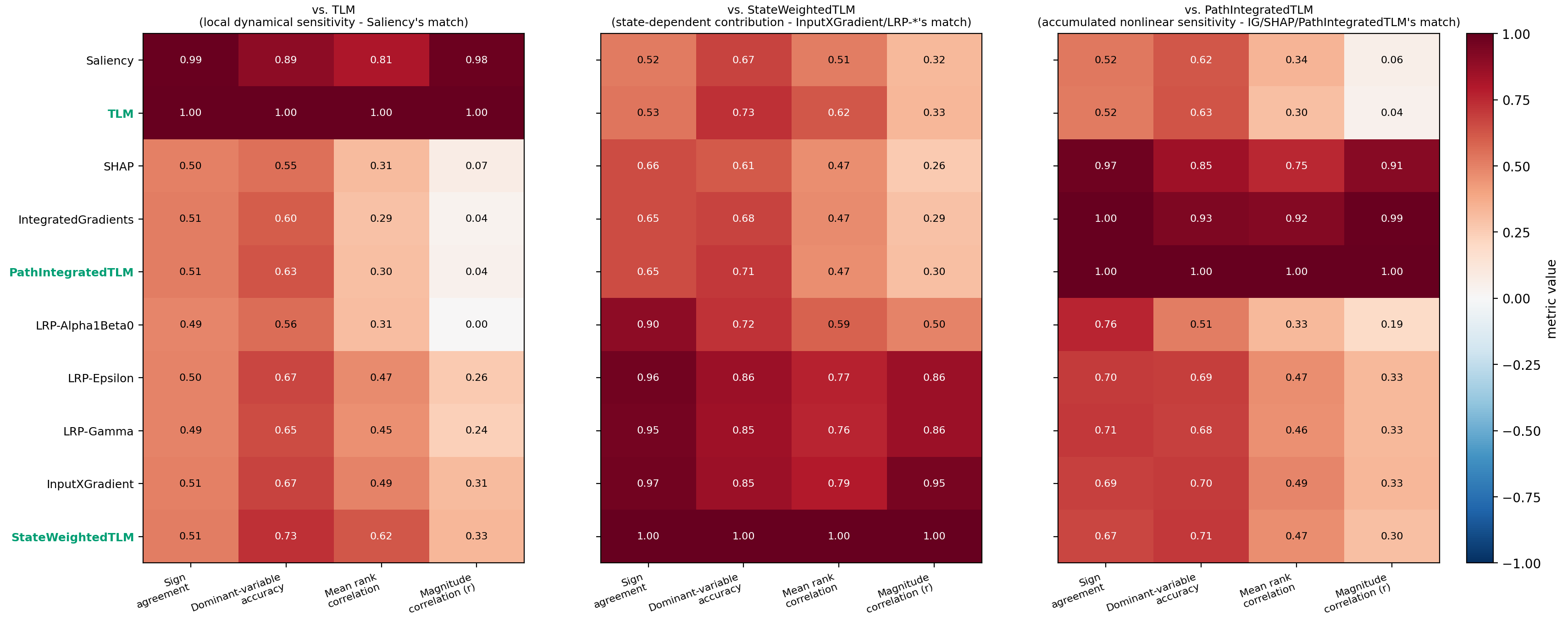}   
    \caption{Multi-criteria comparison of all 10 methods against each of the three TLM-based physical attribution benchmarks (TLM, StateWeightedTLM, PathIntegratedTLM), evaluated over $n=500$ sampled attractor points.
    Each panel reports the four comparison criteria (sign agreement, dominant-variable accuracy, mean rank correlation, magnitude correlation) as columns for one physical benchmark; colour and the annotated value both
    encode the metric score. Rows follow the correlation-clustering order of Fig.~\ref{fig:method-correlation}, with the three physical objects (bold) included as methods in their own right (diagonal self-comparisons are trivially perfect and shown for reference only). }
    \label{fig:faithfulness-3panel}
\end{figure}

Because StateWeightedTLM and PathIntegratedTLM are built from the same TLM Jacobian, just combined with the state more richly, the same three-way grouping reappears whichever is used as the benchmark: each family scores highest against its own matched construction. Matching a family does not make its members interchangeable, however: methods within a family can still differ in their construction, as LRP-Alpha1Beta0 does from the other LRP rules, and exact Shapley values, which average over feature coalitions, do from the path integral of Integrated Gradients (Fig.~\ref{fig:faithfulness-3panel}). 
The gap between matched and mismatched scores depends sharply on which metric is used. Sign agreement is coarse: both families score well against their own benchmark ($0.90$--$1.00$) but also reasonably well against the other's ($0.65$--$0.76$), since the sign of each Jacobian entry is fixed by the Lorenz equations and so is shared across all three constructions. Magnitude correlation is sharper, high within the matched family ($0.86$--$0.99$, excluding LRP-Alpha1Beta0) but dropping to $0.19$--$0.33$ when mismatched, because relative magnitudes, unlike signs, are exactly what state-weighting and path-integration change construction to construction. No single criterion is therefore a reliable summary of physical grounding on its own; the four should be read jointly. 

Dominant-variable accuracy is coarser than it appears: for three variables it is a top-1 hit rate against a 1-in-3 baseline, discarding the ordering information mean rank correlation retains. In higher-dimensional systems it would carry more signal per hit, as the chance baseline falls as $1/n$, and a top-$k$ generalisation would interpolate naturally toward full rank correlation as $k$ grows. The two can diverge already: a method can recover the full ordering well (high rank correlation) while being inconsistent about which variable is dominant when the top two are close in magnitude (lower dominant-variable accuracy), without any real loss of ranking fidelity.

\subsection{An accurate forecast does not imply correctly learned dynamics!}

\begin{figure}[h!]
    \centering
    \includegraphics[width=0.99\linewidth]{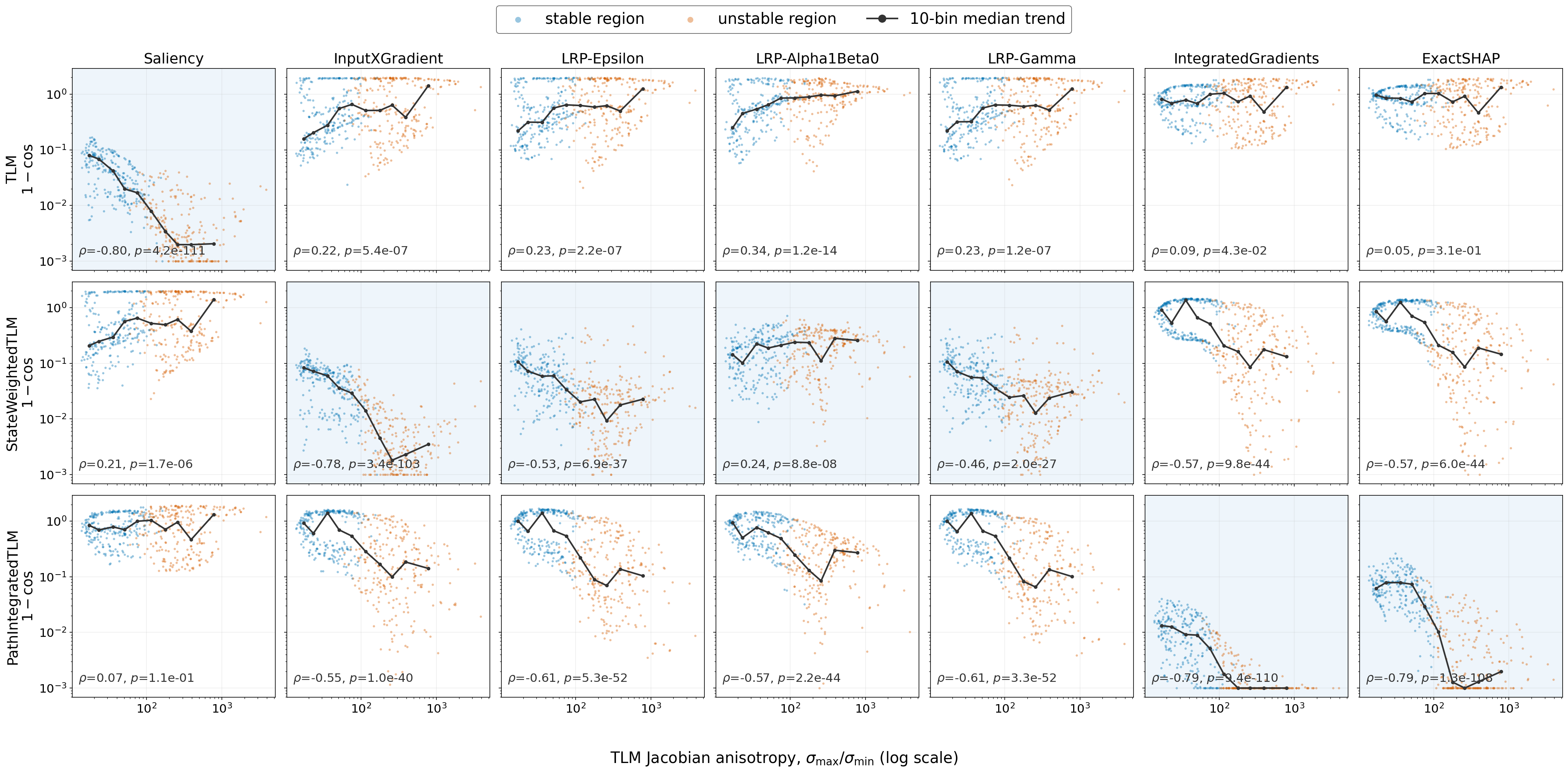}
    \caption{Per-point attribution error, $1-\cos(\hat{a}_i, J_i)$ between the
    signed attribution matrix and each physical benchmark's Jacobian at that
    point, plotted against the local TLM Jacobian anisotropy
    $\sigma_{\max}/\sigma_{\min}$ (both axes log-scaled). Columns are the seven
    XAI methods; rows are the three physical benchmarks (TLM,
    StateWeightedTLM, PathIntegratedTLM). The shaded panel in each column is that
    method's family-matched benchmark. Points are coloured by the
    ensemble-spread-growth stable/unstable split ($n=500$); the black curve is
    the 10-bin median trend, with Spearman's $\rho$ and its $p$-value reported
    per panel.}
    \label{fig:diagnostic_error_scatter_anisotropy}
\end{figure}

\begin{figure}[h!]
    \centering
    \includegraphics[width=1\linewidth]{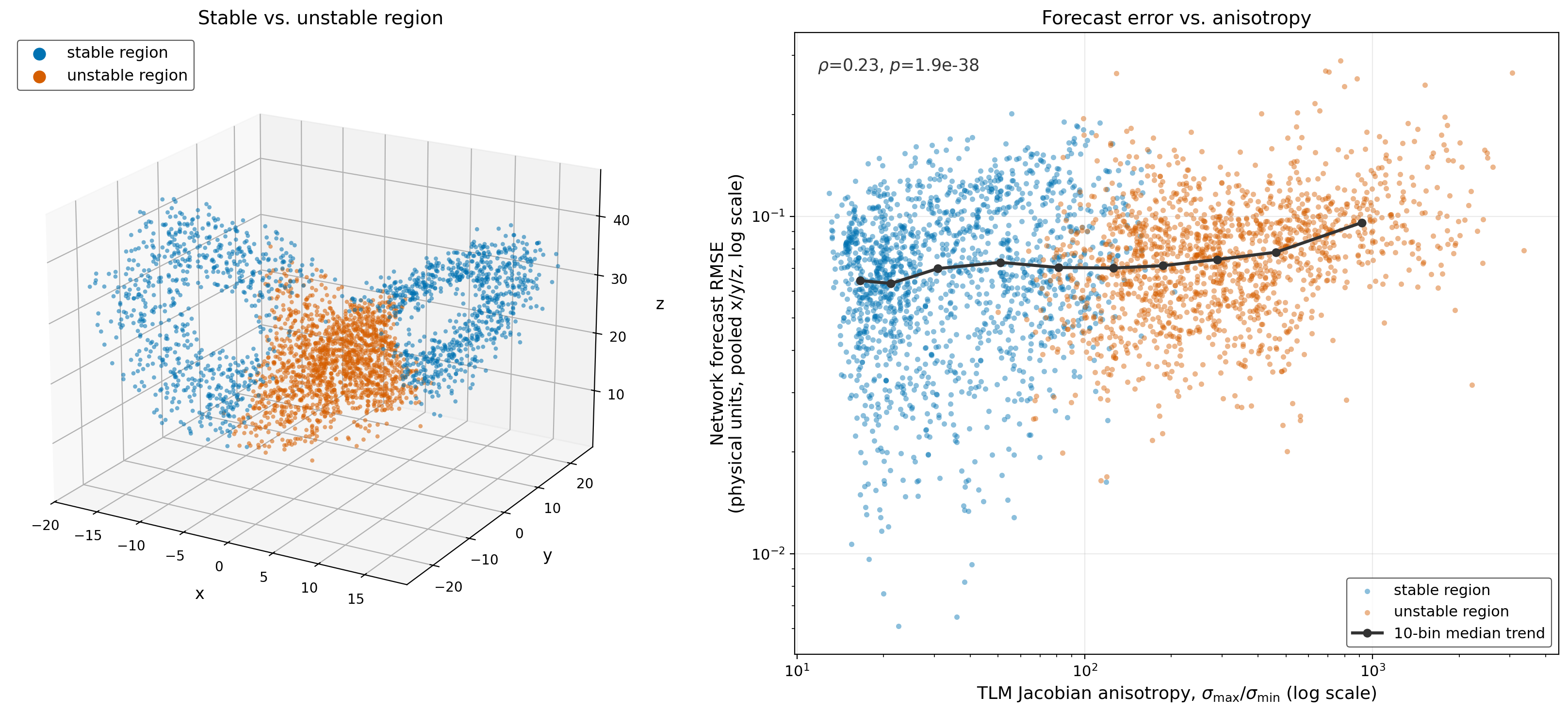}
    \caption{Both panels share the same $n=3000$ points, coloured by the ensemble-spread-growth stable/unstable split. Left: the L63 attractor, showing that the unstable region corresponds to the fast-crossing corridor between the two lobes. Right: the trained network's own forecast RMSE (physical units, pooled across $x,y,z$) against local TLM Jacobian anisotropy at those same points.}
    \label{fig:diagnostic_forecast_error_vs_anisotropy}
\end{figure}

Figure~\ref{fig:diagnostic_error_scatter_anisotropy} plots each method's per-point attribution error against local Jacobian anisotropy, for every method-benchmark pair: in six of seven methods' matched (shaded) panels, error falls as anisotropy rises ($\rho=-0.46$ to $-0.80$, all $p<10^{-26}$). Panels against TLM for the other methods, and Saliency's panels against the other two benchmarks, are flat or weakly positive ($\rho=+0.05$ to $+0.34$). The cross-family panels between StateWeightedTLM and PathIntegratedTLM also fall with anisotropy ($\rho=-0.55$ to $-0.61$), as expected since both are built from the same TLM Jacobian (LRP-Alpha1Beta0 is the one exception with $\rho=0.24$). A natural alternative explanation is that the network simply forecasts better in the unstable corridor where data samples are dense. However, this is ruled out by Fig.~\ref{fig:diagnostic_forecast_error_vs_anisotropy}: forecast RMSE \emph{rises} with anisotropy ($\rho=0.23$, $p=1.9\times10^{-38}$), opposite in sign to the agreement trend. Agreement with the physical references is therefore tied to the geometry of the local Jacobian, not to how well the network predicts there.

This points to a distinction between forecast skill and dynamical realism. High forecast error in the unstable corridor is expected of any model, since chaotic error growth is a property of the flow itself there; that the network's sensitivities also track the true Jacobian's dominant direction there indicates it has genuinely learned the local dynamics. In the stable lobes, low forecast error is weaker evidence of this: errors do not grow quickly there, so many incorrect sensitivity structures can still forecast well, which is consistent with the poor agreement of sensitivities and attributions with their references in those regimes. Forecast accuracy alone would give the opposite impression of where the network's dynamical understanding is strongest, which motivates comparison against physical references and careful interpretation of XAI methods in dynamical regression settings.

\subsection{Further XAI evaluation approaches}

The proposed comparisons above differ in kind from the perturbation-based robustness and faithfulness metrics most commonly used to evaluate XAI methods more generally \citep{alvarez2018towards, huang2024applications, bommer2024finding}, and are not intended as a substitute for that broader evaluation literature, whose assumptions merit separate scrutiny in the Earth science context.

Robustness measures whether small input perturbations lead to small changes in the explanation \citep{bommer2024finding}. In dynamical Earth system models this assumption may be inappropriate: many systems operate in regimes where small perturbations induce large responses, e.g., near tipping points or critical thresholds, where a lack of robustness may reflect physical realism rather than a methodological deficiency.

Faithfulness, in its usual formulation, assesses whether an explanation reflects the model's actual reasoning by testing whether perturbing important features changes the prediction more than perturbing unimportant ones \citep{mamalakis2022investigating, mamalakis2022neural}. In Earth science applications, physical balance constraints and strong spatio-temporal dependencies mean input features generally cannot be perturbed independently without violating physical consistency. Meaningful perturbations should respect correlations and co-variability to remain within the manifold of physically realisable states, perhaps by perturbing ``super-features'', physically coherent groups of model variables (e.g., dynamically coupled fields or spatially contiguous structures), aligning evaluation more closely with the system's underlying physics and reducing misleading conclusions from implausible perturbations. Our own comparison sidesteps this particular difficulty only because L63's entire state is three mutually coupled variables with fully known dynamics, so we can make direct comparisons to physically derived analogues (which while not necessarily a ``truth'', still serve as a physically grounded reference point).

More generally, a given XAI method's suitability depends strongly on the stakeholder's requirements (Sect.~\ref{sec:XAI-review}\ref{sec:stakeholders}). Evaluation can be viewed as two complementary streams: whether an explanation is useful to the intended user, which is inherently application- and stakeholder-dependent; and whether XAI methods can be quantitatively compared on specific properties of their explanation values, whether against an independent ground truth where one is available, as here, or via the perturbation-based properties discussed above where it is not. This latter stream remains relatively immature, and the desirable properties of a ``good'' explanation are typically context dependent.

\section{Overview of XAI applications in Earth system regression contexts}
\label{sec:XAI-review}
\subsection{Sensitivity-based approaches}
\label{sec:apps_sensitivity}
Sensitivity analysis has long been valuable for understanding complex systems, diagnosing error sources, and improving forecasts. Traditionally, gradient-based sensitivities have been computed using adjoint methods.
In modern machine learning, directly analogous gradient-based approaches underpin both training and several XAI methods (i.e., Sect.~\ref{sec:illustrate_XAI_methods}\ref{sec:saliency}-\ref{sec:illustrate_XAI_methods}\ref{sec:integrated_gradients}). This offers a pre-established, unifying framework for sensitivity (and potentially attribution) diagnostics in high-dimensional Earth system applications.

Sensitivity can be illustrated conceptually through perturbation-based approaches, such as ensemble experiments, examining forecast outcomes in response to perturbations in the initial state or model configuration (e.g., Fig.~\ref{fig:l63_combine}). While intuitive and fully non-linear, ensemble approaches are computationally expensive compared to single model runs, subject to sampling error, difficult to tune, and often yield large data volumes from which concise and interpretable sensitivity information is difficult to extract. These limitations motivate more direct alternatives.

Gradient-based techniques offer a direct solution. In particular, adjoint models\footnote{An adjoint model computes the transpose of the Jacobian of a model with respect to its inputs, enabling gradients to be propagated backwards from a scalar output to all inputs in a single pass, analogous to backpropagation in neural networks.} enable efficient computation of a scalar forecast quantity's gradient with respect to a large model input space, with a computational cost that scales primarily with the cost of the forward model simulation \citep{le1986variational}. Adjoint-based sensitivity analysis has thus become established in numerical weather prediction, diagnosing forecast error growth and identifying dynamically influential regions and processes \citep{rabier1996sensitivity}, without needing an ensemble of model runs.

A wide range of applications illustrates the flexibility of adjoint-based sensitivity methods, including forecast error attribution \citep{mahfouf2007adjoint, lorenc2014forecast}, diagnosis of sensitivities to cloud-related variables and environmental fields \citep{benedetti2003variational}, tracking influential features \citep{wang2018tracking}, and designing targeted observing strategies \citep{pu1998forecast}. 
Diagnostics can use tailored response functions: \citet{kleist2005application} defined multiple response functions to assess the sensitivity of a poorly forecast US East Coast snowstorm to initial conditions, including energy-weighted forecast error, lower-tropospheric cyclonic flow, 750-hPa frontogenesis, and vertical motion.
\cite{doyle2012adjoint} define the adjoint sensitivity response function as kinetic energy within a specified region, to quantify how initial-state perturbations influence tropical cyclone formation and intensification.

\begin{figure}[h!]
    \centering
    \includegraphics[width=0.99\textwidth]{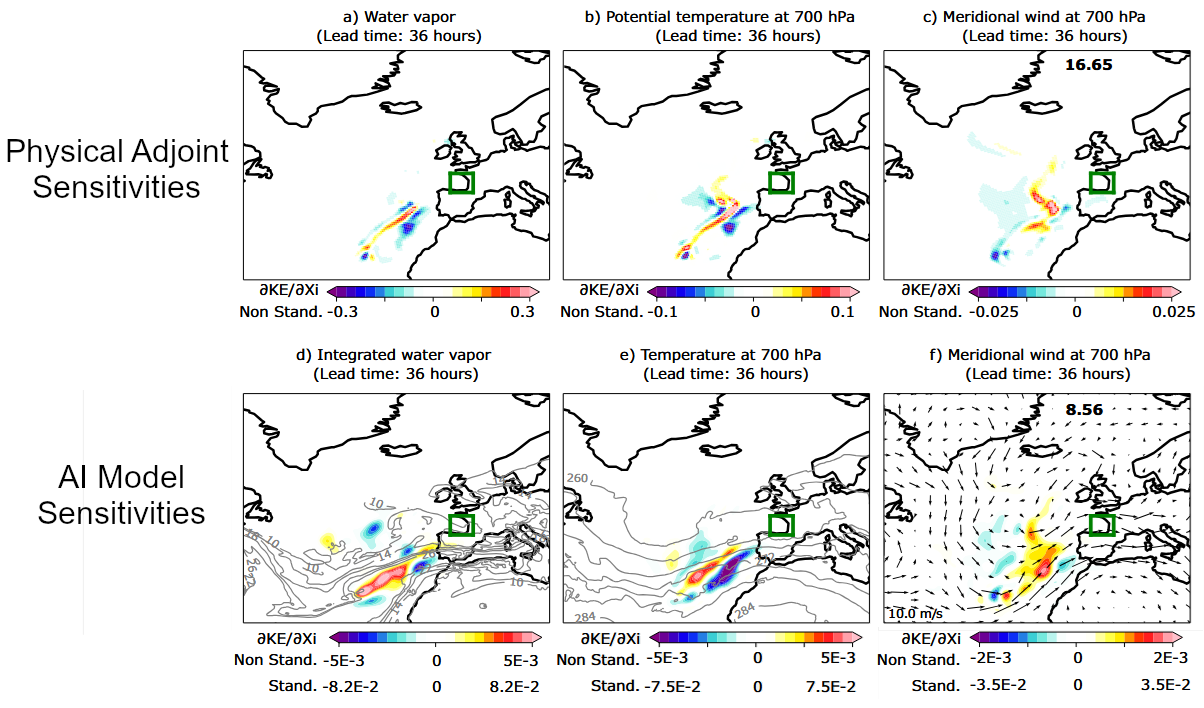}   
    \caption{\citet{bano2025ai}. Top panels: sensitivities of water vapour, temperature, and meridional wind fields to kinetic energy over the Bay of Biscay, from the adjoint of a physics-based model (Coupled Ocean-Atmosphere Mesoscale Prediction System, COAMPS). Bottom panels: sensitivities of equivalent quantities from a gradient-based backpropagation technique of an AI model (Spherical Fourier Neural Operator, SFNO).
    }
    \label{fig:bano_sensitivities}
\end{figure}

Recent work highlights strong conceptual and mathematical connections between these traditional sensitivity techniques and modern ML approaches. 
\citet{geer2021learning} notes the equivalence between tangent-linear/adjoint models, long used in variational data assimilation, and neural network backpropagation, framing both as gradient-based optimisation applied to high-dimensional non-linear systems. 
Building on this, \citet{bano2025ai} demonstrate sensitivity fields from learned models as a diagnostic tool comparing AI-based representations with physically based climate models, which is similar to our sensitivity-based family in our agreement evaluation of Section~\ref{sec:eval_metrics}. As shown in Fig.~\ref{fig:bano_sensitivities}, for cyclone Xynthia both approaches identify the same narrow moisture filament within the atmospheric river as the primary driver of intensification, with very similar spatial structures across water vapour, temperature, and meridional wind fields.
Because these sensitivity fields directly reflect the Jacobian of the model response with respect to the atmospheric state, their close correspondence indicates that the AI model has learned somewhat realistic dynamical relationships consistent with those represented by the physical model (even if they are smoother than their physical counterparts), rather than merely statistical associations that would seem unphysical (e.g., a highly non-local sensitivity).

Further gradient-based approaches have been applied to phenomena such as the El Ni\~no-Southern Oscillation (ENSO), where sensitivity analyses assess how ML models capture key dynamical features relative to established understanding \citep{ chen2025toward, zhou20253d}. \citet{vonich2026atmospheric} demonstrate that gradient-based sensitivity analysis through GraphCast can identify systematic corrections to ERA5 initial conditions, reducing 10-day forecast errors by 86\%, showing how backpropagation-derived insights can expose reanalysis biases and inform future data assimilation schemes.

\citet{laloyaux2025using} examine sensitivity-based explanations for GraphDOP, a graph neural network trained directly on heterogeneous meteorological observations without an explicit physical model or data assimilation cycle. The authors apply gradient-based forecast sensitivity to observations, a diagnostic tool well-established in data assimilation, to assess how different data sources influence forecast error. Crucially, even in the absence of the physical scaffolding on which such diagnostics were originally developed, the method recovers physically coherent influence patterns and sensible rankings of observation types, suggesting GraphDOP has learned a genuine internal representation of the Earth system.

\citet{tian2026evaluating} derived tangent-linear and adjoint models for two ML weather models, a graph neural network (GraphCast) and a hybrid model combining a dynamical core with learned parameterisations (NeuralGCM), and compared their sensitivities against a conventional NWP model (MPAS-A). Although the ML Jacobians were numerically exact, the graph neural network retained spurious sensitivity at the perturbation location, lacked physically expected upstream structure, and overestimated moisture sensitivities by an order of magnitude, while the hybrid model showed noise across vertical levels. Since an ML adjoint is identical to a gradient-based saliency map, this demonstrates that physical adjoints offer an independent benchmark for judging whether model sensitivities, and hence gradient explanations, are physically plausible.

\subsection{Attribution-based approaches}
\label{sec:apps_attribution}
Unlike the sensitivities above, attribution methods decompose a prediction into contributions from individual inputs. Any such decomposition is necessarily relative to a reference: an attribution explains the difference $f(\mathbf{x}) - f(\bar{\mathbf{x}})$ between the prediction and that at some reference input $\bar{\mathbf{x}}$. Only pure gradient sensitivities (Sect.~\ref{sec:XAI-review}\ref{sec:apps_sensitivity}) are genuinely baseline-free, since they describe local sensitivity rather than decomposing an output difference. Attribution methods differ, however, in how explicitly this reference is specified. SHAP and Integrated Gradients require an explicit baseline, a reference input or background distribution chosen by the user, whereas Input$\times$Gradient and LRP fix one implicitly through the method itself. We group applications by this distinction, since the reference determines the question an attribution answers (Sect.~\ref{sec:XAI-review}\ref{sec:apps_synthesis}).

\subsubsection{Explicit-baseline methods: SHAP and Integrated Gradients}
\label{sec:apps_shap}
SHAP and Integrated Gradients both define attributions relative to a user-specified reference (Sect.~\ref{sec:illustrate_XAI_methods}). SHAP averages each feature's marginal contribution over coalitions of the other features, with absent features marginalised over a background dataset (e.g., KernelSHAP, DeepSHAP) or, in TreeSHAP's default mode, over the training distribution encoded in the trees, making the reference arguably implicit in that case. Integrated Gradients accumulates gradients along a path from a reference input to the actual input. SHAP-based attribution is widely used in the Earth science literature, in both regression and classification contexts, typically used to quantify the marginal influence of individual variables on model outputs \citep[e.g.,][]{huang2022analysis, wang2024multiple, kan2025seasonal, schiller2025artificial, higgs2026hybrid}. Previous regression studies have found that Integrated Gradients and SHAP-based methods give almost identical results when computed relative to the same baseline \citep{mamalakis2023carefully}, although the two are only guaranteed to coincide for additive models and can diverge where features strongly interact.

\begin{figure}[h!]
    \centering
    \includegraphics[width=0.99\textwidth]{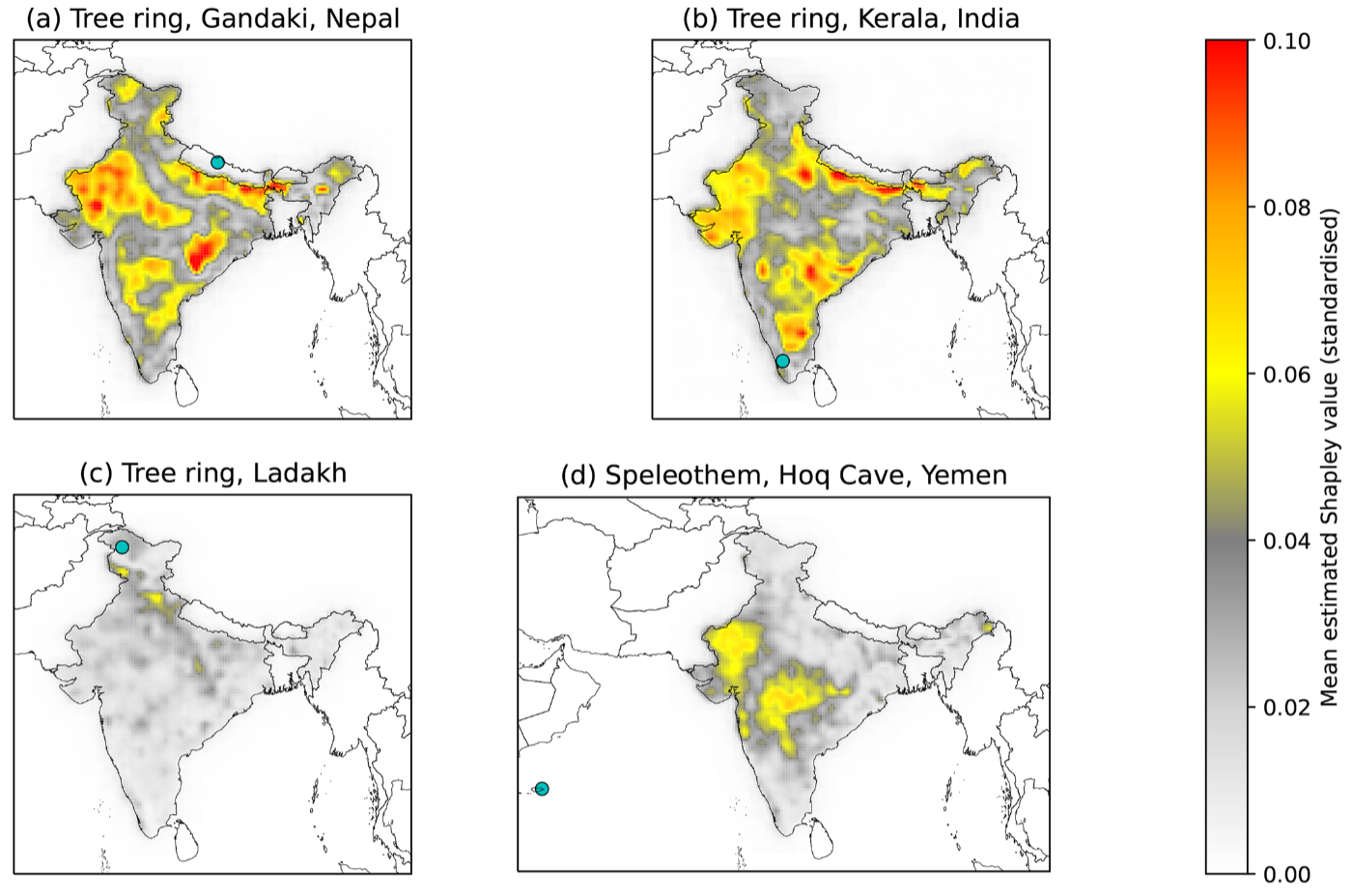}
    \caption{\citet{hunt2025novel}. Mean estimated Shapley value magnitudes for four input palaeoclimate records in the CNN ensemble model.
    Values are standardised: a grid point value of 0.1 means the selected dataset changes the predicted standardised seasonally averaged monsoon value at that grid point by $0.1\sigma$ on average.
    }
    \label{fig:hunt_shap}
\end{figure}

\citet{hunt2025novel} train convolutional neural networks on palaeo-environmental proxy datasets and instrumental rainfall observations to reconstruct South Asian monsoon rainfall anomalies over the past five centuries. They apply KernelSHAP to quantify each proxy's relative influence, both overall and as its mean impact on predicted anomalies at each grid cell (Fig.~\ref{fig:hunt_shap}). Records with a strong local signal appear as spatially concentrated features (e.g., Fig.~\ref{fig:hunt_shap}c), while those tied to large-scale monsoon dynamics show influence extending well beyond their source region (e.g., Fig.~\ref{fig:hunt_shap}d).

\citet{hunt2024using} train an XGBoost decision-tree ensemble to predict characteristics of the low pressure systems that often bring South Asian monsoon rain, using TreeSHAP (an efficient, exact method for decision trees). To mitigate misleading attribution between correlated variables, feature pairs/groups sharing a correlation of at least 0.5 are pruned until none remain. Beyond a transparency exercise, this generates new insight into the system's underlying physics, e.g., mechanisms of low-pressure system intensification, bridging data-driven modelling and traditional physical reasoning.

\citet{abdullah2025explainable} develop an attention-based Temporal Convolutional Network for 3-hourly dew point temperature prediction at 35 Bangladeshi weather stations, outperforming recurrent and convolutional baselines. SHAP is applied globally to rank feature importance and guide feature selection, alongside LIME (a surrogate linear model) for local explanations. SHAP identifies the long-term trend component, from seasonal decomposition, as the dominant driver, followed by day/night indicators and cyclical time features.

\citet{mamalakis2023carefully} train a fully connected ANN to predict ensemble- and global-mean temperature from a yearly mean global temperature field. Comparing integrated gradients and DeepSHAP, both requiring a baseline, they show attributions can vary greatly with the baseline chosen, which they argue can be beneficial if considered carefully, since different baselines answer different questions. A zero baseline asks \textit{``which patterns made the network predict 15.4°C as opposed to 0°C?''}, yielding broad tropical/subtropical attributions reflecting solar radiation and greenhouse-effect physics. A climatological baseline (1850--1880 mean) asks \textit{``which regions made 2022 warmer than the pre-industrial period?''}, highlighting land and midlatitude ocean regions. A nearby future year (2025) asks only \textit{``which regions made 2022 cooler than 2025 by 0.3°C?''}, producing low-magnitude attributions that do not highlight ENSO, consistent with the network treating ENSO as internal rather than forced variability. Each baseline is physically defensible, yet each tells a fundamentally different story.

\citet{ye2025explainable} train an LSTM on physically based model simulations to predict three-dimensional soil water movement at grid scale, and analyse SHAP attributions spatially and temporally. Aggregating these local explanations, they identify emergent patterns of hydrologic connectivity across a watershed, linking threshold-like changes in feature importance to organised runoff responses, illustrating how XAI can move beyond local explainability to support cross-scale process understanding.

\subsubsection{Implicit-baseline methods: Input$\times$Gradient and LRP}
\label{sec:apps_lrp}
Input$\times$Gradient and layer-wise relevance propagation (LRP) require no user-specified baseline, but are not reference-free. A first-order Taylor expansion gives $f(\mathbf{x}) - f(\mathbf{0}) \approx \mathbf{x}\cdot\nabla f(\mathbf{x})$, so Input$\times$Gradient implicitly attributes relative to an all-zero input. LRP instead propagates the prediction back through the network as feature-wise relevance; many rules are designed to conserve relevance, so that it decomposes the prediction itself, although bias terms and stabilisers (e.g., $\epsilon$) make this approximate \citep{montavon2019layer}. Viewed as a deep Taylor decomposition, each rule corresponds to a reference (root point) at which a neuron's output vanishes, reached by moving that neuron's inputs in a rule-specific direction; the reference is therefore defined neuron by neuron, not as a single input state \citep{montavon2019layer}. LRP-0 (and approximately LRP-$\epsilon$) shrinks all inputs towards zero, matching the zero reference of Input$\times$Gradient, which $\epsilon$-LRP approximates for ReLU networks \citep{kindermans2016investigating}. The $\alpha1\beta0$ rule shrinks only the positively contributing inputs, so its reference keeps the opposing (inhibitory) inputs unchanged and they never enter the attribution. The $\gamma$-rule shrinks all inputs but the positively contributing ones more strongly, interpolating between the two. Choosing an LRP rule therefore implicitly chooses the reference. As the regression applications reviewed here predominantly use LRP, we focus on it below.

\citet{hilburn2020development} develop a CNN to transform observed radiances and lightning into synthetic radar reflectivity fields for assimilation in short-term convective-scale forecasts. They use LRP with the $\alpha$1-$\beta$0 rule, which propagates only positive relevance: a natural choice here (and less limiting than in Sect.~\ref{sec:illustrate_XAI_methods}\ref{sec:lrp_illustration}.\ref{sec:lrp_alphabeta}), since the output is bounded and non-negative and the focus is on extreme rather than typical values.

\citet{kim2022untangling} apply LRP to an RNN predicting PM$_{2.5}$ concentrations from land-use, emission-related, and meteorological predictors. A small subset of input time-steps and variables accounts for $\sim80\%$ of total relevance, and retraining on this subset gives no significant loss of skill, demonstrating XAI's utility for feature selection (``which inputs are necessary for a minimal skilful model?'').

\citet{tao2024explainable} combine Boruta feature selection\footnote{Boruta compares each feature's importance against randomly permuted copies of the inputs (shadow features), retaining only features that consistently outperform them \citep{kursa2010boruta}.} with LRP for daily streamflow forecasting, showing that a modified LSTM assigns higher relevance to precipitation than a standard LSTM, indicating the architecture adjustment yields more physically meaningful input-output relationships.

\citet{hoffman2025evaluating} examine the reliability of several XAI techniques (here, we focus on LRP) for a CNN forecasting daily Arctic sea ice velocity, applying local and global (sample-aggregated) LRP with $\alpha1\beta0$ rules for convolutional layers and $\epsilon$ rules elsewhere. LRP highlights patterns consistent with known physical drivers such as wind forcing, suggesting that the CNN relies on relationships that align with established physical understanding.

\subsection{Bespoke explainability and interpretability approaches}
\label{sec:assorted_other}

Although the main scope of this review is input-feature sensitivity and attribution, we briefly consider several adjacent approaches to show the wider XAI landscape in Earth system regression. These answer different explanatory questions and do not always fit neatly into existing XAI taxonomies, being tailored to the application through context-specific information and/or tools.

\citet{jimenez2025ai} combine AIWP predictions with physics-based global climate models to quantify the anthropogenic contribution to heatwave events, comparing AI-generated forecasts under factual (observed) and counterfactual (no anthropogenic forcing) conditions. From an XAI perspective, this improves interpretability by explicitly linking AI outputs to physically meaningful causal contrasts rather than post-hoc feature attributions. \citet{trok2024machine} similarly use convolutional neural networks trained on climate-model simulations to construct counterfactual versions of an observed extreme event under different levels of global mean temperature, holding the meteorological conditions fixed while varying the level of warming.

\citet{spuler2025learning} seek dynamical patterns that are both predictable and informative of a local impact of interest, a combination conventional dimensionality-reduction methods tend to miss. Using a variational autoencoder (VAE) variant trained on ERA5 for targeted dimensionality reduction and probabilistic clustering of large-scale atmospheric fields conditioned on extreme precipitation over Morocco, they obtain latent circulation regimes more informative of local extreme precipitation than linear methods while maintaining subseasonal predictability, interpretable via known teleconnection patterns such as the Madden-Julian Oscillation and stratospheric polar vortex variability. In contrast to the post-hoc techniques above (Sects.~\ref{sec:XAI-review}\ref{sec:apps_sensitivity}-\ref{sec:XAI-review}\ref{sec:apps_attribution}), this embeds interpretability within the learning model itself, suggesting a direction for integrating interpretability with predictive skill in geoscientific ML.

\citet{horinouchi2025statistical} predict tropical cyclone rapid intensification in the western North Pacific using a scheme \citep[based on Wide Learning,][]{iwashita2020efficient} that scores each case by counting how many of twelve predictors (environmental conditions and storm state) fall within learned favourable ranges. Each predictor's contribution can therefore be read off directly, without post-hoc analysis, making the approach well suited to operational decision-making and communicating forecast reasoning to policymakers.

\citet{ham2023anthropogenic} use occlusion sensitivity, systematically masking local patches of the input precipitation field, to interpret a CNN's predictions of annual global mean 2m air temperature. Trends in occlusion sensitivity identify ``hotspot'' regions and daily precipitation variability influencing the prediction, consistent with SHAP and integrated gradients, suggesting the identified drivers are robust to attribution technique.

\citet{behnoudfar2025bridging} combine a high complexity model with an idealised, more interpretable model via latent data assimilation. The high complexity model's latent space is extended to include sparse observations of the idealised model; an LSTM predicts the evolution of this extended latent space and DA corrects it, aligning the two models. The authors argue this extension provides a physically interpretable ``handle'' through which corrective contributions can be traced back to the simpler model's structure, offering insight into how and why the operational model's biases are mitigated.

\begin{figure}[h!]
    \centering
    \includegraphics[width=0.85\textwidth]{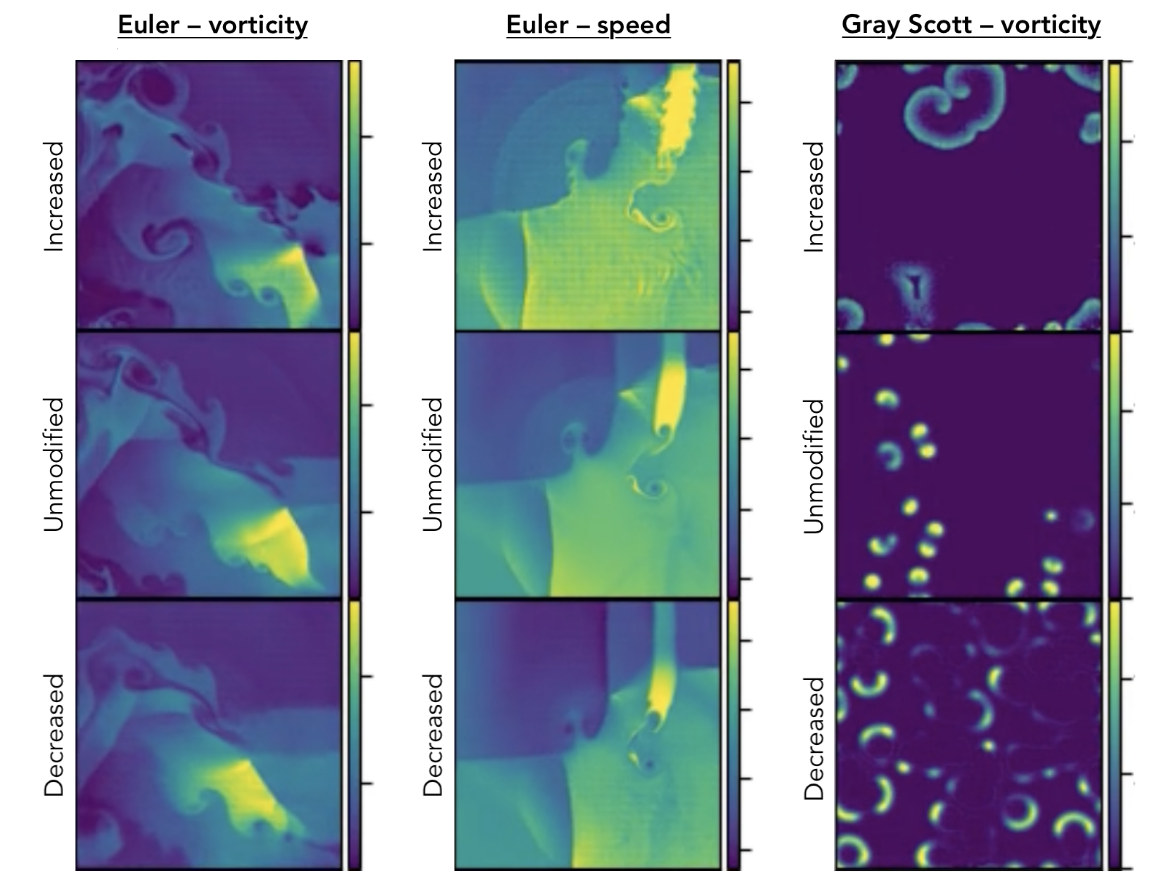}
    \caption{\citet{fear2025physics}. Causal manipulation of physical behaviour via activation steering in a physics foundation model. Concept vectors are identified by contrasting model activations across physical regimes and injected at inference time with steering coefficient $\alpha$, which induces ($\alpha > 0$), preserves ($\alpha = 0$), or suppresses ($\alpha < 0$) the target behaviour. Left and Middle: air density fields from $\Delta$-vortex and $\Delta$-speed injection into Euler quadrants. Right: chemical species B concentration from $\Delta$-vortex injection into the Gray--Scott reaction-diffusion system.
    }
    \label{fig:fear_latent_concepts}
\end{figure}

\citet{fear2025physics} ask whether a physics foundation model learns interpretable, causally meaningful internal representations that can be deliberately manipulated. Using a transformer\footnote{A neural network architecture built around self-attention, which dynamically weights the relevance of different input parts \citep{vaswani2017attention}; originally developed for natural language processing, transformers are now widely used in Earth system modelling.} pre-trained on partial differential equation simulations, they adapt activation steering methods from LLM interpretability \citep{arditi2024refusal}, identifying latent directions corresponding to physical concepts by contrasting model activations across physical regimes. Injecting these concept vectors at inference time systematically induces or suppresses specific physical behaviours (Fig.~\ref{fig:fear_latent_concepts}), demonstrating causal control rather than mere correlation, and the same directions generalise across different physical systems. This suggests scientific foundation models contain linearly accessible, manipulable representations of abstract physical concepts, supporting interpretation, debugging, and counterfactual scientific analysis.

\subsection{Synthesis of recurring limitations}
\label{sec:apps_synthesis}

The studies above show that each family of methods has yielded physically interpretable insight in Earth system regression problems. Several limitations nonetheless recur across them; we synthesise these here, noting which are shared across families and which have been demonstrated rather than hypothesised.

\paragraph{Correlated predictors.}
The strong spatial, temporal, and multivariate correlations typical of geoscientific data violate the feature independence assumptions underlying many SHAP implementations \citep{frye2020shapley}, potentially causing unstable or ambiguous attributions \citep{krell2025influence}. Conditional SHAP variants attempt to address this but introduce distributional assumptions rarely well-constrained in Earth system applications. The concern is most acute for generic KernelSHAP, which relies on a background dataset to marginalise over feature coalitions, and is left unaddressed in \citet{abdullah2025explainable} despite moderate correlations among several engineered temporal features. Correlation-based pruning \citep{hunt2024using} is effective but does not trivially scale to higher-dimensional feature sets. The issue extends beyond SHAP: attribution-guided feature selection \citep[e.g., using LRP in][]{kim2022untangling} can retain skill without reflecting the correct mechanisms, particularly where retained and discarded predictors are correlated. In the reviewed studies, this concern is raised or mitigated rather than quantified.

\paragraph{Design choices.}
In contrast, sensitivity to design choices has been demonstrated directly: \citet{mamalakis2023carefully} show that the baseline determines the question an attribution answers, so it should reflect a physically meaningful null state. Our L63 evaluation shows the same at the level of method families: attributions cluster into three groups by construction (local, state-weighted, and path-integrated derivatives), each closest to its own TLM-based analogue (Fig.~\ref{fig:method-correlation}), so explicit- and implicit-baseline methods effectively answer different questions. For explicit-baseline methods this choice is at least visible; for implicit-baseline methods it is fixed by the method and easily overlooked. The zero reference of Input$\times$Gradient, for example, depends on preprocessing: for standardised anomalies it corresponds to climatology, often physically sensible but still a choice. For LRP, rule selection plays the same role, a flexibility that is both a strength and a limitation, since it can meaningfully alter the resulting attributions: in L63, LRP-Alpha1Beta0 departs from Input$\times$Gradient ($r=0.68$) far more than the epsilon and gamma rules ($r=0.95$); as \citet{hilburn2020development} illustrate, rule choice should be specific and intentional. Counterfactual frameworks such as \citet{jimenez2025ai} and \citet{trok2024machine} likewise depend on the validity of their counterfactual assumptions and the underlying models' accuracy.

\paragraph{Scalability and transferability.}
For high-dimensional inputs or structured outputs, SHAP becomes costly (Eq.~\eqref{eq:shap}) and hard to interpret, often requiring approximation or aggregation that can obscure physical structure (compare Fig.~\ref{fig:app_shap_detail} with Fig.~\ref{fig:attr_KSHAP}). Aggregation to magnitudes is especially costly in regression, since the L63 method families separate far more cleanly under signed than magnitude-only correlation (Fig.~\ref{fig:method-correlation}). TreeSHAP is exact and efficient only for tree-based models, while KernelSHAP trades exactness for tractability. Gradient-based methods and LRP, requiring only a backward pass, are far less constrained; neural networks compute sensitivities via backpropagation without the substantial development and maintenance costs of traditional adjoint models \citep[][discusses some computational aspects]{geer2021learning}. For bespoke methods, the equivalent limitation is transferability: scoring schemes \citep{horinouchi2025statistical} are unlikely to generalise to higher-dimensional problems or capture complex predictor interactions, concept directions identified in simpler settings \citep{fear2025physics} may not scale to operational regimes with higher interaction complexity, and latent-space methods are tightly coupled to specific architectures.

\paragraph{Physical references and expert interpretation.}
Retained skill confirms which inputs are necessary, not whether attributions are physically realistic \citep{kim2022untangling}. In L63 the two even diverge: explanations agree least with their physical references in the stable lobes, where the network forecasts best, since agreement tracks the local Jacobian geometry rather than predictive accuracy (Fig.~\ref{fig:diagnostic_error_scatter_anisotropy}). Most attribution studies \citep[e.g.,][]{tao2024explainable, hoffman2025evaluating} therefore rely on an ``expert-in-the-loop'' to distinguish physically interpretable signals from method artefacts. Sensitivity-based methods are the exception: physical adjoints offer an independent benchmark, against which ML sensitivities have proved physically coherent in some cases \citep{bano2025ai} and physically implausible, despite being numerically exact, in others \citep{tian2026evaluating}. These comparisons have focused on raw sensitivities, which is natural given their direct correspondence with tangent-linear and adjoint models. Our L63 evaluation extends such benchmarks to attribution methods through matched TLM-based analogues (Sect.~\ref{sec:eval_metrics}), but also shows that no single agreement metric suffices (sign agreement, for instance, is coarse because signs are fixed by the governing equations), and the construction requires an exact TLM, unavailable for realistic high-dimensional models. Short of such benchmarks, convergent evidence from multiple methods \citep{ham2023anthropogenic} can reduce method-specific biases.

\paragraph{Method suitability.}
Taken together, these limitations indicate where each family is currently most defensible. Being gradient-free, SHAP is genuinely model-agnostic and its additive, signed attributions are easy to communicate, making it a reasonable choice for tabular or moderately aggregated problems and non-differentiable models, provided correlations are addressed (e.g., through grouping) and baselines are physically meaningful \citep[e.g.,][]{hunt2024using, ye2025explainable}, and aggregation is applied with care: in L63, aggregated SHAP scores hide that most of $y$'s importance acts through output $z$ (compare Figs.~\ref{fig:attr_KSHAP} and~\ref{fig:app_shap_detail}). Gradient-based sensitivity and LRP, being computationally efficient, suit high-dimensional problems: sensitivities are currently the most verifiable, since adjoint-based physical modelling offers an independent reference (in L63, saliency scores highest against the TLM; Fig.~\ref{fig:faithfulness-3panel}), while LRP requires careful justification of rule choices (in L63, the rule alone separates LRP-Alpha1Beta0 from the rest of its family; Sect.~\ref{sec:illustrate_XAI_methods}\ref{sec:lrp_illustration} and Fig.~\ref{fig:method-correlation}) and close collaboration between ML practitioners and domain experts. Bespoke methods can be substantially more precise, answering specific mechanistic questions rather than primarily validating predictive skill, but often trade off the breadth, robustness, and model-agnostic applicability of generic post-hoc approaches.

\subsection{Explanations for whom? A limitation of technical evaluation}
\label{sec:stakeholders}

The analysis in Sections~\ref{sec:illustrate_XAI_methods},~\ref{sec:eval_metrics}~and~\ref{sec:XAI-review} is primarily a technical overview of how XAI methods behave and inspects their agreement with known physical model behaviour. Here we offer a brief perspective on a different point: what counts as a useful explanation will depend on the context of who is asking and why, so an explanation can be technically valid yet only partially meet a given stakeholder's needs. Model developers may want diagnostics or inherently interpretable architectures that expose weaknesses, biases or training artefacts \citep[e.g.,][]{barnes2022looks, achtibat2024attnlrp, fear2025physics, vonich2026atmospheric}. Evaluators may want to understand why models with similar benchmark skill behave differently \citep{rasp2024weatherbench, loegel2025ai}. Domain experts and forecasters may want to test whether learned dependencies are physically plausible, often through targeted case studies or extreme events \citep{bonavita2024some, charlton2024ai, dacre2025northern, singh2026generative}. Decision makers and affected communities are unlikely to engage with attribution maps at all, yet their experience can reveal where models fail to meet practical needs \citep{brocker2026verification}. Needs also differ within groups: a farmer deciding when to plant may need a local explanation of a single forecast, while a government body setting water policy may need a global view of model behaviour.

\section{Conclusions}
\label{sec:trends_and_future_directions}
\subsection{Summary and key takeaways}
XAI aims to link predictive performance with scientific understanding, building the confidence needed for high-stakes use. In this work, we focus on interpreting XAI outputs in regression-focused Earth system settings, in contrast to previous studies that tend to focus on classification or do not make the distinction explicit. Specifically, we present: (i) a novel illustration of XAI methods applied to a regression task, using a forecast emulator of the L63 model (Sect.~\ref{sec:illustrate_XAI_methods}); (ii) physically derived reference explanations and a suite of metrics for interpreting XAI output against them (Sect.~\ref{sec:eval_metrics}); and (iii) a review of XAI applications to regression problems in Earth system science, with a perspective on the distinction between stakeholder needs and technical limitations (Sect.~\ref{sec:XAI-review}). The recurring limitations identified in the literature are synthesised in Sect.~\ref{sec:XAI-review}\ref{sec:apps_synthesis}; here we draw out the broader conclusions, ordered from those established by theory or demonstrated here to those that remain open:
\begin{itemize}
    \item \textbf{The mathematics transfers; the interpretation does not.} XAI methods carry over mathematically from classification, but regression lacks the discrete decision boundaries that often provide context for classification explanations, while objective ground truth for evaluating attributions is rarely available in realistic applications. In practice, the literature therefore relies largely on expert judgement to separate physically interpretable signals from method artefacts.

    \item \textbf{Gradient-based explanations can be matched to physics-derived references.} Saliency is mathematically equivalent to adjoint sensitivity, while our state-weighted and path-integrated constructions provide matched TLM-based references for the corresponding attribution families (against which each method family scores highest in L63; Sect.~\ref{sec:eval_metrics}). For AI weather models, sensitivity comparisons (saliency only) have already revealed both physically coherent \citep{bano2025ai} and implausible \citep{tian2026evaluating} sensitivities.

    \item \textbf{Every attribution answers a question set by its reference.} Attributions are strongly baseline dependent: in L63 they cluster into three families (local, state-weighted, and path-integrated derivatives), each matching its own TLM-based analogue, and the same dependence has been demonstrated for global temperature prediction \citep{mamalakis2023carefully}. Within LRP, rule choice alone separates LRP-Alpha1Beta0 from the rest of its family.

    \item \textbf{Off-manifold evaluation is established; its consequences are not yet quantified.} Because marginal SHAP assumes feature independence and Integrated Gradients interpolates from a baseline \citep[Sects.~\ref{sec:illustrate_XAI_methods}\ref{sec:integrated_gradients}--\ref{sec:illustrate_XAI_methods}\ref{sec:shap_illustration};][]{frye2020shapley, zhang2026baselines}, both evaluate correlated geoscientific inputs off the data manifold by construction. How much this distorts attributions in realistic Earth system models remains unquantified (Sect.~\ref{sec:XAI-review}\ref{sec:apps_synthesis}), but early synthetic experiments indicate significance \citep{krell2025influence}.

    \item \textbf{Aggregation discards information that matters in regression.} Aggregating attribution magnitudes discards the sign of influence by definition; in L63, sign is what most sharply distinguishes the method families (Figs.~\ref{fig:method-correlation}~and~\ref{fig:faithfulness-3panel}), and aggregation can also hide which output a feature acts through (comparing aggregations in Sect.~\ref{sec:illustrate_XAI_methods} to counterparts in Appendix~A\ref{sec:app_complete_XAI_structure}).

    \item \textbf{Accurate forecasts do not imply correctly learned dynamics.} In L63, the emulator’s explanations agree least with their physical references where it forecasts best, because agreement tracks the local Jacobian geometry rather than predictive accuracy (Figs.~\ref{fig:diagnostic_error_scatter_anisotropy}~and~\ref{fig:diagnostic_forecast_error_vs_anisotropy}); likewise, retaining skill on a reduced feature set does not confirm that the selected attributions are physically realistic \citep{kim2022untangling}. Whether the same gap appears in operational models has not yet been tested.

    \item \textbf{Scalability favours gradient-based methods.} Gradient-based sensitivities and LRP require only a backward pass, whereas exact SHAP scales exponentially with feature count. In practice, SHAP is applied mainly to moderately aggregated problems, where correlated predictors are handled by bespoke pruning or grouping, while gradient-based sensitivities readily run on global AI weather and observation-driven models \citep[e.g.,][]{bano2025ai, laloyaux2025using, vonich2026atmospheric}.

    \item \textbf{L63 is a best case.} L63 allows complete tangent-linear model recovery, comprehensive permutation experiments, and a well-mapped phase space, none of which is possible for large-scale Earth system models, where climate change also introduces data drift. Not all applications will have a physical adjoint either, so how to judge and interpret explanations without one is an open challenge.
    
    \item \textbf{Bespoke techniques are promising but untested at scale.} Approaches probing AI models' internal mechanisms show strong potential to support model verification and deployment for domain experts and end-users \citep{brocker2026verification}, but evidence so far comes from specific, well-defined settings, and their transferability to operational models is unknown.
\end{itemize}

\subsection{Future directions}
We group the findings of this survey into three themes. Together, these point towards a future in which XAI provides well-integrated diagnostic information into AI model workflows, supporting trust and transparency, as well as scientific insight and model advancement.

\subsubsection*{Theme 1: Physically grounded references and evaluation.}
The most immediate opportunity is to make physical reference points routine. Wherever tangent-linear or adjoint models exist, as in many NWP and data assimilation systems, explanations should be benchmarked against matched physical analogues, as in our TLM-based framework (Sect.~\ref{sec:eval_metrics}). These provide a reference point rather than a ground truth, but would let the physical realism of what models have learned be compared across models and methods, rather than asserted. Applications without a physical adjoint will need alternative references, such as idealised or intermediate-complexity models with known dynamics, alongside regression-specific evaluation frameworks that combine complementary metrics with physical consistency or stakeholder-defined criteria \citep[e.g., building on][]{mamalakis2022neural}. Baseline choice should likewise become principled rather than ad hoc: defined by the scientific question (e.g., a climatological mean, an observational reference, or a stakeholder-defined counterfactual) and reported with every attribution. Perturbation and attribution methods that respect physical dependencies, for example by keeping perturbed inputs on the data manifold or encoding known relationships between correlated variables, are a further priority. Finally, XAI could complement out-of-sample detection, with unrealistic gradients or attributions during extreme events flagging that a model has failed to learn the underlying physics.

\subsubsection*{Theme 2: Scaling to large systems such as NWP.}
Gradient-based explanations are the natural route to operational scale, since adjoint-like sensitivity information is more readily accessible from an AI model than from traditional adjoint construction \citep{bano2025ai}. Future work should move beyond single scalar response functions towards structured, multi-output diagnostics, for example the leading singular vectors of the learned Jacobian, which connect directly to established predictability analysis. For LRP, a theoretical basis for rule selection in regression is needed, building on the clearer justification of the epsilon rule. Explanations should also be built at physically meaningful scales, for instance from emergent ``super-features'' identified by clustering, and by methods designed to operate across interacting spatial and temporal scales at once. Systematic frameworks for probing internal representations (e.g., latent spaces) could then test directly whether operational models encode known physical concepts, dynamical regimes, and scale interactions, building on early demonstrations \citep{fear2025physics, behnoudfar2025bridging}. 

\subsubsection*{Theme 3: Stakeholder-specific usefulness.}
XAI should become part of the model development cycle rather than a post-hoc check, used routinely to diagnose how model components contribute to skill and biases and to target improvements. This requires two-way engagement: XAI methods adapted to Earth system modelling's specific characteristics, and domain scientists validating XAI outputs against process understanding, for example through shared benchmarks and co-designed case studies. For downstream decision-makers, the goal should be concise, actionable summaries derived from attribution maps rather than the maps themselves, with their design informed by the users who rely on them.

Underlying all three themes is a single principle: refinement of XAI methods must be guided by the questions stakeholders actually need answered, rather than an aspiration to fully characterise model behaviour, which is neither tractable nor, in most operational contexts, useful.

\subsection{Practical recommendations}
For practitioners applying existing XAI methods to regression problems in Earth system science, we distil the key lessons from this review into practical recommendations:

\begin{enumerate}[label=\Roman*.]
\item \textbf{Do not assume interpretations of XAI methods popularised in classification settings transfer directly to regression.} Classification can provide convenient interpretive anchors that regression lacks, such as a discrete decision boundary or class-specific target against which an explanation can be framed (see II). Without these, explanations must characterise behaviour across a continuous output space (see III), where sign, magnitude, and variation carry richer meaning, and where aggregation choices suited to binary outcomes can obscure important structure (see IV).
\item \textbf{Choose your baseline or reference states with intentionality.} Attributions are strongly baseline dependent (\citealp{mamalakis2023carefully}; Sect.~\ref{sec:eval_metrics}), so for methods which require a baseline, ground it in a physically meaningful, justifiable reference, such as a dynamical state of interest, or a quantity relevant to the stakeholder's question, and report it. An arbitrary baseline produces attributions that are difficult to interpret and potentially misleading. Methods without an explicit baseline still have one: for Input$\times$Gradient it is a zero input (for standardised anomalies, climatology), and for LRP it is set by the propagation rule. Asymmetric rules such as $\alpha1\beta0$ discard opposing evidence and should be avoided for signed regression targets (Sect.~\ref{sec:illustrate_XAI_methods}\ref{sec:lrp_illustration}).
\item \textbf{Prioritise inputs ``familiar'' to the model.} Attributions from physically implausible inputs (or which differ significantly from the training data) lose their connection to realistic behaviour, and speak to the model's limitations rather than the phenomenon of interest. Marginal SHAP and Integrated Gradients produce such inputs by construction when predictors are correlated (Sects.~\ref{sec:illustrate_XAI_methods}\ref{sec:integrated_gradients}--\ref{sec:illustrate_XAI_methods}\ref{sec:shap_illustration}), so should not be applied unmodified to strongly correlated, high-dimensional inputs; instead, group correlated features and choose baselines from realistic states, e.g., from training data, a model ensemble member, or an observed state with similar large-scale conditions. Off-manifold explanations may still help diagnose model failures, but should be clearly distinguished from physically meaningful ones.
\item \textbf{Justify aggregation of explanation metrics, or examine in more detail.} Removing sign information, averaging, or grouping features are common but consequential choices that should be explicitly justified; the full attribution values can contain rich information warranting close examination before being discarded. In L63, sign is what most sharply distinguishes the method families, and aggregation hides which output a feature acts through (Sect.~\ref{sec:eval_metrics}; Fig.~\ref{fig:app_shap_detail}). Attribution values might be aggregated over clusters of related features, justified by domain considerations (e.g., attributing to a weather front rather than individual grid cells) or by correlation structure.
\item \textbf{Do not treat predictive skill as evidence of physically realistic learned behaviour.} In L63, explanations agree least with their physical references where the network forecasts best, and retaining skill after attribution-based feature selection does not confirm physical realism \citep{kim2022untangling}. Where tangent-linear or adjoint models exist, benchmark sensitivities and attributions against matched physical analogues (Sect.~\ref{sec:eval_metrics}), using several complementary agreement metrics rather than one.
\item \textbf{Match explanations to stakeholder questions.} Do not attempt to explain a model in its entirety; this is ill-posed, since the most complete description of a model is the model itself. Explanation is inherently lossy, and that loss should be directed by a concrete stakeholder question: define the question first, then select or design an appropriate method. For example, an operational forecaster may need to know which features drive an unusual prediction in a specific event, while a developer may care whether the model has systematically learned physically consistent sensitivities across regimes; conflating such different questions risks explanations that are technically correct but practically useless.
\end{enumerate}

%

%

\clearpage
\acknowledgments
The authors gratefully acknowledge the collaborative support and funding provided by the University of Reading and the UK Met Office, which enabled the work reported in this publication via the Met Office Academic Partnership framework.


%
%
\datastatement All code used for the L63 illustrations in this work can be found at https://github.com/ichiggs/Evaluating-XAI-Methods-L63


%



\appendix[A] 


\appendixtitle{Additional detail on illustrative L63 explanations}
\label{sec:appendix_detail_A}
%



\subsection{Structure of explanations}
\begin{figure}[h!]
    \centering
    \includegraphics[width=0.5\linewidth]{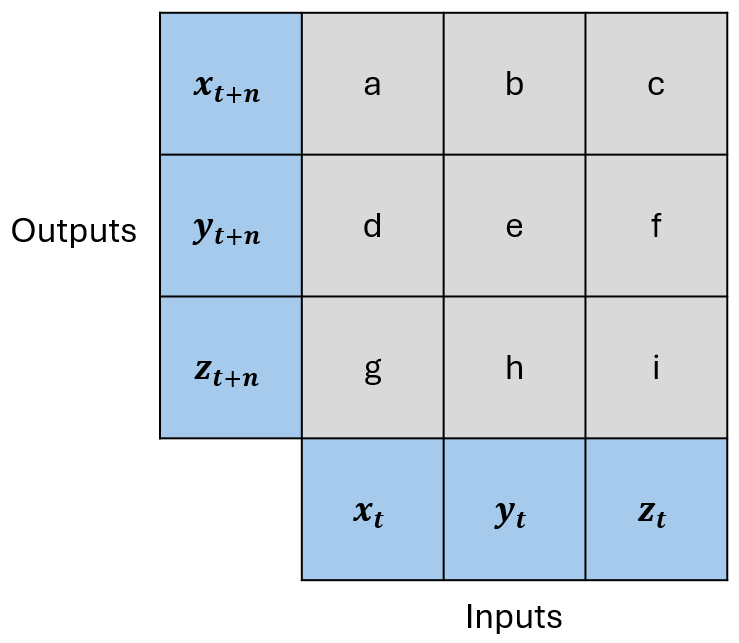}
    \caption{Labelled structure of the matrix of values given by iterating an XAI method over every output target in the L63 state vector (Eq.~\eqref{eq:l63_state}). Each element of the matrix corresponds to a pairwise relationship between an element of the state vector at times $t$ and $t + n$, where $n$ is the lead time of the forecast. }
    \label{fig:app_summarise_matrix}
\end{figure}

Figure~\ref{fig:app_summarise_matrix} shows the structure of the full attribution, sensitivity or relevance matrix described in Sect.~\ref{sec:illustrate_XAI_methods}\ref{sec:scalar_objective}. Each entry quantifies the pairwise relationship between one input feature and one output feature, yielding a matrix of size $N\times M$ for any ML model with $M$ inputs and $N$ outputs; in the L63 system considered here, this gives a $3\times3$ matrix.

The summed result of Eq.~\eqref{eq:aggregation} is the column-wise summation of the absolute values in Fig.~\ref{fig:app_summarise_matrix}. That is to say, our ``Importance'' for $x_t$ is found by: $\text{a} + \text{d} + \text{g}$.

An individual input attribution vector for a single output quantity, given by each of Eqs.~\eqref{eq:saliency}-\eqref{eq:lrp_attribution}, corresponds to a row in Fig.~\ref{fig:app_summarise_matrix} (e.g., if the output target is set to be $z_{t+n}$ then the vector returned consists of elements $[\text{g}, \text{h}, \text{i}]$).

\subsection{A note on the effect of alternative baseline selection}

\begin{figure}[h!]
    \centering
    \includegraphics[width=0.99\textwidth]{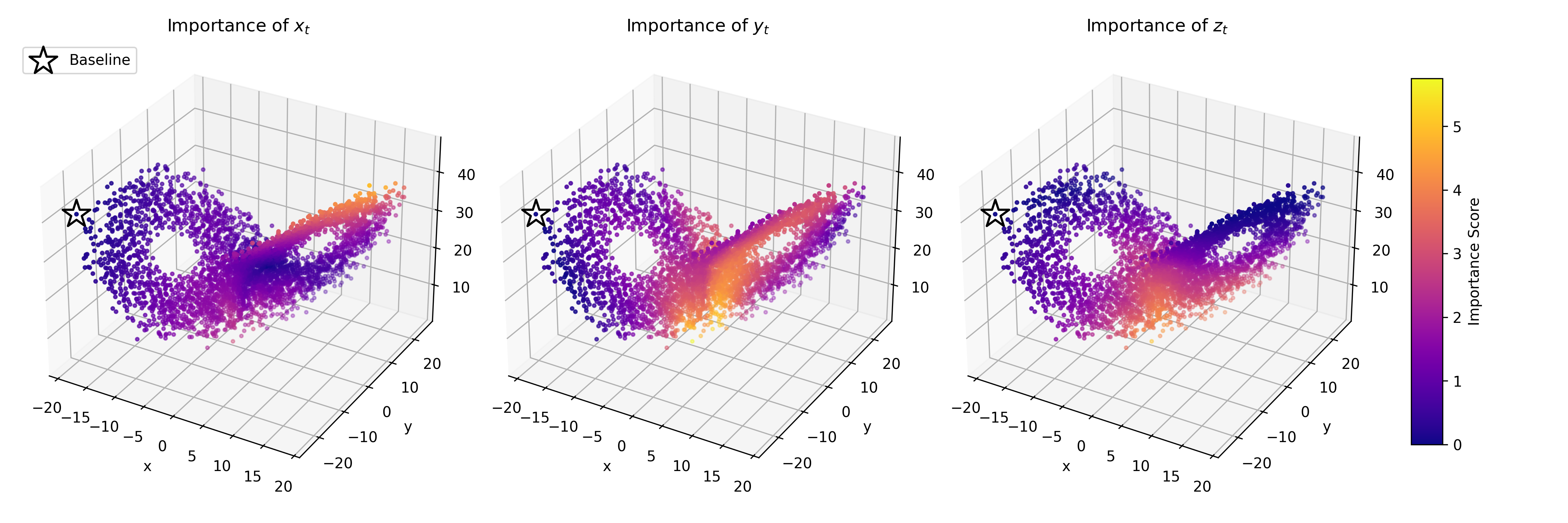}   
    \caption{IntG-derived importance scores $\bar{A}_i(\mathbf{x})$ for each input feature of the L63 system, computed via Eq.~\eqref{eq:aggregation}~and~\eqref{eq:integratedgradients}. Each panel shows the L63 attractor, coloured by the scores attributed to the corresponding input variable. Here, a single baseline state (white star) at the far-left edge of the left lobe is used for every explanation, in place of the climatological mean used in Fig.~\ref{fig:attr_IntG}.
    }
    \label{fig:attr_IntG_time}
\end{figure}

Figure~\ref{fig:attr_IntG_time} repeats the IntG attribution (Fig.~\ref{fig:attr_IntG}, Sect.~\ref{sec:illustrate_XAI_methods}\ref{sec:integrated_gradients}) with an alternative baseline to illustrate how strongly the reference state shapes the resulting explanations, not to advocate for any particular baseline as optimal. 
The mathematical formulation is unchanged, and the baseline is again a single state shared by every explanation. However, instead of the climatological mean, it is now an extreme state at the far-left edge of the left lobe. 
Because IntG attributions satisfy completeness, $\sum_i A_i(\mathbf{x}) = f(\mathbf{x}) - f(\mathbf{x}')$, changing the baseline changes the question being answered. With the mean baseline, IntG explains why the forecast from $\mathbf{x}$ differs from the forecast from the climatological mean state. With the left-lobe baseline, it explains why the forecast from $\mathbf{x}$ differs from the forecast from that extreme left-lobe state.

The resulting importance pattern differs markedly from Fig.~\ref{fig:attr_IntG}. With the mean baseline at the centre of the attractor, the two lobes are treated roughly symmetrically. With the left-lobe baseline, that symmetry is lost: importance is close to zero for states near the baseline on the left lobe and grows toward the central transition region and the right lobe, where the separations $(\mathbf{x}_i - \mathbf{x}'_i)$ are largest. Neither picture is wrong: the network and the attribution method are the same, and only the question differs. A single attribution map therefore cannot be interpreted without stating the reference it is measured against, and an ``important'' variable may simply be one whose value differs most from the chosen baseline.

\begin{figure}[h!]
    \centering
    \includegraphics[width=0.99\textwidth]{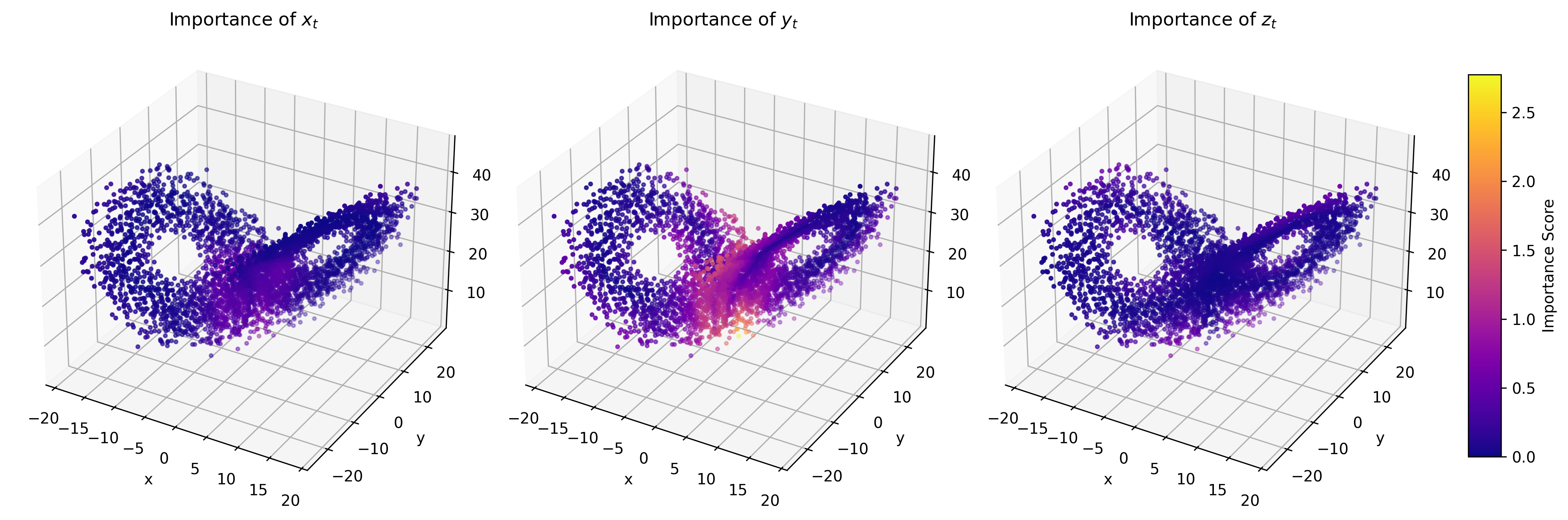}   
    \caption{SHAP-derived importance scores $\bar{A}_i(\mathbf{x})$ for each input feature of the L63 system, computed via Eq.~\eqref{eq:aggregation}~and~\eqref{eq:shap}. Each panel shows the L63 attractor, coloured by the scores attributed to the corresponding input variable. For each explanation, the reference distribution for missing features is the set of real attractor states within a fixed radius of the target state.
    }
    \label{fig:attr_KSHAP_noise}
\end{figure}

Here, we also show an alternative baseline strategy in Fig.~\ref{fig:attr_KSHAP_noise}. Instead of a zero or mean value, replacement values are drawn from a background distribution of other real states local to each target state. 
Here the structure of importance across phase space resembles saliency (Fig.~\ref{fig:attr_Saliency}), showing higher importance in $x$ and $y$ around the diverging part of the attractor. In effect, a local background asks how the forecast depends on small, realistic departures from the current state, which is close to the question saliency answers.

These alternative baselines are not posited as strictly better, but rather to highlight how significantly baseline choice can change the resulting explanation and interpretation of feature importance. Each baseline poses a different counterfactual question, and the most appropriate one depends on what the explanation is meant to reveal. This highlights the need for further investigation into path or kernel dependence, careful baseline selection, and clarity of integrated attribution methods, regarding both their limitations and how they can be most effectively leveraged to answer specific questions.

\subsection{Complete explanation structure of each method}
\label{sec:app_complete_XAI_structure}
\begin{figure}[h!]
    \centering
    \includegraphics[width=1.0\linewidth]{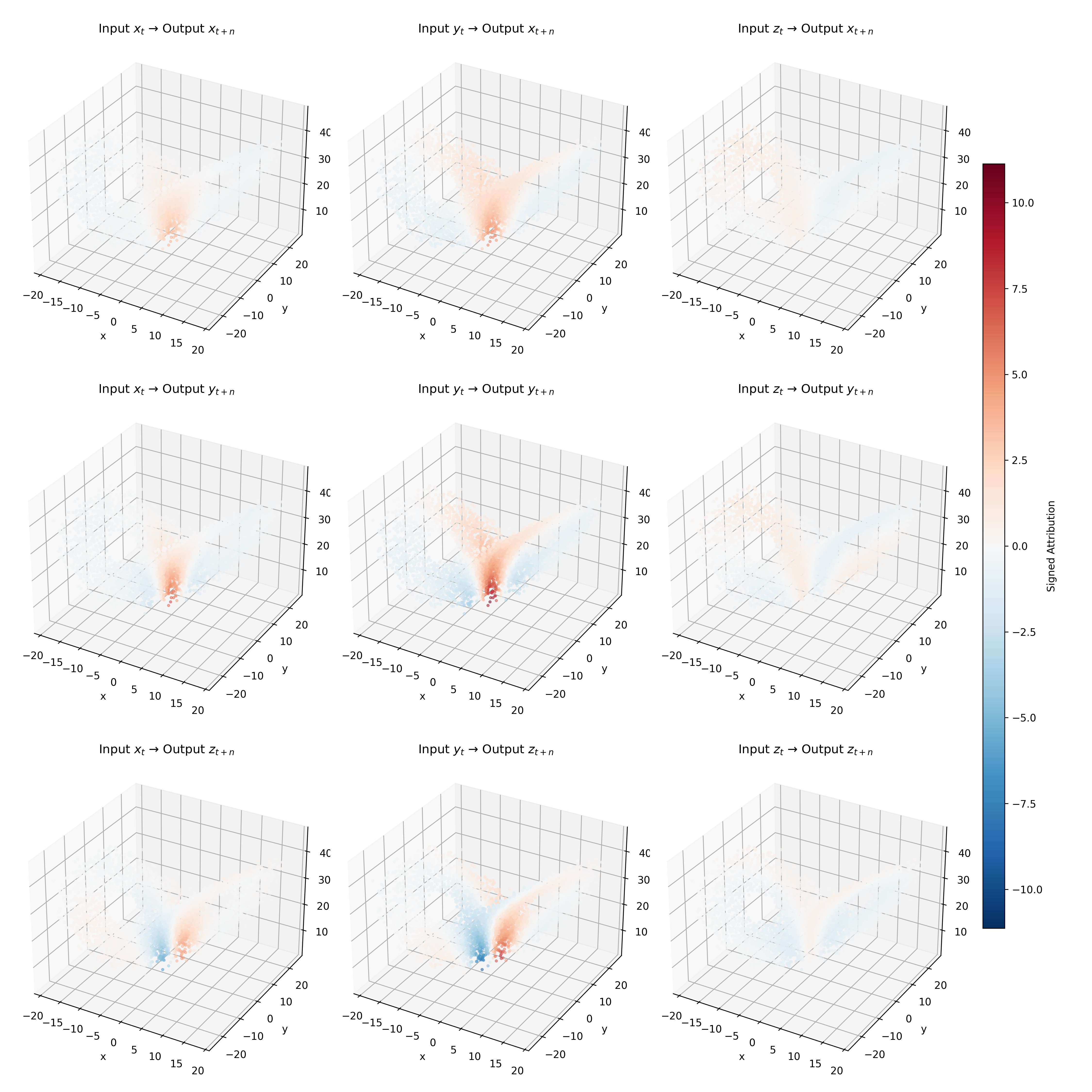}
    \caption{The pairwise, signed explanation values of each input-output pair in our L63 emulator (Sect.~\ref{sec:XAI-L63}), using the saliency method (Sect.~\ref{sec:illustrate_XAI_methods}\ref{sec:saliency}). }
    \label{fig:app_saliency_detail}
\end{figure}

\begin{figure}[h!]
    \centering
    \includegraphics[width=1.0\linewidth]{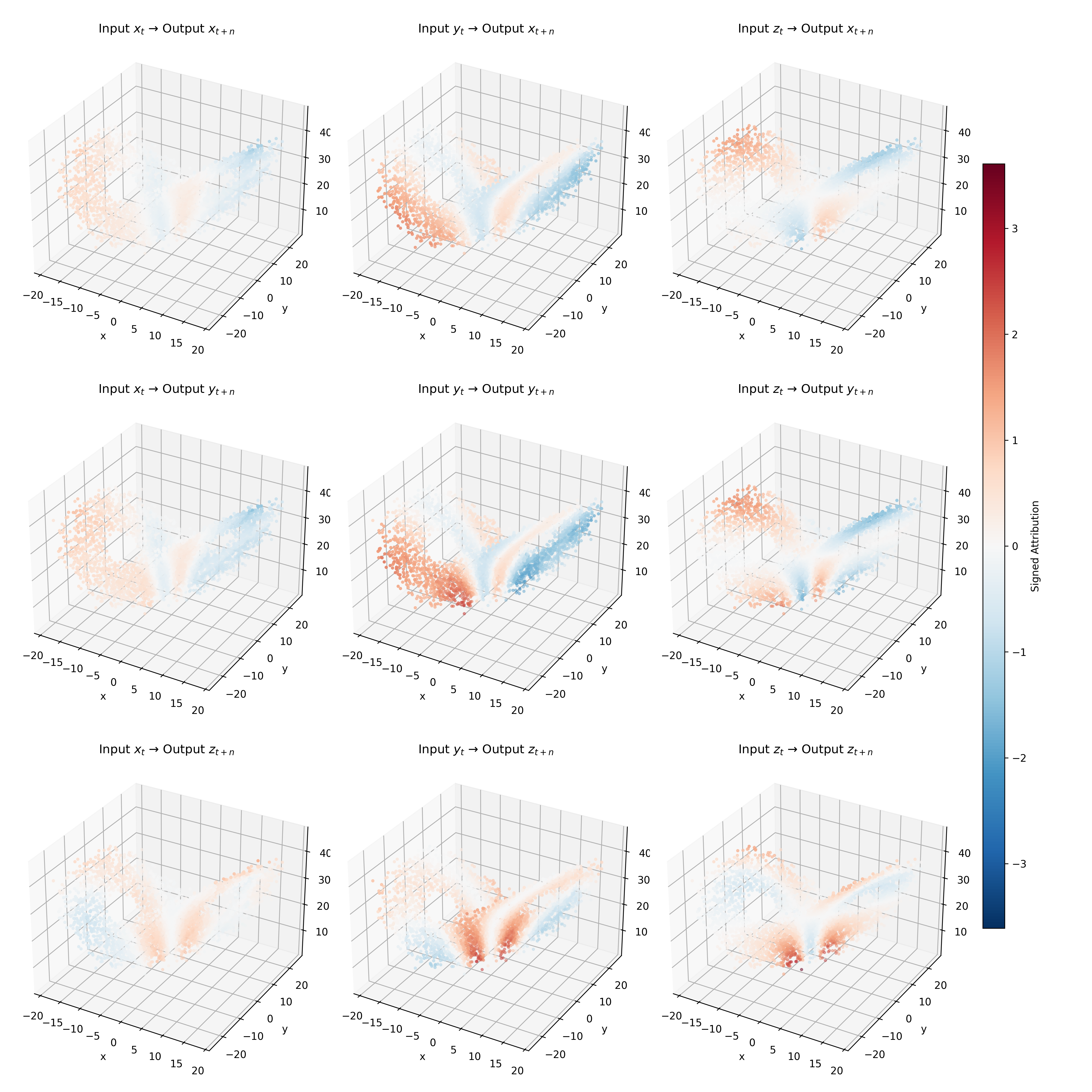}
    \caption{As with Fig.~\ref{fig:app_saliency_detail} but for Input$\times$Gradient (Sect.~\ref{sec:illustrate_XAI_methods}\ref{sec:inputxgradient}).}
    \label{fig:app_inputxgradient_detail}
\end{figure}

\begin{figure}[h!]
    \centering
    \includegraphics[width=1.0\linewidth]{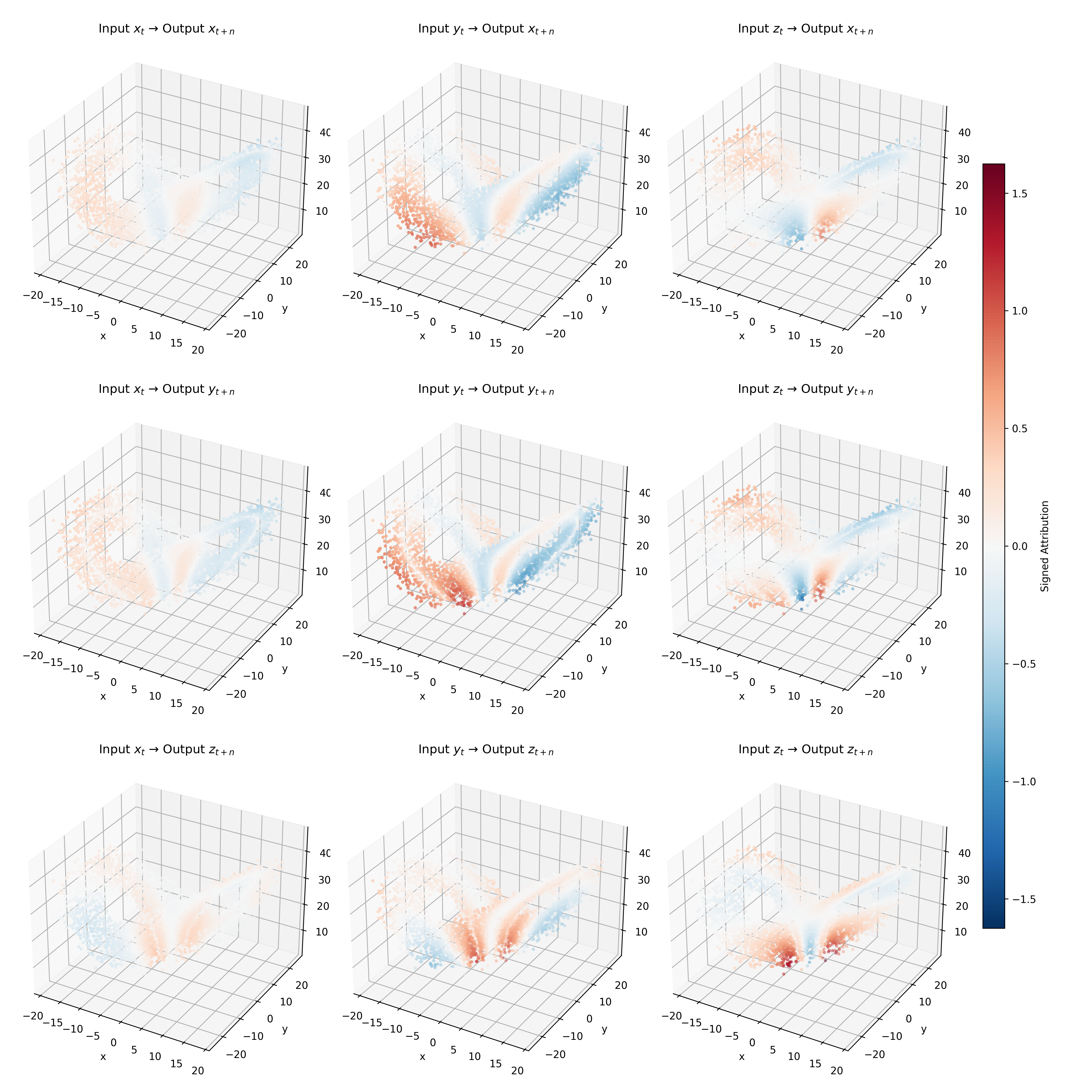}
    \caption{As with Fig.~\ref{fig:app_saliency_detail} but for LRP-$\epsilon$ rule (Sect.~\ref{sec:illustrate_XAI_methods}\ref{sec:lrp_illustration}.\ref{sec:lrp_epsilon}).}
    \label{fig:app_lrp_epsilon_detail}
\end{figure}

\begin{figure}[h!]
    \centering
    \includegraphics[width=1.0\linewidth]{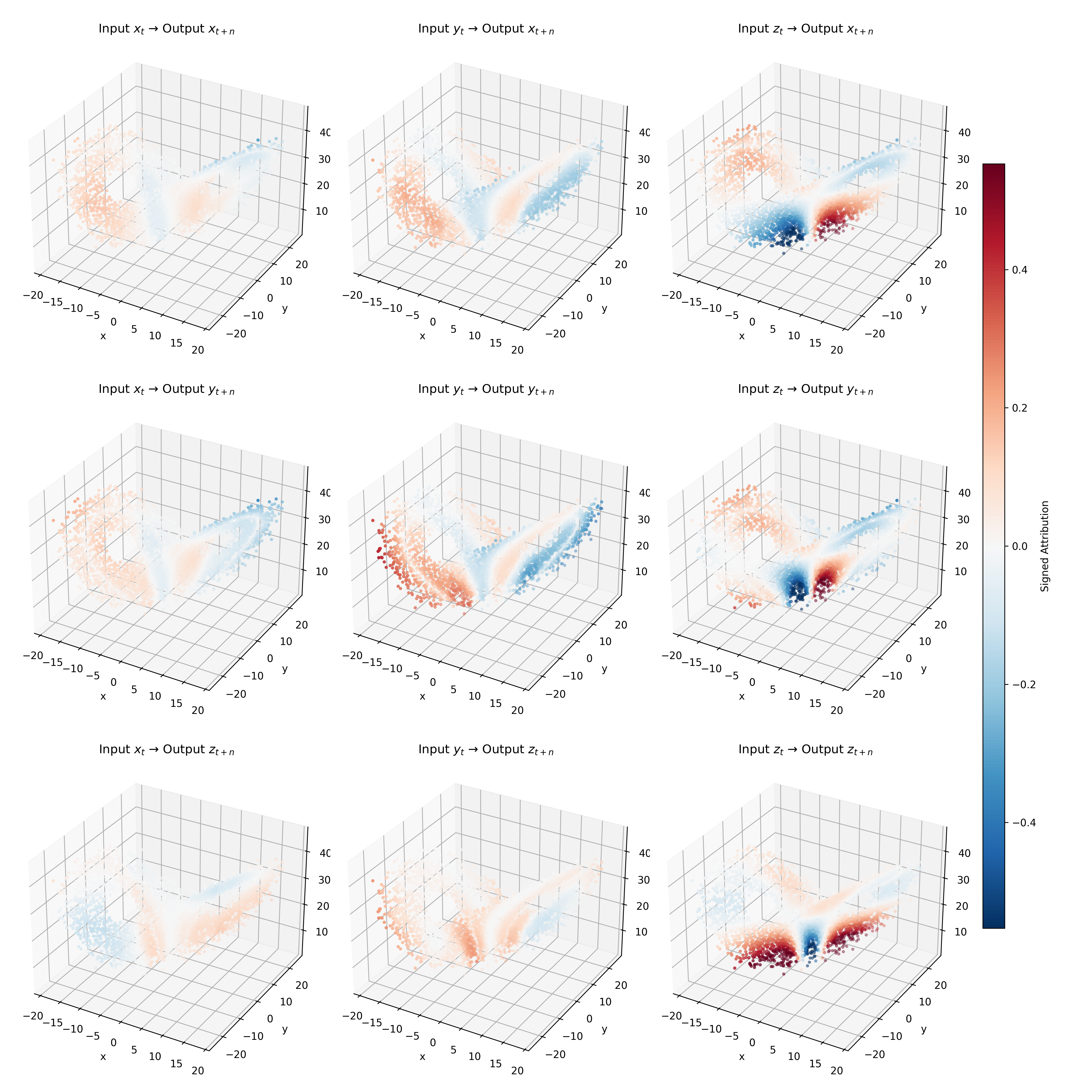}
    \caption{As with Fig.~\ref{fig:app_saliency_detail} but for LRP-$\alpha 1 \beta 0$ rule (Sect.~\ref{sec:illustrate_XAI_methods}\ref{sec:lrp_illustration}.\ref{sec:lrp_alphabeta}).}
    \label{fig:app_lrp_alphabeta_detail}
\end{figure}

\begin{figure}[h!]
    \centering
    \includegraphics[width=1.0\linewidth]{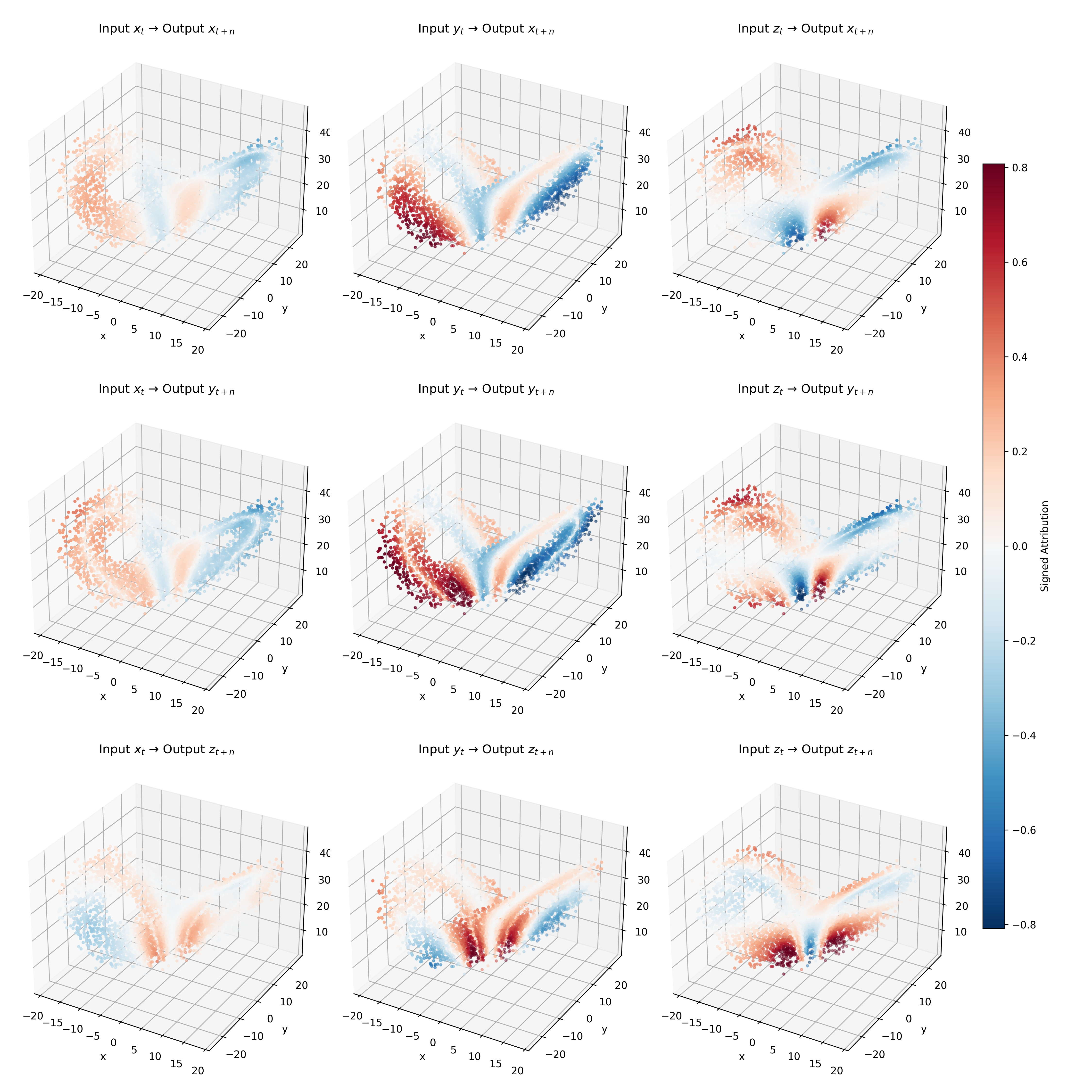}
    \caption{As with Fig.~\ref{fig:app_saliency_detail} but for LRP-$\gamma$ rule (Sect.~\ref{sec:illustrate_XAI_methods}\ref{sec:lrp_illustration}.\ref{sec:lrp_gamma}).}
    \label{fig:app_lrp_gamma_detail}
\end{figure}

\begin{figure}[h!]
    \centering
    \includegraphics[width=1.0\linewidth]{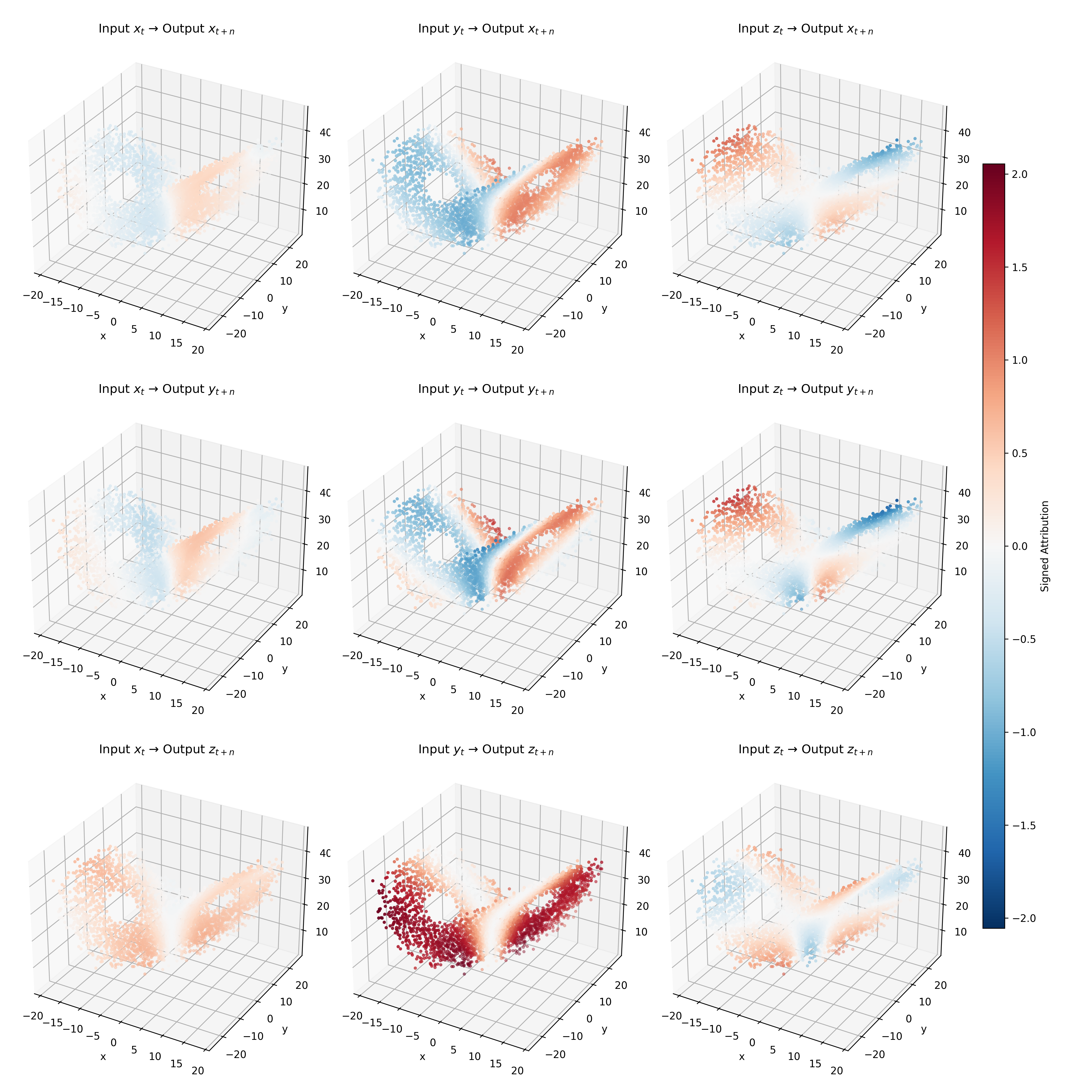}
    \caption{As with Fig.~\ref{fig:app_saliency_detail} but for Integrated Gradients (with mean state as baseline, which is zero for standardised inputs; Sect.~\ref{sec:illustrate_XAI_methods}\ref{sec:integrated_gradients}).}
    \label{fig:app_integratedgradient_detail}
\end{figure}

\begin{figure}[h!]
    \centering
    \includegraphics[width=1.0\linewidth]{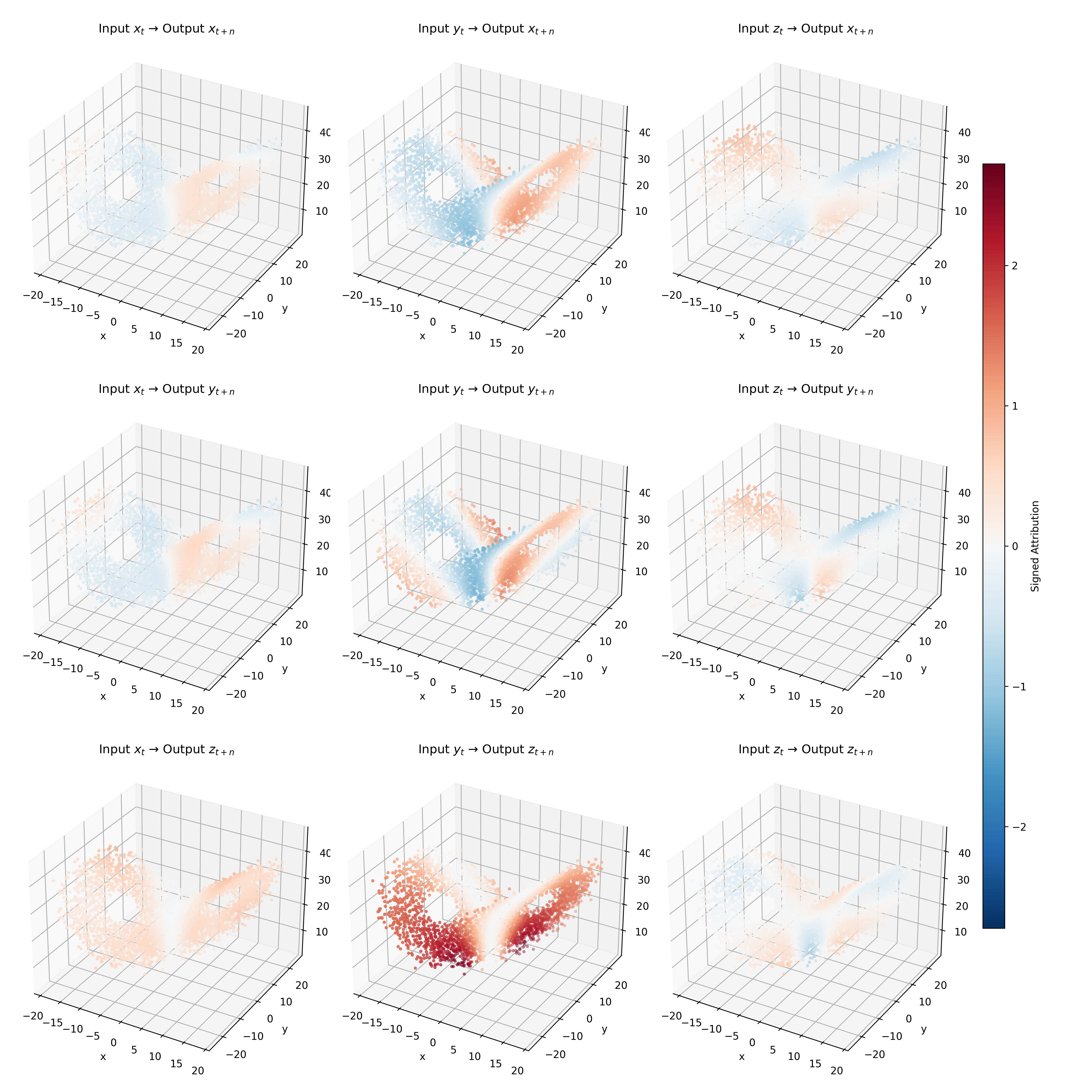}
    \caption{As with Fig.~\ref{fig:app_saliency_detail} but for SHAP (with mean state as baseline, which is zero for standardised inputs; Sect.~\ref{sec:illustrate_XAI_methods}\ref{sec:shap_illustration}).}
    \label{fig:app_shap_detail}
\end{figure}

Figures~\ref{fig:app_saliency_detail}-\ref{fig:app_shap_detail} show the complete explanation structure of the methods shown in Sects.~\ref{sec:illustrate_XAI_methods}\ref{sec:saliency}-~\ref{sec:illustrate_XAI_methods}\ref{sec:shap_illustration}, without summing column-wise (Eq.~\eqref{eq:aggregation}), and retaining the sign of the explanation values. While some finer scale details are present in this complete explanation structure, we argue that the aggregated quantity described by Eq.~\eqref{eq:aggregation} is largely representative of the structures shown here, and so is a useful summary for understanding the overall importance of each input variable on output behaviour. 

We caution that this aggregation structure is suitable for this application because we have observed the more detailed structures shown below. Such aggregation should not be applied blindly, and should be suitable for the given context of the XAI work.


\clearpage




\clearpage

\bibliographystyle{ametsocV6}
\bibliography{references}

@STRING{AIP 	= "Amer. Inst. Phys."}

@STRING{AN        = "Astrophys.\ Norv."}

@STRING{AO        = "Atmos.--Ocean"}

@STRING{CHAOS     = "Chaos"}

@STRING{MA        = "Meteor.\ Appl."}

@STRING{TELLUS    = "Tellus"}

@article{lorenz1963deterministic,
  title={Deterministic nonperiodic flow},
  author={Lorenz, Edward N},
  volume={20},
  number={2},
  pages={130-141},
  year={1963},
  journal={Journal of the Atmospheric Sciences}
}

@article{le1986variational,
  title={Variational algorithms for analysis and assimilation of meteorological observations: theoretical aspects},
  author={Le Dimet, Fran{\c{c}}ois-Xavier and Talagrand, Olivier},
  journal={Tellus A: Dynamic Meteorology and Oceanography},
  volume={38},
  number={2},
  pages={97--110},
  year={1986},
  publisher={Taylor \& Francis}
}

@article{trevisan1995transient,
  title={Transient error growth and local predictability: A study in the {L}orenz system},
  author={Trevisan, Anna and Legnani, Roberto},
  journal={Tellus A},
  volume={47},
  number={1},
  pages={103--117},
  year={1995},
  publisher={Wiley Online Library}
}

@article{rabier1996sensitivity,
  title={Sensitivity of forecast errors to initial conditions},
  author={Rabier, Florence and Klinker, E and Courtier, Ph and Hollingsworth, A},
  journal={Quarterly Journal of the Royal Meteorological Society},
  volume={122},
  number={529},
  pages={121--150},
  year={1996},
  publisher={Wiley Online Library},
  doi={10.1002/qj.49712252906}
}

@article{pu1998forecast,
  title={Forecast sensitivity with dropwindsonde data and targeted observations},
  author={Pu, Zhaoxia and Lord, Stephen J and Kalnay, Eugenia},
  journal={Tellus A},
  volume={50},
  number={4},
  pages={391--410},
  year={1998},
  publisher={Wiley Online Library}
}

@article{boffetta1998extension,
  title={An extension of the {L}yapunov analysis for the predictability problem},
  author={Boffetta, Guido and Giuliani, P and Paladin, G and Vulpiani, A},
  journal={Journal of the Atmospheric Sciences},
  volume={55},
  number={23},
  pages={3409--3416},
  year={1998}
}

@article{benedetti2003variational,
  title={Variational assimilation of radar reflectivities in a cirrus model. {I}: Model description and adjoint sensitivity studies},
  author={Benedetti, Angela and Stephens, Graeme L and Vuki{\'c}evi{\'c}, Tomislava},
  journal={Quarterly Journal of the Royal Meteorological Society: A journal of the atmospheric sciences, applied meteorology and physical oceanography},
  volume={129},
  number={587},
  pages={277--300},
  year={2003},
  publisher={Wiley Online Library}
}

@article{kleist2005application,
  title={Application of adjoint-derived forecast sensitivities to the 24--25 {J}anuary 2000 {US} {E}ast {C}oast snowstorm},
  author={Kleist, Daryl T and Morgan, Michael C},
  journal={Monthly weather review},
  volume={133},
  number={11},
  pages={3148--3175},
  year={2005}
}

@article{mahfouf2007adjoint,
  title={Adjoint sensitivity of surface precipitation to initial conditions},
  author={Mahfouf, Jean-Fran{\c{c}}ois and Bilodeau, Bernard},
  journal={Monthly weather review},
  volume={135},
  number={8},
  pages={2879--2896},
  year={2007}
}

@article{kursa2010boruta,
  title={Boruta--a system for feature selection},
  author={Kursa, Miron B and Jankowski, Aleksander and Rudnicki, Witold R},
  journal={Fundamenta informaticae},
  volume={101},
  number={4},
  pages={271--285},
  year={2010},
  publisher={SAGE Publications Sage UK: London, England}
}

@article{baehrens2010explain,
  title={How to explain individual classification decisions},
  author={Baehrens, David and Schroeter, Timon and Harmeling, Stefan and Kawanabe, Motoaki and Hansen, Katja and M{\"u}ller, Klaus-Robert},
  journal={The Journal of Machine Learning Research},
  volume={11},
  pages={1803--1831},
  year={2010},
  publisher={JMLR. org}
}

@book{coiffier2011fundamentals,
  title     = {Fundamentals of Numerical Weather Prediction},
  author    = {Coiffier, Jean},
  year      = {2011},
  publisher = {Cambridge University Press},
  address   = {Cambridge},
  doi       = {10.1017/CBO9780511734458}
}

@article{doyle2012adjoint,
  title={Adjoint sensitivity and predictability of tropical cyclogenesis},
  author={Doyle, James D and Reynolds, Carolyn A and Amerault, Clark and Moskaitis, Jonathan},
  journal={Journal of the atmospheric sciences},
  volume={69},
  number={12},
  pages={3535--3557},
  year={2012}
}

@book{sparrow2012lorenz,
  title={The {L}orenz equations: bifurcations, chaos, and strange attractors},
  author={Sparrow, Colin},
  volume={41},
  year={2012},
  publisher={Springer Science \& Business Media}
}

@article{simonyan2013deep,
  title={Deep inside convolutional networks: Visualising image classification models and saliency maps},
  author={Simonyan, Karen and Vedaldi, Andrea and Zisserman, Andrew},
  journal={arXiv preprint arXiv:1312.6034},
  year={2013}
}

@article{kingma2014adam,
  title={Adam: A method for stochastic optimization},
  author={Kingma, Diederik P},
  journal={arXiv preprint arXiv:1412.6980},
  year={2014}
}

@article{lorenc2014forecast,
  title={Forecast sensitivity to observations in the {M}et {O}ffice global numerical weather prediction system},
  author={Lorenc, Andrew C and Marriott, Richard T},
  journal={Quarterly Journal of the Royal Meteorological Society},
  volume={140},
  number={678},
  pages={209--224},
  year={2014},
  publisher={Wiley Online Library}
}

@article{feng2014application,
  title={The application of nonlinear local {L}yapunov vectors to ensemble predictions in {L}orenz systems},
  author={Feng, Jie and Ding, Ruiqiang and Liu, Deqiang and Li, Jianping},
  journal={Journal of the Atmospheric Sciences},
  volume={71},
  number={9},
  pages={3554--3567},
  year={2014}
}

@article{bach2015pixel,
  title={On pixel-wise explanations for non-linear classifier decisions by layer-wise relevance propagation},
  author={Bach, Sebastian and Binder, Alexander and Montavon, Gr{\'e}goire and Klauschen, Frederick and M{\"u}ller, Klaus-Robert and Samek, Wojciech},
  journal={PloS one},
  volume={10},
  number={7},
  pages={e0130140},
  year={2015},
  publisher={Public Library of Science San Francisco, CA USA}
}

@article{coughlan2016action,
  title={Action-based flood forecasting for triggering humanitarian action},
  author={Coughlan de Perez, Erin and Van den Hurk, Bart and Van Aalst, Maarten K and Amuron, Irene and Bamanya, Deus and Hauser, Tristan and Jongma, Brenden and Lopez, Ana and Mason, Simon and Mendler de Suarez, Janot and others},
  journal={Hydrology and Earth System Sciences},
  volume={20},
  number={9},
  pages={3549--3560},
  year={2016},
  publisher={Copernicus GmbH}
}

@article{economou2016use,
  title={On the use of {B}ayesian decision theory for issuing natural hazard warnings},
  author={Economou, Theodoros and Stephenson, David B and Rougier, Jonathan C and Neal, RA and Mylne, Ken R},
  journal={Proceedings of the Royal Society A: Mathematical, Physical and Engineering Sciences},
  volume={472},
  number={2194},
  pages={20160295},
  year={2016},
  publisher={The Royal Society}
}

@article{kindermans2016investigating,
  title={Investigating the influence of noise and distractors on the interpretation of neural networks},
  author={Kindermans, Pieter-Jan and Sch{\"u}tt, Kristof and M{\"u}ller, Klaus-Robert and D{\"a}hne, Sven},
  journal={arXiv preprint arXiv:1611.07270},
  year={2016}
}

@article{smilkov2017smoothgrad,
  title={{SmoothGrad}: removing noise by adding noise},
  author={Smilkov, Daniel and Thorat, Nikhil and Kim, Been and Vi{\'e}gas, Fernanda and Wattenberg, Martin},
  journal={arXiv preprint arXiv:1706.03825},
  year={2017}
}

@inproceedings{sundararajan2017axiomatic,
  title={Axiomatic attribution for deep networks},
  author={Sundararajan, Mukund and Taly, Ankur and Yan, Qiqi},
  booktitle={International conference on machine learning},
  pages={3319--3328},
  year={2017},
  organization={PMLR}
}

@article{lundberg2017unified,
  title={A unified approach to interpreting model predictions},
  author={Lundberg, Scott M and Lee, Su-In},
  journal={Advances in neural information processing systems},
  volume={30},
  year={2017}
}

@inproceedings{shrikumar2017learning,
  title={Learning important features through propagating activation differences},
  author={Shrikumar, Avanti and Greenside, Peyton and Kundaje, Anshul},
  booktitle={International conference on machine learning},
  pages={3145--3153},
  year={2017},
  organization={PMlR}
}

@article{vaswani2017attention,
  title={Attention is all you need},
  author={Vaswani, Ashish and Shazeer, Noam and Parmar, Niki and Uszkoreit, Jakob and Jones, Llion and Gomez, Aidan N and Kaiser, {\L}ukasz and Polosukhin, Illia},
  journal={Advances in neural information processing systems},
  volume={30},
  year={2017}
}

@article{carrassi2018data,
  title={Data assimilation in the geosciences: An overview of methods, issues, and perspectives},
  author={Carrassi, Alberto and Bocquet, Marc and Bertino, Laurent and Evensen, Geir},
  journal={Wiley Interdisciplinary Reviews: Climate Change},
  volume={9},
  number={5},
  pages={e535},
  year={2018},
  publisher={Wiley Online Library}
}

@article{wang2018tracking,
  title={Tracking sensitive source areas of different weather pollution types using {GRAPES-CUACE} adjoint model},
  author={Wang, Chao and An, Xingqin and Zhai, Shixian and Hou, Qing and Sun, Zhaobin},
  journal={Atmospheric Environment},
  volume={175},
  pages={154--166},
  year={2018},
  publisher={Elsevier}
}

@article{lipton2018mythos,
  title={The mythos of model interpretability: In machine learning, the concept of interpretability is both important and slippery.},
  author={Lipton, Zachary C},
  journal={Queue},
  volume={16},
  number={3},
  pages={31--57},
  year={2018},
  publisher={ACM New York, NY, USA}
}

@article{alvarez2018towards,
  title={Towards robust interpretability with self-explaining neural networks},
  author={Alvarez Melis, David and Jaakkola, Tommi},
  journal={Advances in neural information processing systems},
  volume={31},
  year={2018}
}

@article{montavon2019layer,
  title={Layer-wise relevance propagation: an overview},
  author={Montavon, Gr{\'e}goire and Binder, Alexander and Lapuschkin, Sebastian and Samek, Wojciech and M{\"u}ller, Klaus-Robert},
  journal={Explainable AI: interpreting, explaining and visualizing deep learning},
  pages={193--209},
  year={2019},
  publisher={Springer}
}

@article{paszke2019pytorch,
  title={Pytorch: An imperative style, high-performance deep learning library},
  author={Paszke, Adam and Gross, Sam and Massa, Francisco and Lerer, Adam and Bradbury, James and Chanan, Gregory and Killeen, Trevor and Lin, Zeming and Gimelshein, Natalia and Antiga, Luca and others},
  journal={Advances in Neural Information Processing Systems},
  volume={32},
  year={2019}
}

@article{rudin2019stop,
  title={Stop explaining black box machine learning models for high stakes decisions and use interpretable models instead},
  author={Rudin, Cynthia},
  journal={Nature machine intelligence},
  volume={1},
  number={5},
  pages={206--215},
  year={2019},
  publisher={Nature Publishing Group UK London}
}

@inproceedings{mathews2019explainable,
  title={Explainable artificial intelligence applications in {NLP}, biomedical, and malware classification: a literature review},
  author={Mathews, Sherin Mary},
  booktitle={Intelligent Computing-proceedings of the computing conference},
  pages={1269--1292},
  year={2019},
  organization={Springer}
}

@article{mcgovern2019making,
  title={Making the black box more transparent: Understanding the physical implications of machine learning},
  author={McGovern, Amy and Lagerquist, Ryan and John Gagne, David and Jergensen, G Eli and Elmore, Kimberly L and Homeyer, Cameron R and Smith, Travis},
  journal={Bulletin of the American Meteorological Society},
  volume={100},
  number={11},
  pages={2175--2199},
  year={2019}
}

@article{bonavita2020machine,
  title={Machine learning for model error inference and correction},
  author={Bonavita, Massimo and Laloyaux, Patrick},
  journal={Journal of Advances in Modeling Earth Systems},
  volume={12},
  number={12},
  pages={e2020MS002232},
  year={2020},
  publisher={Wiley Online Library}
}

@article{frye2020shapley,
  title={Shapley explainability on the data manifold},
  author={Frye, Christopher and de Mijolla, Damien and Begley, Tom and Cowton, Laurence and Stanley, Megan and Feige, Ilya},
  journal={arXiv preprint arXiv:2006.01272},
  year={2020}
}

@inproceedings{mamalakis2020explainable,
  title={Explainable artificial intelligence in meteorology and climate science: Model fine-tuning, calibrating trust and learning new science},
  author={Mamalakis, Antonios and Ebert-Uphoff, Imme and Barnes, Elizabeth A},
  booktitle={International Workshop on Extending Explainable AI Beyond Deep Models and Classifiers},
  pages={315--339},
  year={2020},
  organization={Springer}
}

@article{pintelas2020explainable,
  title={Explainable machine learning framework for image classification problems: case study on glioma cancer prediction},
  author={Pintelas, Emmanuel and Liaskos, Meletis and Livieris, Ioannis E and Kotsiantis, Sotiris and Pintelas, Panagiotis},
  journal={Journal of Imaging},
  volume={6},
  number={6},
  pages={37},
  year={2020},
  publisher={MDPI}
}

@inproceedings{bodria2020explainability,
  title={Explainability methods for natural language processing: Applications to sentiment analysis},
  author={Bodria, Francesco and Panisson, Andr{\'e} and Perotti, Alan and Piaggesi, Simone and others},
  booktitle={CEUR workshop proceedings},
  volume={2646},
  pages={100--107},
  year={2020},
  organization={CEUR-WS}
}

@article{iwashita2020efficient,
  title={Efficient constrained pattern mining using dynamic item ordering for explainable classification},
  author={Iwashita, Hiroaki and Takagi, Takuya and Suzuki, Hirofumi and Goto, Keisuke and Ohori, Kotaro and Arimura, Hiroki},
  journal={arXiv preprint arXiv:2004.08015},
  year={2020}
}

@article{lundberg2020local,
  title={From local explanations to global understanding with explainable {AI} for trees},
  author={Lundberg, Scott M and Erion, Gabriel and Chen, Hugh and DeGrave, Alex and Prutkin, Jordan M and Nair, Bala and Katz, Ronit and Himmelfarb, Jonathan and Bansal, Nisha and Lee, Su-In},
  journal={Nature machine intelligence},
  volume={2},
  number={1},
  pages={56--67},
  year={2020},
  publisher={Nature Publishing Group UK London}
}

@article{hilburn2020development,
  title={Development and interpretation of a neural-network-based synthetic radar reflectivity estimator using {GOES-R} satellite observations},
  author={Hilburn, Kyle A and Ebert-Uphoff, Imme and Miller, Steven D},
  journal={Journal of Applied Meteorology and Climatology},
  volume={60},
  number={1},
  pages={3--21},
  year={2020},
  publisher={American Meteorological Society}
}

@article{ebert2020evaluation,
  title={Evaluation, tuning, and interpretation of neural networks for working with images in meteorological applications},
  author={Ebert-Uphoff, Imme and Hilburn, Kyle},
  journal={Bulletin of the American Meteorological Society},
  volume={101},
  number={12},
  pages={E2149--E2170},
  year={2020}
}

@article{boukabara2021outlook,
  title={Outlook for exploiting artificial intelligence in the {E}arth and environmental sciences},
  author={Boukabara, Sid-Ahmed and Krasnopolsky, Vladimir and Penny, Stephen G and Stewart, Jebb Q and McGovern, Amy and Hall, David and Ten Hoeve, John E and Hickey, Jason and Allen Huang, Hung-Lung and Williams, John K and others},
  journal={Bulletin of the American Meteorological Society},
  volume={102},
  number={5},
  pages={E1016--E1032},
  year={2021}
}

@article{geer2021learning,
  title={Learning earth system models from observations: machine learning or data assimilation?},
  author={Geer, Alan J},
  journal={Philosophical Transactions of the Royal Society A},
  volume={379},
  number={2194},
  pages={20200089},
  year={2021},
  publisher={The Royal Society Publishing}
}

@article{mamalakis2022neural,
  title={Neural network attribution methods for problems in geoscience: A novel synthetic benchmark dataset},
  author={Mamalakis, Antonios and Ebert-Uphoff, Imme and Barnes, Elizabeth A},
  journal={Environmental Data Science},
  volume={1},
  pages={e8},
  year={2022},
  publisher={Cambridge University Press}
}

@ARTICLE{letzgus2022towards,
  author={Letzgus, Simon and Wagner, Patrick and Lederer, Jonas and Samek, Wojciech and Müller, Klaus-Robert and Montavon, Grégoire},
  journal={IEEE Signal Processing Magazine}, 
  title={Toward Explainable Artificial Intelligence for Regression Models: A methodological perspective}, 
  year={2022},
  volume={39},
  number={4},
  pages={40-58},
  doi={10.1109/MSP.2022.3153277}
}

@article{christensen2022parametrization,
  title={Parametrization in weather and climate models},
  author={Christensen, Hannah and Zanna, Laure},
  year={2022},
  publisher={Oxford University Press}
}

@article{sun2022review,
  title={A review of {E}arth artificial intelligence},
  author={Sun, Ziheng and Sandoval, Laura and Crystal-Ornelas, Robert and Mousavi, S Mostafa and Wang, Jinbo and Lin, Cindy and Cristea, Nicoleta and Tong, Daniel and Carande, Wendy Hawley and Ma, Xiaogang and others},
  journal={Computers \& Geosciences},
  volume={159},
  pages={105034},
  year={2022},
  publisher={Elsevier}
}

@article{kondylatos2022wildfire,
  title={Wildfire danger prediction and understanding with deep learning},
  author={Kondylatos, Spyros and Prapas, Ioannis and Ronco, Michele and Papoutsis, Ioannis and Camps-Valls, Gustau and Piles, Mar{\'\i}a and Fern{\'a}ndez-Torres, Miguel-{\'A}ngel and Carvalhais, Nuno},
  journal={Geophysical Research Letters},
  volume={49},
  number={17},
  pages={e2022GL099368},
  year={2022},
  publisher={Wiley Online Library}
}

@article{bacsaugaouglu2022review,
  title={A review on interpretable and explainable artificial intelligence in hydroclimatic applications},
  author={Ba{\c{s}}a{\u{g}}ao{\u{g}}lu, Hakan and Chakraborty, Debaditya and Lago, Cesar Do and Gutierrez, Lilianna and {\c{S}}ahinli, Mehmet Arif and Giacomoni, Marcio and Furl, Chad and Mirchi, Ali and Moriasi, Daniel and {\c{S}}eng{\"o}r, Sema Sevin{\c{c}}},
  journal={Water},
  volume={14},
  number={8},
  pages={1230},
  year={2022},
  publisher={MDPI}
}

@article{mamalakis2022investigating,
  title={Investigating the fidelity of explainable artificial intelligence methods for applications of convolutional neural networks in geoscience},
  author={Mamalakis, Antonios and Barnes, Elizabeth A and Ebert-Uphoff, Imme},
  journal={Artificial Intelligence for the Earth Systems},
  volume={1},
  number={4},
  pages={e220012},
  year={2022}
}

@article{gevaert2022explainable,
  title={Explainable {AI} for earth observation: A review including societal and regulatory perspectives},
  author={Gevaert, Caroline M},
  journal={International Journal of Applied Earth Observation and Geoinformation},
  volume={112},
  pages={102869},
  year={2022},
  publisher={Elsevier}
}

@article{huang2022analysis,
  title={Analysis of the atmospheric duct existence factors in tropical cyclones based on the SHAP interpretation of extreme gradient boosting predictions},
  author={Huang, Lang and Zhao, Xiaofeng and Liu, Yudi and Yang, Pinglv},
  journal={Remote Sensing},
  volume={14},
  number={16},
  pages={3952},
  year={2022},
  publisher={MDPI}
}

@article{kim2022untangling,
  title={Untangling the contribution of input parameters to an artificial intelligence {PM}2.5 forecast model using the layer-wise relevance propagation method},
  author={Kim, Dasol and Ho, Chang-Hoi and Park, Ingyu and Kim, Jinwon and Chang, Lim-Seok and Choi, Min-Hyeok},
  journal={Atmospheric Environment},
  volume={276},
  pages={119034},
  year={2022},
  publisher={Elsevier}
}

@article{barnes2022looks,
  title={This looks like that there: Interpretable neural networks for image tasks when location matters},
  author={Barnes, Elizabeth A and Barnes, Randal J and Martin, Zane K and Rader, Jamin K},
  journal={Artificial Intelligence for the Earth Systems},
  volume={1},
  number={3},
  pages={e220001},
  year={2022}
}

@misc{eu_ai_act_2023,
  author       = {{European Union}},
  title        = {Regulation ({EU}) 2023/{XXXX} of the {E}uropean {P}arliament and of the {C}ouncil on artificial intelligence ({AI} {A}ct)},
  year         = {2023},
  howpublished = {\url{https://artificialintelligenceact.eu/}},
  note         = {Official Journal of the European Union}
}

@article{pirone2023short,
  title={Short-term rainfall forecasting using cumulative precipitation fields from station data: a probabilistic machine learning approach},
  author={Pirone, Dina and Cimorelli, Luigi and Del Giudice, Giuseppe and Pianese, Domenico},
  journal={Journal of Hydrology},
  volume={617},
  pages={128949},
  year={2023},
  publisher={Elsevier}
}

@article{corcoran2023current,
  title={Current data and modeling bottlenecks for predicting crop yields in the {U}nited {K}ingdom},
  author={Corcoran, Evangeline and Afshar, Mehdi and Curceac, Stelian and Lashkari, Azam and Raza, Muhammad Mohsin and Ahnert, Sebastian and Mead, Andrew and Morris, Richard},
  journal={Frontiers in Sustainable Food Systems},
  volume={7},
  pages={1023169},
  year={2023},
  publisher={Frontiers Media SA}
}

@article{nguyen2023climax,
  title={Climax: A foundation model for weather and climate},
  author={Nguyen, Tung and Brandstetter, Johannes and Kapoor, Ashish and Gupta, Jayesh K and Grover, Aditya},
  journal={arXiv preprint arXiv:2301.10343},
  year={2023}
}

@article{jones2023ai,
  title={{AI} for climate impacts: applications in flood risk},
  author={Jones, Anne and Kuehnert, Julian and Fraccaro, Paolo and Meuriot, Ophelie and Ishikawa, Tatsuya and Edwards, Blair and Stoyanov, Nikola and Remy, Sekou L and Weldemariam, Kommy and Assefa, Solomon},
  journal={Npj Climate and Atmospheric Science},
  volume={6},
  number={1},
  pages={63},
  year={2023},
  publisher={Nature Publishing Group UK London}
}

@article{ham2023anthropogenic,
  title={Anthropogenic fingerprints in daily precipitation revealed by deep learning},
  author={Ham, Yoo-Geun and Kim, Jeong-Hwan and Min, Seung-Ki and Kim, Daehyun and Li, Tim and Timmermann, Axel and Stuecker, Malte F},
  journal={Nature},
  volume={622},
  number={7982},
  pages={301--307},
  year={2023},
  publisher={Nature Publishing Group UK London}
}

@article{liu2023evaluation,
  title={Evaluation of tropical cyclone disaster loss using machine learning algorithms with an explainable artificial intelligence approach},
  author={Liu, Shuxian and Liu, Yang and Chu, Zhigang and Yang, Kun and Wang, Guanlan and Zhang, Lisheng and Zhang, Yuanda},
  journal={Sustainability},
  volume={15},
  number={16},
  pages={12261},
  year={2023},
  publisher={MDPI}
}

@article{mamalakis2023carefully,
  title={Carefully choose the baseline: Lessons learned from applying {XAI} attribution methods for regression tasks in geoscience},
  author={Mamalakis, Antonios and Barnes, Elizabeth A and Ebert-Uphoff, Imme},
  journal={Artificial Intelligence for the Earth Systems},
  volume={2},
  number={1},
  pages={e220058},
  year={2023}
}

@article{trok2024machine,
  title={Machine learning--based extreme event attribution},
  author={Trok, Jared T and Barnes, Elizabeth A and Davenport, Frances V and Diffenbaugh, Noah S},
  journal={Science Advances},
  volume={10},
  number={34},
  pages={eadl3242},
  year={2024},
  publisher={American Association for the Advancement of Science}
}

@article{jiang2024interpretable,
  title={How interpretable machine learning can benefit process understanding in the geosciences},
  author={Jiang, Shijie and Sweet, Lily-belle and Blougouras, Georgios and Brenning, Alexander and Li, Wantong and Reichstein, Markus and Denzler, Joachim and Shangguan, Wei and Yu, Guo and Huang, Feini and others},
  journal={Earth's Future},
  volume={12},
  number={7},
  pages={e2024EF004540},
  year={2024},
  publisher={Wiley Online Library}
}

@article{cheng2024deep,
  title={Deep learning surrogate models of {JULES-INFERNO} for wildfire prediction on a global scale},
  author={Cheng, Sibo and Chassagnon, Hector and Kasoar, Matthew and Guo, Yike and Arcucci, Rossella},
  journal={IEEE Transactions on Emerging Topics in Computational Intelligence},
  year={2024},
  publisher={IEEE}
}

@article{zhao2024artificial,
  title={Artificial intelligence for geoscience: Progress, challenges and perspectives},
  author={Zhao, Tianjie and Wang, Sheng and Ouyang, Chaojun and Chen, Min and Liu, Chenying and Zhang, Jin and Yu, Long and Wang, Fei and Xie, Yong and Li, Jun and others},
  journal={The Innovation},
  year={2024},
  publisher={Elsevier}
}

@article{hunt2024using,
  title={Using interpretable gradient-boosted decision-tree ensembles to uncover novel dynamical relationships governing monsoon low-pressure systems},
  author={Hunt, Kieran MR and Turner, Andrew G},
  journal={Quarterly Journal of the Royal Meteorological Society},
  volume={150},
  number={758},
  pages={1--24},
  year={2024},
  publisher={Wiley Online Library}
}

@article{lang2024aifs,
  title={{AIFS--ECMWF}'s data-driven forecasting system},
  author={Lang, Simon and Alexe, Mihai and Chantry, Matthew and Dramsch, Jesper and Pinault, Florian and Raoult, Baudouin and Clare, Mariana CA and Lessig, Christian and Maier-Gerber, Michael and Magnusson, Linus and others},
  journal={arXiv preprint arXiv:2406.01465},
  year={2024}
}

@article{bocquet2024accurate,
  title={Accurate deep learning-based filtering for chaotic dynamics by identifying instabilities without an ensemble},
  author={Bocquet, Marc and Farchi, Alban and Finn, Tobias S and Durand, Charlotte and Cheng, Sibo and Chen, Yumeng and Pasmans, Ivo and Carrassi, Alberto},
  journal={Chaos: An Interdisciplinary Journal of Nonlinear Science},
  volume={34},
  number={9},
  year={2024},
  publisher={AIP Publishing}
}

@article{wang2024multiple,
  title={Multiple spatio-temporal scale runoff forecasting and driving mechanism exploration by {K}-means optimized {XGBoost} and {SHAP}},
  author={Wang, Shuo and Peng, Hui},
  journal={Journal of Hydrology},
  volume={630},
  pages={130650},
  year={2024},
  publisher={Elsevier}
}

@article{talaat2024toward,
  title={Toward interpretable credit scoring: integrating explainable artificial intelligence with deep learning for credit card default prediction},
  author={Talaat, Fatma M and Aljadani, Abdussalam and Badawy, Mahmoud and Elhosseini, Mostafa},
  journal={Neural Computing and Applications},
  volume={36},
  number={9},
  pages={4847--4865},
  year={2024},
  publisher={Springer}
}

@article{arditi2024refusal,
  title={Refusal in language models is mediated by a single direction},
  author={Arditi, Andy and Obeso, Oscar and Syed, Aaquib and Paleka, Daniel and Panickssery, Nina and Gurnee, Wes and Nanda, Neel},
  journal={Advances in Neural Information Processing Systems},
  volume={37},
  pages={136037--136083},
  year={2024}
}

@article{charlton2024ai,
  title={Do {AI} models produce better weather forecasts than physics-based models? {A} quantitative evaluation case study of {S}torm {C}iar{\'a}n},
  author={Charlton-Perez, Andrew J and Dacre, Helen F and Driscoll, Simon and Gray, Suzanne L and Harvey, Ben and Harvey, Natalie J and Hunt, Kieran MR and Lee, Robert W and Swaminathan, Ranjini and Vandaele, Remy and others},
  journal={npj Climate and Atmospheric Science},
  volume={7},
  number={1},
  pages={93},
  year={2024},
  publisher={Nature Publishing Group UK London}
}

@article{rasp2024weatherbench,
  title={Weather{B}ench 2: A benchmark for the next generation of data-driven global weather models},
  author={Rasp, Stephan and Hoyer, Stephan and Merose, Alexander and Langmore, Ian and Battaglia, Peter and Russell, Tyler and Sanchez-Gonzalez, Alvaro and Yang, Vivian and Carver, Rob and Agrawal, Shreya and others},
  journal={Journal of Advances in Modeling Earth Systems},
  volume={16},
  number={6},
  pages={e2023MS004019},
  year={2024},
  publisher={Wiley Online Library}
}

@article{awosika2024transparency,
  title={Transparency and privacy: the role of explainable {AI} and federated learning in financial fraud detection},
  author={Awosika, Tomisin and Shukla, Raj Mani and Pranggono, Bernardi},
  journal={IEEE access},
  volume={12},
  pages={64551--64560},
  year={2024},
  publisher={IEEE}
}

@article{bommer2024finding,
  title={Finding the right {XAI} method—a guide for the evaluation and ranking of explainable {AI} methods in climate science},
  author={Bommer, Philine Lou and Kretschmer, Marlene and Hedstr{\"o}m, Anna and Bareeva, Dilyara and H{\"o}hne, Marina M-C},
  journal={Artificial Intelligence for the Earth Systems},
  volume={3},
  number={3},
  pages={e230074},
  year={2024},
  publisher={American Meteorological Society}
}

@article{achtibat2024attnlrp,
  title={{AttnLRP}: attention-aware layer-wise relevance propagation for transformers},
  author={Achtibat, Reduan and Hatefi, Sayed Mohammad Vakilzadeh and Dreyer, Maximilian and Jain, Aakriti and Wiegand, Thomas and Lapuschkin, Sebastian and Samek, Wojciech},
  journal={arXiv preprint arXiv:2402.05602},
  year={2024}
}

@article{tao2024explainable,
  title={An explainable multiscale {LSTM} model with wavelet transform and layer-wise relevance propagation for daily streamflow forecasting},
  author={Tao, Lizhi and Cui, Zhichao and He, Yufeng and Yang, Dong},
  journal={Science of The Total Environment},
  volume={929},
  pages={172465},
  year={2024},
  publisher={Elsevier}
}

@article{bonavita2024some,
  title={On some limitations of current machine learning weather prediction models},
  author={Bonavita, Massimo},
  journal={Geophysical Research Letters},
  volume={51},
  number={12},
  pages={e2023GL107377},
  year={2024},
  publisher={Wiley Online Library}
}

@article{flora2024machine,
  title={A machine learning explainability tutorial for atmospheric sciences},
  author={Flora, Montgomery L and Potvin, Corey K and McGovern, Amy and Handler, Shawn},
  journal={Artificial Intelligence for the Earth Systems},
  volume={3},
  number={1},
  pages={e230018},
  year={2024},
  publisher={American Meteorological Society}
}

@article{huang2024applications,
  title={Applications of Explainable artificial intelligence in {E}arth system science},
  author={Huang, Feini and Jiang, Shijie and Li, Lu and Zhang, Yongkun and Zhang, Ye and Zhang, Ruqing and Li, Qingliang and Li, Danxi and Shangguan, Wei and Dai, Yongjiu},
  journal={arXiv preprint arXiv:2406.11882},
  year={2024}
}

@article{silva2024limitations,
  title={Limitations of {XAI} methods for process-level understanding in the atmospheric sciences},
  author={Silva, Sam J and Keller, Christoph A},
  journal={Artificial Intelligence for the Earth Systems},
  volume={3},
  number={1},
  pages={e230045},
  year={2024},
  publisher={American Meteorological Society}
}

@article{seth2025bridging,
  title={Bridging the Gap in {XAI}-Why Reliable Metrics Matter for Explainability and Compliance},
  author={Seth, Pratinav and Sankarapu, Vinay Kumar},
  journal={arXiv preprint arXiv:2502.04695},
  year={2025}
}

@article{kan2025seasonal,
  title={Seasonal heatwave forecasting with explainable machine learning and remote sensing data},
  author={Kan, Jung-Ching and Vieira Passos, Marlon and Destouni, Georgia and Barquet, Karina and Ferreira, Carla SS and Kalantari, Zahra},
  journal={Stochastic Environmental Research and Risk Assessment},
  volume={39},
  number={8},
  pages={3333--3352},
  year={2025},
  publisher={Springer}
}

@article{jimenez2025ai,
  title={A{I}-driven weather forecasts to accelerate climate change attribution of heatwaves},
  author={Jim{\'e}nez-Esteve, B and Barriopedro, D and Johnson, JE and Garc{\'\i}a-Herrera, R},
  journal={Earth's Future},
  volume={13},
  number={8},
  pages={e2025EF006453},
  year={2025},
  publisher={Wiley Online Library}
}

@article{hunt2025novel,
  title={A novel explainable deep learning framework for reconstructing {S}outh {A}sian palaeomonsoons},
  author={Hunt, Kieran MR and Harrison, Sandy P},
  journal={Climate of the Past},
  volume={21},
  number={1},
  pages={1--26},
  year={2025},
  publisher={Copernicus Publications G{\"o}ttingen, Germany}
}

@article{abdullah2025explainable,
  title={Explainable {AI}-driven dew point forecasting with attention-based temporal convolutional networks},
  author={Abdullah, Md and Waheed, Sajjad and Hasan, Md Mahmodul and Morshed, Monir},
  journal={IEEE Access},
  year={2025},
  publisher={IEEE}
}

@article{zhang2025machine,
  title={Machine learning methods for weather forecasting: a survey},
  author={Zhang, Huijun and Liu, Yaxin and Zhang, Chongyu and Li, Ningyun},
  journal={Atmosphere},
  volume={16},
  number={1},
  pages={82},
  year={2025},
  publisher={MDPI}
}

@article{bano2025ai,
  title={Are {AI} weather models learning atmospheric physics? {A} sensitivity analysis of cyclone {X}ynthia},
  author={Ba{\~n}o-Medina, Jorge and Sengupta, Agniv and Doyle, James D and Reynolds, Carolyn A and Watson-Parris, Duncan and Monache, Luca Delle},
  journal={npj Climate and Atmospheric Science},
  volume={8},
  number={1},
  pages={92},
  year={2025},
  publisher={Nature Publishing Group UK London}
}

@article{chen2025toward,
  title={Toward long-range {ENSO} prediction with an explainable deep learning model},
  author={Chen, Qi and Cui, Yinghao and Hong, Guobin and Ashok, Karumuri and Pu, Yuchun and Zheng, Xiaogu and Zhang, Xuanze and Zhong, Wei and Zhan, Peng and Wang, Zhonglei},
  journal={npj Climate and Atmospheric Science},
  volume={8},
  number={1},
  pages={259},
  year={2025},
  publisher={Nature Publishing Group UK London}
}

@article{horinouchi2025statistical,
  title={Statistical prediction of tropical cyclone rapid intensification with explainable {AI}},
  author={Horinouchi, Takeshi and Yanase, Takashi and Ohta, Yuiko and Matsuoka, Daisuke and Kitamoto, Asanobu and Shimada, Udai and Yoshida, Ryuji and Fudeyasu, Hironori},
  journal={Weather and Forecasting},
  volume={40},
  number={10},
  pages={1859--1875},
  year={2025},
  publisher={American Meteorological Society}
}

@article{behnoudfar2025bridging,
  title={Bridging Idealized and Operational Models: An Explainable {AI} Framework for {E}arth System Emulators},
  author={Behnoudfar, Pouria and Moser, Charlotte and Bocquet, Marc and Cheng, Sibo and Chen, Nan},
  journal={arXiv preprint arXiv:2510.13030},
  year={2025}
}

@article{laloyaux2025using,
  title={Using data assimilation tools to dissect {GraphDOP}},
  author={Laloyaux, Patrick and Alexe, Mihai and Boucher, Eulalie and Lean, Peter and Pinnington, Ewan and Lang, Simon and Necker, Tobias and McNally, Anthony},
  journal={arXiv preprint arXiv:2510.27388},
  year={2025}
}

@article{zhou20253d,
  title={The 3{D}-Geoformer for {ENSO} studies: a Transformer-based model with integrated gradient methods for enhanced explainability},
  author={Zhou, Lu and Zhang, Rong-Hua},
  journal={Journal of Oceanology and Limnology},
  pages={1--21},
  year={2025},
  publisher={Springer}
}

@article{fear2025physics,
  title={Physics Steering: Causal Control of Cross-Domain Concepts in a Physics Foundation Model},
  author={Fear, Rio Alexa and Mukhopadhyay, Payel and McCabe, Michael and Bietti, Alberto and Cranmer, Miles},
  journal={arXiv preprint arXiv:2511.20798},
  year={2025}
}

@article{spuler2025learning,
  title={Learning predictable and informative dynamical drivers of extreme precipitation using variational autoencoders},
  author={Spuler, Fiona R and Kretschmer, Marlene and Balmaseda, Magdalena Alonso and Kovalchuk, Yevgeniya and Shepherd, Theodore G},
  journal={Weather and Climate Dynamics},
  volume={6},
  number={3},
  pages={995--1014},
  year={2025},
  publisher={Copernicus Publications G{\"o}ttingen, Germany}
}

@article{dacre2025northern,
  title={Northern hemisphere midlatitude cyclone intensity biases in machine learning weather prediction models},
  author={Dacre, HF and Charlton-Perez, AJ and Driscoll, S and Gray, SL and Harvey, B and Harvey, NJ and Hodges, KI and Hunt, KMR and Volont{\'e}, A},
  journal={Bulletin of the American Meteorological Society},
  pages={BAMS--D},
  year={2025},
  publisher={American Meteorological Society}
}

@article{loegel2025ai,
  title={The {AI} {W}eather {Q}uest: an international competition for sub-seasonal forecasting with {AI}},
  author={Loegel, Olga and Talib, Joshua and Vitart, Frederic and Hoffmann, J{\"o}rn and Chantry, Matthew},
  journal={Machine Learning: Earth},
  volume={1},
  number={1},
  pages={010701},
  year={2025},
  publisher={IOP Publishing}
}

@article{hoffman2025evaluating,
  title={Evaluating the Trustworthiness of Explainable Artificial Intelligence {(XAI)} Methods Applied to Regression Predictions of {A}rctic Sea Ice Motion},
  author={Hoffman, Lauren and Mazloff, Matthew R and Gille, Sarah T and Giglio, Donata and Heimbach, Patrick},
  journal={Artificial Intelligence for the Earth Systems},
  volume={4},
  number={1},
  pages={e240027},
  year={2025},
  publisher={American Meteorological Society}
}

@article{ye2025explainable,
  title={Explainable artificial intelligence ({XAI}) for scaling: An application for deducing hydrologic connectivity at watershed scale},
  author={Ye, Sheng and Li, Jiyu and Chai, Yifan and Liu, Lin and Sivapalan, Murugesu and Ran, Qihua},
  journal={arXiv preprint arXiv:2509.02127},
  year={2025}
}

@article{krell2025influence,
  title={The influence of correlated features on neural network attribution methods in geoscience},
  author={Krell, Evan and Mamalakis, Antonios and King, Scott A and Tissot, Philippe and Ebert-Uphoff, Imme},
  journal={Environmental Data Science},
  volume={4},
  pages={e29},
  year={2025},
  publisher={Cambridge University Press}
}

@article{schiller2025artificial,
  title={Artificial intelligence in environmental and {E}arth system sciences: explainability and trustworthiness},
  author={Schiller, Josepha and Stiller, Stefan and Ryo, Masahiro},
  journal={Artificial Intelligence Review},
  volume={58},
  number={10},
  pages={316},
  year={2025},
  publisher={Springer}
}

@article{higgs2026hybrid,
  title={Hybrid machine learning data assimilation for marine biogeochemistry},
  author={Higgs, Ieuan and Bannister, Ross and Sk{\'a}kala, Jozef and Carrassi, Alberto and Ciavatta, Stefano},
  journal={Biogeosciences},
  volume={23},
  number={1},
  pages={315--344},
  year={2026},
  publisher={Copernicus Publications G{\"o}ttingen, Germany}
}

@article{brocker2026verification,
  title={Verification of {AI}--based environmental forecasting systems: what can we do, what do we need to do, and what are the challenges?},
  author={Br{\"o}cker, Jochen and Driscoll, Simon and Necker, Tobias and Rodr{\'\i}guez, Jos{\'e} and Dacre, Helen and Harvey, Natalie and Bouall{\`e}gue, Zied Ben},
  journal={Journal of the European Meteorological Society},
  volume={4},
  pages={100032},
  year={2026},
  publisher={Elsevier}
}

@article{zhang2026baselines,
  title={From Baselines to Transport Geodesics: Axiomatic Attribution via Optimal Generative Flows},
  author={Zhang, Cenwei and Zhu, Lin and Lin, Manxi and You, Lei},
  journal={arXiv preprint arXiv:2603.05093},
  year={2026}
}

@article{moldovan2026aifs,
  title={{AIFS S}ingle 1.1.0: an update to {ECMWF}'s machine-learned weather forecast model {AIFS}},
  author={Moldovan, Gabriel and Pinnington, Ewan and Prieto Nemesio, Ana and Lang, Simon and Ben Bouall{\`e}gue, Zied and Dramsch, Jesper and Alexe, Mihai and Santa Cruz, Mario and Hahner, Sara and Cook, Harrison and others},
  journal={Geoscientific Model Development},
  volume={19},
  number={10},
  pages={4703--4724},
  year={2026},
  publisher={Copernicus Publications G{\"o}ttingen, Germany}
}

@article{vonich2026atmospheric,
  title={Atmospheric Predictability Beyond 30 Days with Machine Learning},
  author={Vonich, P Trent and Hakim, Gregory J},
  journal={Artificial Intelligence for the Earth Systems},
  pages={e260009},
  year={2026},
  publisher={American Meteorological Society}
}

@article{singh2026generative,
  title={How Do Generative Downscaling Models Use Noise? Insights from Conditional Precipitation Downscaling},
  author={Singh, Shivam and Papalexiou, Simon Michael and Abdelmoaty, Hebatallah M and Hartvigsen, Tom and Mamalakis, Antonios},
  journal={Artificial Intelligence for the Earth Systems},
  pages={e260044},
  year={2026},
  publisher={American Meteorological Society}
}

@article{tian2026evaluating,
  title={Evaluating machine learning weather models for data assimilation: Fundamental limitations in tangent linear and adjoint properties},
  author={Tian, Xiaoxu and Holdaway, Daniel and Kleist, Daryl},
  journal={Geophysical Research Letters},
  volume={53},
  number={2},
  pages={e2025GL119402},
  year={2026},
  publisher={Wiley Online Library}
}

\end{document}